\documentclass[12pt,twoside]{report}
\usepackage{amsmath}
\usepackage{amscd}
\usepackage{amssymb}
\usepackage{amsfonts}
\usepackage{eufrak}
\usepackage{fancyheadings}
\usepackage[mathscr]{euscript}
\usepackage{pstricks}
\usepackage{pst-node}
\usepackage[dvips]{pstcol}
\def \quattrova{$(\v^a, c^a,\l_a,\bc_a)~$}
\def \Qb{ Q_{\scriptscriptstyle BRS}}
\def \QBb{{\overline {Q}}_{\scriptscriptstyle BRS}}
\def \HT{{\widetilde{\mathcal H}}}
\def \MT{{\widetilde{\mathcal M}}}
\def \KT{{\widetilde{\mathcal K}}}
\def \DT{{\widetilde{\mathcal D}}}
\def \LT{{\widetilde{\mathcal L}}}
\def \Box{\p^2}
\def \HCT{\widetilde{\mathscr H}}
\def \KCT{\widetilde{\mathscr K}}
\def \DCT{\widetilde{\mathscr D}}
\def \QH{{Q_{\scriptscriptstyle H}}}
\def \QK{{Q_{\scriptscriptstyle K}}}
\def \QD{{Q_{\scriptscriptstyle D}}}
\def \QS{{Q_{\scriptscriptstyle S}}}
\def \e{\epsilon}
\def \ov{\overline}

\def \t{\theta}
\def \tb{\overline{\theta}}
\def \z{\zeta}
\def \zb{\overline{\zeta}}
\def \l{\lambda}
\def \si{\sigma}
\def \r{\rho}
\def \a{\alpha}
\def \k{\kappa}
\def \kb{\overline{\kappa}}
\def \ve{\varepsilon}
\def \b{\beta}
\def \dq{\dot{q}(\tau)}
\def \q{q(\tau)}
\def \qp{q^{\prime}(\tau^{\prime})}
\def \dqp{\dot{q}^{\prime}(\tau^{\prime})}
\def \bc{\ov{c}}
\def \w{\omega}
\def \s{\scriptscriptstyle}
\def \v{\varphi}
\def \p{\partial}
\def \QDT{Q^t_{\scriptscriptstyle D}}
\def \QBDT{\overline{Q}^t_{\scriptscriptstyle D}}
\def \QBH{{\overline{Q}}_{\scriptscriptstyle H}}
\def \QBK{{\overline{Q}}_{\scriptscriptstyle K}}
\def \QBD{{\overline{Q}}_{\scriptscriptstyle D}}
\def \QCBK{\ov{\mathscr{Q}}_{\s K}}
\def \QCBD{\ov{\mathscr{Q}}_{\s D}}

\def \QCK{\mathscr{Q}_{\s K}}
\def \QCD{\mathscr{Q}_{\s D}}
\def \QKT{Q^t_{\scriptscriptstyle K}}
\def \QBKT{\overline{Q}^t_{\scriptscriptstyle K}}
\def \NK{ N_{\scriptscriptstyle K}}
\def \ND{ N_{\scriptscriptstyle D}}
\def \NH{ N_{\scriptscriptstyle H}}
\def \NDT{ N^{t}_{\scriptscriptstyle D}}
\def \NKT{ N^{t}_{\scriptscriptstyle K}}
\def \NHB{\overline{N}_{\s H}}
\def \HS{H_{\scriptscriptstyle Susy}}

\def \scite{ \cite}

\def \LTloc{\LT_{Susy}}
\def \pslsh{\diagup{\hspace*{-.32cm} p}}
\def \bpsi{\ov{\psi}}
\def \pipsi{\Pi_{\scriptscriptstyle \psi}}
\def \pipsibar{\Pi_{\scriptscriptstyle\ov \psi}}
\def \pigi{\Pi_{\scriptscriptstyle g}}
\def \gaug{(\psi,{\ov\psi},g)}
\def \HTloc{\HT_{Susy}}
\makeatletter
\@addtoreset{equation}{section}
\makeatother

\renewcommand{\chaptermark}[1]%
               {\markboth{#1}{#1}}
\renewcommand{\sectionmark}[1]%
               {\markright{\thesection\ #1}}
\begin{document}
\thispagestyle{empty}
\begin{center}
{\large Universit\`{a} degli Studi di Trieste} \\
\vspace{.4cm}
{\large Dottorato di Ricerca in Fisica - Ciclo XIII} \\
\vspace{.4cm}
{\large a.a. 2000/2001} \\
\vspace{5cm}
{\scshape\Huge
\begin{tabular}{cc}
Supersymmetry \\
in \\
1-Dimensional Systems
\end{tabular}
}\\
\vspace{4cm}
{\Large Dottorando: dott. {\scshape Enrico Deotto}} \\
\vspace{2cm}
{\Large Relatore: prof. {\scshape Ennio Gozzi}}\\
\vspace{.15cm}
{\large (Universit\`a degli Studi di Trieste)} \\
\vspace{1.5cm}
{\Large Coordinatore: prof. {\scshape Nello Paver}}\\
\vspace{.15cm}
{\large (Universit\`a degli Studi di Trieste)}
\end{center}
\newpage
\pagestyle{fancy}\pagenumbering{roman}
\tableofcontents
\newpage
\pagenumbering{arabic}
\pagestyle{fancy}
\markboth{}{Introduction}
\setcounter{page}{1}
\chapter*{Introduction}
\addcontentsline{toc}{chapter}{\numberline{}Introduction}
\noindent
Supersymmetry is one of the most important concepts introduced in Theoretical 
Physics in the last 30 years. Despite the fact that no experimental evidence for it has been found 
so far, intensive theoretical studies on supersymmetric theories have continued for 
all these years. It is therefore reasonable to wonder what can be the main motivations
for such an amount of efforts.

The answer is twofold. Firstly Supersymmetry is  
very fascinating from the theoretical point of view, because it unifies fermions
(i.e. matter) with bosons (i.e. interaction) both in flat space ({\it
Supersymmetry}) and in curved space-time ({\it Supergravity}). 
Secondly, Supersymmetry is still a promising solution to many problems in
particle Physics, as we will see in a while.
The main goal of particle Physics in the last 50 years has been the unification of the
four interactions of Nature. Many successful models have been constructed, but it is
a matter of fact that many problems are still unsolved. In fact, while the Grand Unification 
Theories provide a satisfactory solution for the unification of electromagnetic, weak and strong forces,
the problem of how to incorporate gravity in this scheme is still an open and intriguing question.
One of the main motivations for supersymmetric theories is the hope that Supersymmetry may solve
the non-renormalizability problem of quantum gravity. In fact, as is well known, gravity is not a 
renormalizable interaction and therefore it can not be combined with the other forces which,
on the contrary, are renormalizable. An important feature of Supersymmetric Field Theories 
is that they exhibit a better ultraviolet behaviour than 
standard Quantum Field Theories.    
The reason is basically that fermions and bosons come in pairs and contribute with 
opposite sign to higher order corrections. Therefore, in order to insert gravity into the unification
picture, the strategy is to make it supersymmetric ({\it Supergravity}) in order to obtain a renormalizable 
theory. This hope is still alive even if many models of supergravity have failed to reach this goal.
However, many other promising models have been put forward, such as Superstring theories, 
M-theory, higher-dimensional models etc. The last word in this direction has not been said yet
and therefore the hope is still alive that these new supersymmetry models be renormalizable or even
finite.  

Another problem for which Supersymmetry turns out to be useful is the so-called {\it hierarchy problem}, that is
the huge disparity between the energy scale at which the GUT symmetry is broken ($\sim 10^{15}$ GeV) and
the energy scale characterizing the breaking of the electroweak symmetry (Higgs mass: $10^2\div
10^3$ GeV).  The advantage of a supersymmetric theory is that these two scales may be built into the
tree level effective potential and then, thanks to Supersymmetry, there are some 
(non-renormalization) theorems ensuring that higher-order corrections do not destroy the hierarchy. 

As we said at the beginning, the most fascinating feature of Supersymmetry is that it unifies
fermions (i.e. matter) and bosons (i.e. interaction). However the realization of such a symmetry 
was proven to be impossible in the context of standard relativistic Quantum Field Theories by a 
number of no-go theorems which forbid a direct
symmetry transformation between fields of different spin. Among these theorems, the most famous is the
celebrated Coleman-Mandula theorem \cite{ColMand}, which claims that --- in the framework of relativistic field
theories --- all the charge operators whose eigenvalues represent internal symmetries (electric
charge, isospin, hypercharge) must be Lorentz-invariant, and therefore must commute with the energy, 
the momentum and the angular momentum operators. This in turn means that irreducible multiplets of
symmetry groups cannot contain particles with different mass or different spins. The first
possibility to circumvent the Coleman-Mandula theorem was proposed by Gel'fand and 
Likhtman \cite{Gelfand} and Sakita and Gervais \cite{Gerv} in 1971, followed by Akulov and Volkov \cite{Volkov} and then by
Wess and Zumino \cite{Wess}. The way out was to give up one of the hypotheses of Coleman and Mandula, that is
the requirement that only the symmetries forming Lie groups with real parameters were admitted.
The new idea was to introduce an algebra whose elements obey {\it anticommutation relations} or,
equivalently, symmetry transformations with {\it Grassmannian parameters}.
This is precisely the case of the Susy algebra. In this manner it became 
possible to collect in the same multiplet particles of different spins provided one introduces, for each particle, 
its supersymmetric partner with equal mass
and different spin (for example the {\it photino}, the partner of the photon, with spin 
$\frac{1}{2}$ and the {\it selectron}, the partner of the electron,  with spin 0).

Unfortunately, as the experiments have shown, no boson with the mass of an electron has been found and the
same holds also for all the known particles of the Standard Model. This means that if Supersymmetry
is a true symmetry of Nature, it must be broken at some energy scale. Therefore many models have
been put forward, since the mid seventies, to provide satisfactory mechanisms to break Susy in a manner consistent
with phenomenology. It was in this context that people tried to test the breaking
mechanism on models simpler than the phenomenological ones. 
In particular Supersymmetry was first studied in a nonrelativistic framework by Witten \cite{Witten}
who invented a toy model called Supersymmetric Quantum Mechanics (Susy-QM).  

It became clear soon
that this model was interesting in its own right, not just as a toy model for testing field theory
methods. For example, 
the introduction by Witten \cite{Witten2} of a topological invariant (the Witten index) to study the
breaking of Susy gave rise to a lot of interest in the geometrical and topological aspects of Susy-QM. Many
properties of the Witten index were analyzed: for instance it was possible \cite{Alvarez} to give 
a proof of the
Atiyah-Singer index theorem by use of the supersymmetric representation of the index theorem.

In Atomic Physics, Kosteleck\'y at al. \cite{Kostel} discussed the relationship between the
physical spectra of different atoms and ions by use of Susy-QM. For example they suggested that the
helium and hydrogen spectra come from Susy partner potentials.

In Nuclear Physics Supersymmetry was applied \cite{Iachello} to obtain relations between the spectra of
even-even and neighbouring even-odd nuclei; it was shown \cite{Baye} that the deep and shallow
nucleus-nucleus potentials are supersymmetric partners. 

In the general context of QM it was shown \cite{GozSusy}\cite{Cooper} that the factorization method of
Schroedinger, Infeld and Hull \cite{Hull} has its roots in a sort of SUSY-QM which one can build out of any
QM model.
SUSY-QM made its appearance also in the context of stochastic processes \cite{Parisi} where it was shown to
be linked to the interplay between forward and backward Fokker-Planck dynamics \cite{GozEn}. 

Therefore SUSY-QM has turned out to be a model which makes its appearance in a large variety of areas
in physics. It somewhat 
indicates a sort of {\it universality} of this 1-dimensional supersymmetry. However, a really universal supersymmetry 
is obtained in a model introduced some time ago \cite{Ennio} which is related to Classical
Mechanics (CM) and not to Quantum Mechanics. 
At first sight this is very surprising because one could ask which are the fermionic variables
in CM. The issue appears less ``bizarre" if one remembers that in differential geometry some objects
have been introduced, whose character is indeed anticommuting. They are the basis $dx^i$ 
of the cotangent space $T^*{M}$ to a given manifold $M$, and it is well known that the product between
these objects is the {\it wedge}-product which is anticommuting. 
It can be shown that the geometric object which determines the evolution of a generic function of $x^i$ 
and $dx^i$ is the {\it Lie derivative} along the Hamiltonian flow. This object somehow treats
at the same level both $x$ and $dx$ and exhibits a symmetry which exchanges these variables. This
symmetry obviously turns a bosonic variable $x$ into a fermionic variable $dx$ and in this sense it 
is a (classical) supersymmetry. Moreover it is a {\it universal} supersymmetry, as we have
already said above, because any classical evolution is described via a Lie derivative and this
object always exhibits this 1-dimensional supersymmetry. This universality is something we consider very important 
because it may tell us that supersymmetry is a wider phenomenon than we previously thought. It is anyhow
a phenomenon strictly intertwined with geometry and which may be difficult to detect at the pure
phenomenological level of particle physics.

Originally the formalism we have just mentioned was born out of an attempt to provide a {\it path integral} formulation
for Classical Mechanics \cite{Ennio}, which turned out to be the classical counterpart of the operatorial
approach to CM pioneered by Koopman and Von Neumann \cite{Koop}.
This path integral, which we will indicate as CPI\footnote{``CPI" is the acronym of ``Classical Path Integral".}, exhibits 
a lot of universal symmetries, some of which were understood geometrically
\cite{Ennio} thanks to the tools provided by the Cartan Calculus of symplectic spaces \cite{Marsd}. 

A symmetry which is present in the model, but not thoroughly analyzed geometrically is 
supersymmetry. In this thesis we will first do that and we will show that the universal 
supersymmetry of Classical Mechanics is strictly related to the geometrical concept of 
{\it equivariant cohomology}, which is a concept first introduced by Cartan \cite{Cartan} and developed by 
other people \cite{Stora}\cite{Berline}\cite{Blau}.   
The strategy we will use is to make local the global Classical Susy present in the CPI-Lagrangian
and then study the physical state conditions associated to this local invariance.
The {\it physical} states turn out to be those belonging to the equivariant cohomological classes. The
analysis we perform will also be useful as the starting point for the study of the geometry of the space
of classical orbits.

In this thesis we shall also present other modifications of the original CPI formalism.
Some of them  will help us to make contact with structures (like the $\kappa$-symmetry 
of Siegel) found in many supersymmetric theories like the relativistic superparticles, 
strings and branes.  Other modifications will allow to study structures typical of Classical Mechanics
like the surfaces of constant energy.

Besides these extensions of the CPI, in this thesis we have also tried to find new universal symmetries
not previously discovered in the literature \cite{DG}. One symmetry, whose generator will 
be indicated by $\mathscr{Q}_{\s S}$, has the effect of rescaling the overall action $\widetilde{S}$ 
which enters the CPI. It is a symmetry because the equations of motion are left invariant by a 
rescaling of $\widetilde{S}$. However, it is a symmetry which cannot be implemented by 
canonical commutators on the phase space $\MT$ of the CPI. The reason is that a rescaling 
of the action $\widetilde{S}$ is not a canonical transformation. However, 
its action is better understood if we make use of the concept of superspace.
In fact, like any 
supersymmetric theory, the CPI formalism can be represented in a suitable superspace, which is 
composed by the time $t$ and by two Grassmannian partners $\t$ and $\tb$. This superspace $(t,\t,\tb)$
is like the base space for the target space $\MT$ of our theory. The charge $\mathscr{Q}_{\s S}$ performs
a rescaling of $(\t,\tb)$ while its action on $\MT$ is non-canonical. This transformation seems to
be the counterpart, at the level of the CPI, of a symmetry discovered before \cite{GozMSA} in the context of the 
quantum path integral. That transformation, which we denote by MSA\footnote{From 
``Mechanical Similarity Anomaly". The name ``Mechanical Similarity" was first introduced, even if in a different
context, by Landau in Ref.\cite{Landau}
and the term ``Anomaly" refers to the fact that it is broken in the passage from the classical to
the quantum level.}, is a transformation of the coordinates of the standard phase space $\cal M$ and was shown
to be a non-integrable one because it is {\it path dependent} (and not reducible to a diffeomorphism).
It was also shown that at
the quantum mechanical level this classical symmetry is anomalous because of the presence of $\hbar$.
In the context of the CPI we shall show that also the transformation induced by $\mathscr{Q}_{\s S}$ cannot be reduced to 
a diffeomorphism in the phase space $\MT$. 
The transformation which is closer to the $\mathscr{Q}_{\s S}$-rescaling is that known as {\it superconformal}
transformation.
In the last part of the thesis we shall present a model which exhibits a kind of superconformal 
invariance derived from the classical Susy of the CPI. Our hope is that this particular model 
may be a playground to tackle the more general problems of the transformations induced by $\mathscr{Q}_{\s S}$.   

This thesis is organized as follows.

In Section 1 we shall introduce the formalism of the Path Integral for Classical Mechanics. We will 
show how to derive it starting from the Hamilton equations, and we will describe all the 
symmetries it possesses and their geometrical meaning by use of the Cartan Calculus.

In Section 2 we shall focus on the classical supersymmetry of the CPI and in particular we will 
analyze its geometrical meaning. We will modify the CPI-Lagrangian and we will obtain a gauge theory which
exhibits the Classical Susy as a local symmetry. Next we will show how the physical Hilbert space
is connected with the geometric concept of
equivariant cohomology and with the geometry of the space of the classical trajectories. 
Besides the meaning of the classical supersymmetry, we shall discuss also another topic: the problem of
constraining the CPI-formalism on the hypersurfaces of the phase space $\cal M$ characterized by a
fixed value $E$ of the energy. In other words we will insert in the CPI the constraint $H(\v)-E=0$ and
we will analyze the associate gauge theory. We will see that the original $N=2$ classical
supersymmetry reduces to a $N=1$.

In Section 3 we will continue the analysis of the symmetries of the CPI. We shall show that, if we make local
the symmetries generated by the operators associated to the covariant
derivatives of the SUSY, we obtain a local (super)symmetry which is very similar to
the $\kappa$-symmetry of Siegel. 

In Section 4 we will introduce the symmetry generated by $\mathscr{Q}_{\s S}$ and we will analyze it both at the level
of superspace and at the level of the phase space $\MT$. We shall show that this symmetry is somewhat
similar to the MSA-transformation introduced above, which is also analyzed in detail.  

In Section 5 we will introduce 
the superconformal invariance in the CPI. To do that, we will build the Lie derivative of a model
introduced long time ago by De Alfaro, Fubini and Furlan \cite{DFF} which exhibits a
particular kind of conformal invariance. Since the Lie derivative is automatically
supersymmetric, we will get a model which combines the invariance under conformal symmetries with
the invariance under susy: in other words we will obtain the invariance under the so-called
superconformal transformations. 
To obtain all the symmetry group at the level of $\MT$, we will build the Lie derivatives associated to
all the conformal generators of the theory. Moreover we will construct also the ``square root" of the charges
associated to each of the previous generators and we will calculate the algebra realized by all these
operators. They close on a new kind of superconformal algebra.

At the end some conclusions summarize the work done and indicate the open problems ahead. 
Few appendices contain some detailed calculations omitted in the previous sections. 
 
\pagestyle{fancy}
\chapter*{1. The Path Integral for Classical Mechanics}
\addcontentsline{toc}{chapter}{\numberline{1}The Path Integral for Classical Mechanics}
\setcounter{chapter}{1}
\markboth{1. The Path Integral for Classical Mechanics}{}

\vspace{1cm}
\noindent
In this chapter we shall describe the formalism of the Classical Path Integral (CPI) which was originally 
developed in \cite{Ennio}, where the interested reader can find all the details omitted in the present 
account. The idea originated from the fact that whenever a theory has an operatorial
formulation, it must also possess a corresponding path integral. Now Classical Mechanics (CM) does have 
an operatorial formulation which was proposed long ago by Koopman and von-Neumann\scite{Koop}. This 
operatorial approach describes the time-evolution of phase-space distributions by means of the {\it Liouville} 
operator $\widehat{L}=-\w^{ab}\p_bH\p_a$ where $\w^{ab}$ is the symplectic matrix. 
Therefore, it is reasonable to look for the corresponding path integral formalism and the strategy 
to follow is described in the next section.  

\section{The Classical Kernel}

\noindent Classical Hamiltonian Mechanics describes physical systems in a $2n$-dimensional\break phase
space ${\cal M}$, whose coordinates we will denote by $\varphi^{a}$ $(a=1,\ldots 2n)$,\break i.e.: 
$\varphi^{a}=(q^1\cdots  q^n,p^1\cdots p^n)$. If we call 
$H(\varphi)$ the Hamiltonian of the system, the equations of motion (i.e. the Hamilton equations) are:
	\begin{equation}
	\label{CPI1}
	{\dot\varphi }^{a}=\omega^{ab}{\partial H\over\partial\varphi ^{b}}\equiv\w^{ab}\p_b H(\v) ~~~~~
	\text{$\w^{ab}=$ symplectic matrix}. 
	\end{equation}
\noindent 
We can define the {\it classical} probability $ K_{cl}(f|i)$ to reach a final configuration $\v_f$ at time 
$t_f$ given an initial configuration $\v_i$ at time $t_i$ in the following way:
	\begin{equation}
	\label{CPI2}
	K_{cl}(f|i)=\delta(\varphi^{a}_{f}-\phi^{a}_{cl}(t_{f}|\varphi_{i},t_{i}))
	\end{equation}
\noindent where $\phi^{a}_{cl}(t_{f}|\varphi_{i},t_{i})$ is the classical trajectory (that is the
solution of the Hamilton equations) having $\varphi_{i}$ as initial condition at time $t_i$. 
Since $K_{cl}(f|i)$ is a classical
probability, we can rewrite it as a sum over all the possible intermediate configurations:
	\begin{equation}
	\label{CPI3}
	\begin{array}{rl}
	K_{cl}(f|i) & =\displaystyle\sum_{k_{i}} K_{cl}(f|k_{N-1})K_{cl}(k_{N-1}|k_{N-2})\cdot...\cdot 
	K_{cl}(k_{1}|i)
	\vspace{.2cm} \\
	&=\displaystyle\prod^N_{j=1}\int
	d^{2n}\varphi_{j}~\delta^{(2n)}[\varphi^{a}_{j}-\phi^{a}_{cl}(t_{j} 
	|\varphi_{j-1},t_{j-1})] \vspace{.2cm}\\
	\xrightarrow{N\rightarrow\infty} &=\displaystyle\int{\mathscr
	D}\varphi~\tilde{\delta}[\varphi^{a}(t)-\phi^{a}_{cl}(t)]
	\end{array} 
	\end{equation}
\noindent where in the first equality $k_i$ denotes formally an intermediate configuration $\v_{k_i}$ between 
$\v_{i}$ and $\v_{f}$ and the symbol $\tilde{\delta}({\s\ldots})$ represents a {\it functional} Dirac delta, that is a 
product of an infinite number of Dirac deltas, each one referring to a different time $t$ along the classical trajectory.
The last formula in (\ref{CPI3}) is already a path integral but it does not have the usual form (the integral 
of the exponential of an action) we are used to see in Quantum Mechanics. Anyway we can give it that
form if we rewrite the functional Dirac delta as:
	\begin{equation}
	\label{CPI4}
	{\tilde\delta}[\varphi -\varphi _{cl}]={\tilde\delta}[{\dot\varphi 
	^{a}-\omega^{ab}
	\partial_{b}H]~\det [\delta^{a}_{b}\partial_{t}-	\omega^{ac}\partial_{c}\partial
	_{b}H}]
	\end{equation}
\noindent where we have used the functional analog of the relation
$\delta[f(x)]=\frac{\delta[x-x_i]}{\Bigm|\frac{\partial f}{\partial 
x}\Bigm|_{x_i}}$. 
Then (see Refs.\cite{Ennio} for the details) we can exponentiate both the two terms of the RHS of Eq.(\ref{CPI4}):  
the first via a Lagrange multiplier $\lambda_a$ and the second (the determinant) by use of two families of Grassmannian 
variables ($c^a$,$\bc_a$), as one usually does in the Faddeev-Popov technique in gauge theories
\cite{Teitel}. We finally obtain the following expression:
	\begin{equation}
	\label{CPI5}
	K_{cl}(f|i)=\int_{\v_i,t_i}^{\v_f,t_f}{\mathscr D}\varphi ^{a}{\mathscr D}\lambda_{a}{\mathscr D}
	c^{a}{\mathscr D}{\ov c}_{a}~\exp\biggl[i\int dt~{\widetilde{\cal L}}\biggr]
	\end{equation} 
\noindent where $\widetilde{\cal L}$ is the Lagrangian characterizing the CPI:
	\begin{equation}
	\label{CPI6}
	{\widetilde{\cal L}}=\lambda_{a}[{\dot\varphi }^{a}-	\omega^{ab}\partial_{b}H]+
	i{\ov c}_{a}[\delta^{a}_{b}\partial_{t}-	\omega^{ac}\partial_{c}\partial_{b}H]
	c^{b}.
	\end{equation}
\noindent From (\ref{CPI5}) we can pass to the classical generating functional $Z_{\s CM}$
 	\begin{equation}
	\label{CPI7}
	Z_{\s CM}[J_{\v}, J_{\l}, J_c, J_{\bc}] =\int{\mathscr D}\varphi ^{a}{\mathscr D}\lambda_{a}{\mathscr D}
	c^{a}{\mathscr D}{\ov c}_{a}~\exp\biggl[i\int dt~\big(\widetilde{\cal L}
	+J_{\v}\v + J_{\l}\l + J_c c + J_{\bc}\bc\big)\biggr],
	\end{equation}
\noindent 
where $J_{\v}, J_{\l}, J_c, J_{\bc}$ are the currents associated to the $8n$ variables $(\v^a,\l_a,c^a,\bc_a)$
which characterize 
the new {\it enlarged} phase space, which from now on we will denote by $\widetilde{\cal M}$. We will come back in a
while on these things: for our purposes here, the important thing is the form 
of the Lagrangian $\LT$ appearing in (\ref{CPI6}): 
we can easily Legendre-transform it obtaining the corresponding Hamiltonian:
	\begin{equation}
	\label{CPI8}
	\widetilde{\cal H}=\lambda_a\omega^{ab}\partial_bH+i\ov{c}_a\omega^{ac}
	(\partial_c\partial_bH)c^{b}.
	\end{equation}
\noindent The Lagrangian $\LT$ and the Hamiltonian $\HT$ will be extensively used in the following sections, where we 
will analyze some nice geometrical features they possess.
\section{The enlarged phase space $\MT$}
\noindent 
In the previous section we have seen that the phase space which characterizes the Classical Path Integral is larger 
than the original one (whose variables are $\v^a$), and is composed by the following $8n$ variables: \quattrova . This phase space is  
endowed with both a symplectic and a commutator structure. 
The first one is defined as:
	\begin{equation}
	\label{CPI9}
	\{\v^{a},\l_{b}\}_{\s EPB}
	=\delta^{a}_{b};~~~~\{ c^a, \bc_{b}\}_{\s EPB}=-i\delta^{a}_{b} ~~~~~\text{(all others are zero)}
	\end{equation}
\noindent
where the subscript ``$EPB$" means {\it Extended Poisson Brackets} and the name has been chosen 
to emphasize the difference with the usual 
Poisson Brackets. The EPB allow to express the equations of motion\footnote{Obtained from the Euler equations
associated to the Lagrangian $\LT$.} of the \quattrova variables
in a Hamiltonian form  (where $\HT$ is the {\it extended} Hamiltonian). 
The second structure mentioned above is based on the definition of commutator 
given by Feynman in the quantum case: given a path integral characterized by the generating functional $Z$, we
define the 
commutator between two functions $O_1$ and $O_2$ as the following (graded) expectation value:
	\begin{equation}
	\label{CPI10}
	\langle[O_{1}(t),O_{2}(t)]\rangle_{\s Z}\equiv  \lim_{\epsilon\rightarrow 0}
	\langle O_{1}(t+\epsilon)O_{2}(t)\pm O_{2}(t+\epsilon)O_{1}(t)\rangle_{\s Z}. 
	\end{equation} 
\noindent If we use the same strategy with the $Z_{\s CM}$ in (\ref{CPI7}), we get that the only commutators
different 
from zero are:
	\begin{equation}
	\label{CPI11}
	\langle\big[\v^{a},\l_{b}\big]\rangle_{\s Z}
	=i\delta^{a}_{b};~~~~\langle\big[ c^a, \bc_{b}\big]\rangle_{\s Z}=\delta^{a}_{b} 
	\end{equation} 
\noindent which are clearly isomorphic to the $EPB$ in (\ref{CPI9}).
The information we get either from (\ref{CPI9}) or from (\ref{CPI11}) is that the two pairs $(\v,\l)$ and $(c,\bc)$
are 
couples of conjugate variables. In the following we choose to consider $(\v,c)$ as forming the {\it enlarged 
configuration space} while we consider $(\l,\bc)$ as the conjugate momenta.
This interpretation allows us to realize the commutators (\ref{CPI11}) via the following operatorial
representations:
	\begin{align}
	& \widehat{\v^a}\,\rho(\v,c)=\v^a\cdot\rho(\v,c); & 	\widehat{\l_a}\,\rho(\v,c)=-i\displaystyle
	\frac{\p}{\p\v^a}\rho(\v,c); \vspace{.15cm} \label{CPI12}\\ 
	& \widehat{c^a}\,\rho(\v,c)=c^a\cdot\rho(\v,c); &
	\widehat{\bc_a}\,\rho(\v,c)=\displaystyle\frac{\p}{\p c^a}\rho(\v,c);
	\label{CPI13}
	\end{align}
\noindent where $\rho(\v,c)$ is the analog of the wave function in the quantum case and, as we will see in a 
while, represents a generalization of the Koopman-Von Neumann wavefunction. If we substitute the
differential 
representations (\ref{CPI12}) and (\ref{CPI13}) in the Hamiltonian $\HT$ in (\ref{CPI8}), we get that
$\rho(\v,c)$ evolves according to the following equation:
	\begin{equation}
	\label{CPI14} 
	{\widehat{\widetilde{\cal H}}}\rho(\v,c;t)=i\frac{\p}{\p t}\rho(\v,c;t).
	\end{equation}
\noindent 
where $\widehat\HT$ is:
	\begin{equation}
	\label{CPI15}
	{\widehat{\widetilde{\cal H}}}\equiv -	i\omega^{ab}\partial_{b}H\frac{\p}{\p\v^a}
	-i\w^{ab}\p_b\p_d H c^{d}\frac{\p}{\p c^a}.
	\end{equation} 
\noindent 
Because of the Grassmannian character of the $c$'s, we can 
expand $\rho(\v,c;t)$ as:
	\begin{multline}
	\label{CPI16} 
	\rho(\v,c;t)= \rho_{\s 0}(\v;t)+ \rho_a(\v;t)c^a +
	\frac{1}{2}\rho_{ab}(\v;t)c^a c^b + \\ 
	+\ldots + \frac{1}{(2n)!}\rho_{ab\ldots d}(\v;t)c^ac^b\ldots c^d.
	\end{multline}
\noindent
Now we can substitute Eq.(\ref{CPI16}) in (\ref{CPI14}) and equate separately all the terms with the same number of
$c$'s.  
The first term in this expansion, $\rho_{\s 0}(\v)$, turns out to be the solution of the Liouville equation 
	\begin{equation}
	\label{CPI17} 
	\frac{\p}{\p t}\rho_{\s 0} = -\w^{ab}\p_bH\p_a\rho_{\s 0},
	\end{equation}
\noindent
because the bosonic part of $\widehat{\HT}$ is precisely (apart from a factor $i$) the Liouville operator
	\begin{equation}
	\label{CPI18}
	{\widehat{\widetilde{\cal H}}}_{\scriptscriptstyle B}\equiv 
	-i\omega^{ab}\partial_{b}H\partial_{a}.
	\end{equation}
\noindent  
Thus we see that we recover the Koopman-Von Neumann formalism as the zero-ghost sector of the CPI-formalism. The
next 
step is then to understand the remaining (i.e. with a non-zero number of ghosts $c^a$) terms. 
To do that we must give the ghosts $c^a$ a 
physical meaning and the best strategy is focusing on the equations of motion of these variables:
\begin{equation}
\label{CPI19} 
\partial_{t}c^a-\omega^{ab}\partial_{b}\partial_{d}H\,c^{d}=0.
\end{equation}
\noindent
This is precisely the equation of motion of the Jacobi fields\footnote{ The Jacobi fields are also called 
``first variations". They describe the evolution in time of the difference between two close classical
trajectories in the phase space $\MT$. See Ref.\cite{Schulman} for further details.} $\delta\v^a$. This leads us to
an interpretation 
which may seem a little heuristic at first sight, but which will be very illuminating in the sequel: we interpret 
the ghosts $c^a$ as the basis $d\v^a$ of the cotangent space $T^*{\cal M}$. According to this prescription the 
anticommuting product between the ghosts becomes the $\wedge$-product between the 1-forms $d\v^a$ and the function 
$\rho(\v,c;t)$ in (\ref{CPI16}) becomes       
	\begin{equation}
	\label{CPI20}
	\rho(\v,c;t)\longrightarrow \rho(\v;t) = \rho_{\s 0}(\v;t)+ \rho_a(\v;t)d\v^a +
 	\frac{1}{2}\rho_{ab}(\v;t)d\v^a 
	\wedge d\v^b + \ldots 
	\end{equation}
\noindent which is an inhomogeneous differential form on the phase space $\cal M$. As we will see better in the 
following section, the Hamiltonian $\widehat\HT$ in (\ref{CPI15}) can then be interpreted as the {\it
Lie-derivative} along
the Hamiltonian vector field $h=\w^{ab}\p_bH\p_a$, and the Schr\"odinger-type equation (\ref{CPI14}) describes
the 
time evolution (induced by the Hamilton equations) of a generic inhomogeneous differential form.  

We conclude this section with few words about the geometry of the enlarged phase
space $\MT$. We have just seen that
the ghosts $c^a$ evolve as the basis of the cotangent space $T^*{\cal M}$, but they also transform as the
components of the tangent vectors:
$V^{a}(\varphi){\partial\over\partial\v^{a}}$.
The space $(\varphi^{a}, c^{a})$, which we have simply denoted by {\it enlarged configuration space} is called more 
properly in Ref.\cite{Schw} 
the {\it reversed-parity} tangent bundle and it is indicated as $\Pi T{\cal M}$.
The ``{\it reversed-parity}" specification
is because the $c^{a}$ are Grassmannian variables. Since the $(\lambda_{a}, 
{\ov c}_{a})$ are the ``momenta" of the previous 
variables (see Eq.(\ref{CPI11})) we conclude that the $8n$ variables
\quattrova~span the cotangent bundle to the
reversed-parity tangent bundle:~$T^{*}(\Pi T{\cal M})$; in other words we can write $\MT= T^{*}(\Pi T{\cal M})$.
So $\MT$ is a cotangent bundle and this is the reason why it has a 
Poisson structure. For more details about this we refer the interested
reader to Ref.\cite{Geom}.
\section{Symmetries of the CPI and the Cartan calculus}
\label{sec:sym}

\noindent 
In this section we analyze the symmetries of the Hamiltonian $\HT$ and we focus in particular on their geometrical 
meaning. In fact the Hamiltonian $\HT$ is invariant under some {\it universal global} symmetries: {\it universal} 
because the transformations which leave $\HT$ invariant do not depend on the particular form of the original 
Hamiltonian $H(\v)$, and therefore are symmetries of {\it all} the classical systems; {\it global}, on
the other side, is used in 
opposition to {\it local}, because the symmetries we are going to describe are not gauge symmetries. We will 
gauge some of these symmetries in the next chapter and we will see how we can gain an insight into their 
meaning; anyway for the moment no gauge invariance is present.

We can divide the generators (through the commutators in (\ref{CPI11})) of these symmetries in the following two
groups:
	\begin{gather}
	\text{\bf Geometrical Symmetries} \nonumber\vspace{.15cm} \\
	\Qb \equiv i c^{a}\lambda_{a} \label{CPI21} \\
	\QBb\equiv i {\ov c}_{a}\omega^{ab}\lambda_{b} \label{CPI22} \\  
 	Q_{g} \equiv c^{a}{\ov c}_{a}  \label{CPI23}\\
 	K\equiv {1\over 2}\omega_{ab}c^{a}c^{b} \label{CPI24}\\
 	{\ov K} \equiv {1\over 2}\omega^{ab}{\ov c}_{a}{\ov c}_{b} \label{CPI25}  
	\end{gather}
	\begin{gather}
	\text{\bf Dynamical Symmetries} \nonumber\vspace{.15cm} \\
	\NH\equiv c^{a}\partial_{a}H \label{CPI26} \\
	\NHB\equiv \bc_a\w^{ab}\p_b H \label{CPI27}  \\
	\QH \equiv\Qb - \beta\NH = c^a (i\l_a-\beta\p_aH) \label{CPI28} \\
	\QBH \equiv\QBb + \beta\NHB = \bc_a\w^{ab}(i\l_b+\beta\p_bH)  \label{CPI29} \\
	\text{($\beta$=dimensional parameter)}\nonumber
	\end{gather}
\noindent
We called the first group ``Geometrical" because the functional form of all its components does not contain $H(\v)$:
we will see in a while that they correspond to some precise operations in differential geometry. On the other side, 
the second group is named ``Dynamical" because its components depend formally on $H(\v)$; nevertheless they 
remain {\it universal}, because their form is the same for any classical system. 
Consider the first group: it is not difficult to check that the last three charges $(Q_g, K, \ov{K})$ realize the 
algebra $\mathfrak{sp}(2)$ while the first two $(\Qb,\QBb)$ provide the inhomogeneous part (this will be clearer in 
the following section) in such a way that the overall algebra of the first set of charges is $\mathfrak{isp}(2)$. 

About the second group the most interesting charges are the last two, namely $\QH$ and $\QBH$. In fact they are 
both nilpotent, they commute with $\HT$ and their (graded) commutator is:
	\begin{equation}
	\label{CPI30}
	\big[\QH,\QBH \big]=2i\beta\HT;
	\end{equation}     
	\noindent
this means that we have an $N=2$ real (i.e. $N=1$ complex) non-relativistic supersymmetry. This is precisely the 
classical supersymmetry we have mentioned in the Introduction.

In order to shed some light on the geometrical meaning of the charges defined in (\ref{CPI21})-(\ref{CPI29}), we 
can exploit the correspondence $c^a\longleftrightarrow d\v^a$ introduced in the previous section together
with the differential representations of \quattrova given in Eqs.(\ref{CPI12})(\ref{CPI13}). 
For example it is not difficult to see that:
	\begin{align}
	&\widehat{\Qb}=i\hat{c}^a\hat{\l}_a && \longleftrightarrow && d=d\v^a\p_a && (\text{exterior differential})
	\label{CPI31}\\
	&\widehat{K}=\displaystyle\frac{1}{2}\w_{ab}\,\hat{c}^a\hat{c}^b && \longleftrightarrow 
	&&\w=\displaystyle\frac{1}{2}\w_{ab}\,d\v^a\wedge d\v^b 
	&&(\text{symplectic 2-form}) \\
	&\widehat{\NHB}=\hat{\bc}_a\w^{ab}\p_bH(\hat{\v}) && \longleftrightarrow && \iota_h &&(\text{interior
	contraction w.r.t. $h$}) \label{CPI33}
	\end{align}
\noindent
In particular, since the Hamiltonian $\HT$ can be written as:
	\begin{equation}
	\label{CPI34}
	\HT=-i\big[\Qb,\NHB\big]
	\end{equation}
\noindent
we have the following interpretation for $\HT$:
	\begin{equation}
	\label{CPI35}
	i\HT=\big[\Qb,\NHB\big]\longleftrightarrow d\iota_h+\iota_hd = \mathscr{L}_h
	\end{equation}
\noindent
where $\mathscr{L}_h$ is the {\it Lie-derivative} along the Hamiltonian vector field
$h$, as we had anticipated in the previous section\footnote{It had already been
noted in Ref.\cite{WittenDG} that the Lie-derivative has a supersymmetric structure. What was missing
there was the detailed Cartan Calculus we present here.}.

We end this section giving a more rigorous formalization of the correspondence between the charges in $\MT$ and 
the Cartan Calculus \cite{Cartan}\cite{Marsd}.  

First of all, since under the Hamiltonian evolution $c^{a}$ transforms  
as the basis of 1-forms $d\varphi^{a}$ and  $\ov{c}_a$  transforms  as the basis 
of  vector fields\footnote{Note that $\lambda_{a}$
does not seem to transform as a vector field even if it 
can be  interpreted as $\frac{\partial}{\partial \varphi^{a}}$. The
explanation of this fact is given in  the second paper of Ref.\cite{Geom}.},
(see Eqs.(\ref{CPI12})-(\ref{CPI13})), we can build the following map, 
called ``hat" map $\wedge$:
	\begin{align}
	\label{CPI36}
	\alpha=\alpha_{a}d\varphi ^{a}~~~~&\hat{\longrightarrow}~~~~{\widehat\alpha}\equiv
	\alpha_{a}c^{a}\\
	\label{CPI37}
	V=V^{a}\partial_{a} ~~~~&\hat{\longrightarrow}~~~~{\widehat V}\equiv V^{a}{\ov
	c}_{a}
	\end{align}  
\noindent It is  actually a much more general map between forms $\alpha$, 
antisymmetric tensors $V$ and functions of $\varphi, c, \ov{c}$:
	\begin{align}
	\label{CPI38}
	F^{(p)}={1\over p !}F_{a_{1}\cdots a_{p}}d\varphi ^{a_{1}}\wedge\cdots\wedge
	d\varphi ^{a_{p}} ~~ & \hat{\longrightarrow}~~~{\widehat F}^{(p)}\equiv {1\over p!}
	F_{a_{1}\cdots a_{p}}c^{a_{1}}\cdots c^{a_{p}}\\
	\label{CPI39}
	V^{(p)}={1\over p!}V^{a_{1}\cdots a_{p}}\partial_{a_{1}}\wedge\cdots\wedge 
	\partial_{a_{p}} ~~& \hat{\longrightarrow}~~~{\widehat V}\equiv {1\over 
	p!}V^{a_{1}
	\cdots a_{p}}{\ov c}_{a_{1}}\cdots {\ov c}_{a_{p}}
	\end{align} 
\noindent Once the correspondence (\ref{CPI36})-(\ref{CPI39}) is 
established,  we can easily find  out what corresponds to the various Cartan operations 
like  the exterior derivative ~$d$~
of a form, the  interior contraction $\iota_{{\scriptscriptstyle V}}$
between a vector field $V$ and a form $F$ and
the multiplication of a form by its form number \scite{Ennio}  :
	\begin{eqnarray}
	\label{CPI40}
	dF^{(p)} & \hat{\longrightarrow} & [Q_{\scriptscriptstyle BRS},{\widehat 
	F}^{(p)}] \\
	\label{CPI41}
	\iota_{{\scriptscriptstyle V}}F^{(p)} & \hat{\longrightarrow} & [{\widehat V},
	{\widehat F}^{(p)}]
	\\
	\label{CPI42}
	pF^{(p)} & \hat{\longrightarrow} & [Q_{g}, {\widehat 
	F}^{(p)}]
	\end{eqnarray} 
\noindent where $Q_{\scriptscriptstyle BRS}, \, Q_g$ are the charges in 
(\ref{CPI21})-(\ref{CPI23}).
In the same manner we can translate in our language the usual 
mapping\scite{Marsd} between vector fields $V$ and forms $V^{\flat}$ realized by the 
symplectic 2-form $\omega(V,.)\equiv V^{\flat}$,
or the inverse operation of building a vector field $\alpha^{\sharp}$ out of a 
form:
$\alpha=(\alpha^{\sharp})^{\flat}$. These operations can be translated in our
formalism as follows:
	\begin{eqnarray}
	\label{CPI43}
	V^{\flat} & \hat{\longrightarrow} & [K,{\widehat V}]\\
	\label{eq:trentatre}
	\alpha^{\sharp} & \hat{\longrightarrow} & [{\ov
	K},{\widehat\alpha}]
	\end{eqnarray}  
\noindent where again $K, \ov{K}$ are the charges 
(\ref{CPI24}) and (\ref{CPI25}) respectively.
We can also translate the standard operation of building a
Hamiltonian vector field, indicated as $(df)^{\sharp}$, out of a
function ~$f(\varphi)$, and also the Poisson Brackets between two 
functions $f$ and $g$:
	\begin{eqnarray}
	\label{CPI44}
	(df)^{\sharp} & \hat{\longrightarrow} & [{\ov Q}_{\scriptscriptstyle 
	BRS},f]\\
	\label{CPI45}
	\{f,g\}_{\scriptscriptstyle PB}=df[(dg)^{\sharp}] & \hat{\longrightarrow} & 
	[[[f,Q_{\scriptscriptstyle BRS}],{\ov K}],
	[[[g,Q_{\scriptscriptstyle BRS}],{\ov K}],K]]
	\end{eqnarray}  
\noindent Finally we can translate the concept of Lie derivative $\mathscr{L}_V$ 
along a generic vector field $V$; from the definition 
	\begin{equation}
	\label{CPI45b} 
	{\mathscr L}_{\scriptscriptstyle V}=d\iota_{\scriptscriptstyle V}
	+\iota_{\scriptscriptstyle V}d
	\end{equation}
\noindent it is easy to prove that:
	\begin{equation}
	\label{CPI46}
	{\mathscr L}_{\scriptscriptstyle V}F^{(p)} \;\; \hat{\longrightarrow} \;\; 
	i[{\widetilde {\cal H}}_{\scriptscriptstyle V},{\widehat F}
	^{(p)}]
	\end{equation} 
\noindent where ${\widetilde {\cal H}}_{\scriptscriptstyle 
V}=\lambda_aV^{a}+i\ov{c}_a
\partial_bV^{a}c^{b}$. Note that,
for $V^{a}=\omega^{ab}\partial_bH$,  the ${\widetilde {\cal
H}}_{\scriptscriptstyle V}$ becomes the ${\widetilde {\cal H}}$ of (\ref{CPI8}) and (\ref{CPI15}).
\section{The CPI-formalism in superspace}
\label{sec:superspace}

\noindent In the previous section we have seen that the formalism of the Classical
Path Integral involves a universal $N=2$ supersymmetry. It is very interesting to
give a superspace formulation of the model because we can see that in this context
there is a nice correspondence between the classical and the quantum domain of the
same physical system. 

First of all we must remember that we are dealing with a {\it non relativistic}
supersymmetry and therefore our base space $t$ enlarges to a superspace $(t,\t,\tb)$
where $\t$ and $\tb$ are two (scalar) Grassmannian partners of the time $t$. It is
then possible to introduce a superfield which collects all the $8n$-variables of
$\MT$:
	\begin{equation}
	\label{CPI47}
	\Phi^{a}(t,\theta,{\bar \theta})=\v^{a}(t)+\theta c^{a}(t)+
	{\bar \theta}~\omega^{ab}{\bar
	c}_{b}(t)+i{\bar\theta}\theta~\omega^{ab}\lambda_{b}(t).
	\end{equation} 
\noindent The reason for the $i$ in the last term is due to the choice of reality of
the ghosts $(c^a,\bc_a)$: see Ref.\cite{Ennio} for the details. 
Then it is easy to obtain the representations in superspace of 
all the operators listed in (\ref{CPI21})-(\ref{CPI29}). We make use of the
definition:
	\begin{equation}
	\label{CPI48}
	\delta\Phi^{a}=[\varepsilon O,\Phi^{a}]
	\equiv-\varepsilon {\mathscr O}\Phi^{a}
	\end{equation}
\noindent
where ${\mathscr O}$ is the representation on superspace of the operator $O$ acting
on $\MT$ (the commutator in (\ref{CPI48}) is the usual commutator in $\MT$).
According to (\ref{CPI48}) it is not difficult to realize that the representations we
are looking for are the following:
	\begin{gather}
	{\mathscr Q}_{\s BRS}=-\displaystyle\frac{\p}{\p\t};~~~~
	{\ov{\mathscr Q}}_{\s BRS}=\displaystyle\frac{\p}{\p\tb}; \label{CPI49}\\
	{\mathscr K}=\tb\displaystyle\frac{\p}{\p\t};~~~~
	{\ov{\mathscr K}}=\t\displaystyle\frac{\p}{\p\tb}; \\
	{\mathscr Q}_{g}=\tb\displaystyle\frac{\p}{\p\tb}-
	\t\displaystyle\frac{\p}{\p\t}; \\
	{\mathscr N}_{\s H}=\tb\displaystyle\frac{\p}{\p t};~~~~
	{\ov{\mathscr N}}_{\s H}=\t\displaystyle\frac{\p}{\p t}; \\
	{\mathscr Q}_{\s H}=-\displaystyle\frac{\p}{\p\t}-
	\tb\displaystyle\frac{\p}{\p t};~~~~
	{\ov{\mathscr Q}}_{\s H}=\displaystyle\frac{\p}{\p\tb}+
	\t\displaystyle\frac{\p}{\p t}; \\
	\widetilde{\mathscr H}=i\displaystyle\frac{\p}{\p t}. \label{CPI54}
	\end{gather}
\noindent
It is easy to check that the algebra among all the operators above is the same as
before, the only difference being that here the commutators are given by ordinary
derivation with respect to the variables of superspace. 

Following the analogy with the relativistic formulation of supersymmetry, we can also
construct the covariant derivatives associated to the Susy charges ${\mathscr Q}_{\s
H}$ and $\ov{\mathscr Q}_{\s H}$:
	\begin{align}
	\mathscr{D}_{\s H} & =-i\frac{\p}{\p\t}+i\tb\frac{\p}{\p t} \label{CPI55}\\
	\ov{\mathscr{D}}_{\s H} & =i\frac{\p}{\p\tb}-i\t\frac{\p}{\p t}.\label{CPI56}
	\end{align}	
\noindent which correspond (in $\widetilde{\cal M}$) to the following operators: 
	\begin{align}
	& D_{\s H} =i\Qb+i\beta\NH \label{CPIcov1}\\
	& \ov{D}_{\s H} =i\QBb-i\beta\NHB. \label{CPIcov2}
	\end{align}
\noindent 
Both ${\mathscr D}_{\s H}$, $\ov{\mathscr D}_{\s H}$ commute with 
${\mathscr Q}_{\s H}$, $\ov{\mathscr Q}_{\s H}$ and
$\mathscr H$ and they allow to constrain the general superfield (\ref{CPI47}) in
a Susy-invariant manner. In fact the superfield
(\ref{CPI47}) is not an irreducible representation of the Susy algebra. On the other
hand, if we constrain it imposing that it be annihilated by either $\ov{\mathscr{D}}_{\s
H}$ or $\mathscr{D}_{\s H}$, we obtain a {\it chiral} or ---respectively--- 
{\it antichiral} superfield:
	\begin{align}
	\Phi^a_{ch}(t,\t,\tb)&=\v^a+\t c^a +\tb\t\dot{\v}^a; \label{CPI57} \\  
	\Phi^a_{ach}(t,\t,\tb)&=\v^a+\tb\,\w^{ab}\bc_b -\tb\t\dot{\v}^a; 	\label{CPI58} 	
	\end{align}
\noindent which constitute two inequivalent irreducible representations of the Susy
algebra. 

The last topic of this section concerns the rewriting of the Hamiltonian $\HT$ and 
the Lagrangian $\LT$ in terms of the superfields $\Phi^a(t,\t,\tb)$. In fact every supersymmetric
theory admits a representation of its action in terms of an integral over the superspace variables
of a function of some superfields. This is the case also for the CPI theory. The first thing we
notice is that if we rewrite the classical Hamiltonian $H(\v)$ in terms of the superfields
$\Phi^a(t,\t,\tb)$, we obtain:
	\begin{equation}
	\label{CPI59}
	H\big[\Phi(t,\t,\tb)\big]=H(\v)+\t\NH-\tb\NHB+i\t\tb\HT.
	\end{equation}
\noindent
This in turn implies that:
	\begin{equation}
	\label{CPI60}
	\HT=i\int d\t d\tb\, H\big[\Phi(t,\t,\tb)\big].
	\end{equation}
\noindent 
The same thing (modulo boundary terms\footnote{The problem of how to properly handle
these boundary terms was analyzed in Ref.\cite{Planck}.}) happens for the
Lagrangian (understood as $\frac{1}{2}\v^a\w_{ab}\dot{\v}^b-H(\v)$):
	\begin{equation}
	\label{CPI61}
	L\big[\Phi(t,\t,\tb)\big]=L(\v)+\t(\ldots)-\tb(\ldots)+i\t\tb\LT;
	\end{equation}
\noindent which implies:
	\begin{equation}
	\label{CPI62}
	\LT=i\int d\t d\tb\, L\big[\Phi(t,\t,\tb)\big] + (\text{boundary terms}).
	\end{equation}
\noindent 
This a nice remark because we can see from Eq.(\ref{CPI61}) that both $L(\v)$ and $\LT(\v,c,\l,\bc)$ 
belong to the same superfield, i.e. to the same multiplet. But on the other hand we can think of
$\displaystyle\int dt\, L(\v)$ as the {\it quantum weight} (because it is the weight of the
trajectories in the Feynman path integral) while $\displaystyle\int dt\,\LT(\v,c,\l,\bc)$ 
is the {\it classical
weight} (because it plays the same role in the CPI). Therefore, in our context, we can conclude
that both Quantum Mechanics and Classical Mechanics belong to the same supermultiplet. How can we
pass from one domain to the other? The answer to this question is strictly connected to a limit of
the kind $\t\rightarrow 0$ and $\tb\rightarrow 0$, but we will come back on this topic in
future works.
\chapter*{2. The Classical Susy}
\addcontentsline{toc}{chapter}{\numberline{2}The Classical Susy}
\setcounter{chapter}{2}
\setcounter{section}{0}
\markboth{2. The Classical Susy}{}

\vspace{1cm}
\noindent
In the previous chapter we have introduced the Classical Path Integral (CPI)
and the universal symmetries exhibited by its Lagrangian. We have described 
the geometrical meaning of some of the symmetry charges, such as the 
$\Qb$ and $\QBb$. In this chapter we want to focus on the meaning of the 
universal classical Supersymmetry. An attempt in this direction was already
made in the past\scite{Ergo}, where the authors introduced a nice interplay 
between the classical Supersymmetry and the concept of ergodicity of a classical
system. Here our goal is twofold. First we try to shed some light on the 
geometrical meaning of the classical Susy (which was not explored in Ref.\scite{Ergo}),
and second we develop a little more the ideas put forward in Ref.\scite{Ergo}. 

To tackle the first issue, i.e. the geometrical aspects of our Susy,
the direction we shall take
is to make the Susy local and study in detail what we
obtain. We say ``study in detail" because in the literature
there are some strange statements\scite{Alv} claiming to show that,
at least for the supersymmetric QM of Witten\scite{Witten}, the Lagrangian
with local-Susy is equivalent to the one with global Susy. We shall
show that it is not so. One should actually perform very carefully the full
Dirac\scite{Vinc} procedure or, via path-integrals, 
the Faddeev procedure\scite{Pavao}
or apply the BFV methodology \scite{Teitel} of handling systems
with constraints. If one does that carefully, it is easy to realize that the
system with local Susy has a different number of degrees of
freedom than the one with global Susy. The states themselves are
restricted to the so called physical states by the presence of the 
local symmetry. It was this last step that was missing in Ref.\cite{Alv}
and which led to the wrong conclusion. 

The physical states condition and the BFV procedure
are  what will lead us to understand the geometrical 
meaning of the Susy charges.
They will turn out to be an essential ingredient to
restrict the forms to the so called equivariant ones
\scite{Stora}\scite{Duit}\scite{Berline}. The business of 
equivariant cohomology has popped out
recently in the literature in connection with  topological field theories
\scite{Topol}. Some attempts\scite{Niem} had been done in the past of
cooking up a BRS-BFV charge which would produce as physical states
the equivariant ones but without showing from which local symmetry
this BRS-BFV charge was coming from. Here we shall fill that gap, that means 
we shall show in detail which local symmetry gives rise to a BFV charge 
whose physical states are the equivariant ones.

Next, we shall turn to the other aspect
of our supersymmetry, that is its interplay with the concept of
ergodicity\scite{Ergo}. To get a better grasp of this
problem, it was realized long ago\scite{Ergo} that we had to formulate our
functional approach on constant energy surfaces.
Therefore we shall modify the CPI-Lagrangian constraining by hand the system to move on some 
fixed hypersurfaces, the constant energy ones, and check what happens.
We will realize that in this formulation the energy
plays the role of a coupling and it turns out to be associated to a 
tadpole term of the new Lagrangian. Moreover  we will find that by
constraining the system on constant energy surfaces we gain  a {\it local} 
graded symmetry which is not anyhow a local Susy. Regarding
the  {\it global} symmetries we lose part of the original global N=2 Susy 
which is  now reduced  to an N=1. This, which may appear as a bad feature of the
procedure, may actually  turn out to be a  virtue. In fact we shall  have
to study the interplay of ergodicity and Susy by means of only one Susy charge
and not two as before. Anyhow the detailed study of this interplay  
will be left to future works, where we will concentrate more on dynamical
issues and not on geometrical ones as we do here.
\section{Gauging the Global Susy Invariance.}
\label{sec:gaugeSusy}

\noindent
The direction we take to study the geometrical structures behind the
supersymmetric charges above is to build a Lagrangian where these symmetries
are local. The standard procedure we use is known in the literature\scite{Brin} 
as Noether method.
It basically consists in finding the exact form of the extra terms
generated by {\it local} variations of the original Lagrangian 
which had only the  global invariances. These extra terms, by Noether
theorem, are basically  the derivatives of the infinitesimal parameters
multiplied by the generators. The trick then is to add to the original
Lagrangian a piece made of an {\it auxiliary field} multiplied by
the generator. We can then impose that this auxiliary field  transform
in such a manner as to cancel the extra variations of the Lagrangian
mentioned above.

As the Susy charges are built out of the $\Qb, \QBb, \NH, \NHB$
let us build the {\it local} variations generated by each of these charges 
on the variables \quattrova. If we indicate with $X$ one of those four 
operators and with $({\s\ldots})$ any of the variables \quattrova, 
then by a local variation $\delta^{\s loc}_{\scriptscriptstyle X}({\s\ldots})$ we indicate
the operation:\break $\delta^{\s loc}_{\scriptscriptstyle X}({\s\ldots})\equiv [\varepsilon(t)X,({\s\ldots})]$
where now the Grassmannian parameter $\varepsilon(t)$ is dependent on~$t$.
These four variations are indicated below:

\begin{align}
\label{LS1}
	&
	\delta^{\s loc}_{\scriptscriptstyle Q}\equiv
	\begin{cases}
	\delta\varphi^{a} = \epsilon(t)c^{a}\\
	\delta c^{a} = 0\\
	\delta \ov{c}_{a} = i\epsilon(t)\l_{a}\\
	\delta \l_{a} = 0\\
	\end{cases} 
	&&
	{\delta}^{\s loc}_{\ov{\scriptscriptstyle Q}}\equiv
	\begin{cases}
	\delta\varphi^{a} = -{\ov\epsilon}(t)\w^{ab}\ov{c}_{b}\\
	\delta c^{a} = i{\ov\epsilon}(t)\w^{ab}\l_{b}\\
	\delta \bc_{a} = 0\\
	\delta\l_{a} = 0
	\end{cases}
	\vspace{.3cm}\\
	\label{LS2}
	&
	\delta^{\s loc}_{\scriptscriptstyle N}\equiv
	\begin{cases}
	\delta\varphi^{a} = 0\\
	\delta c^{a} = 0\\
	\delta \bc_{a} = \epsilon(t)\partial_{a}H\\
	\delta \l_{a} = i\epsilon(t)c^{b}\partial_{b}\partial_{a}H\\
	\end{cases}
	&&
	{\delta}^{\s loc}_{\ov{\scriptscriptstyle N}}\equiv
	\begin{cases}
	\delta\varphi^{a} = 0\\
	\delta c^{a} = {\ov\epsilon}(t)\w^{ab}\partial_{b}H\\
	\delta \bc_{a} = 0\\
	\delta\l_{a} = i{\ov\epsilon}(t)\bc_{d}\w^{db}\partial_{b}\partial_{a}H.
	\end{cases}
\end{align}

\noindent We could have used four different parameters for the four
different charges (as we will do later on)
 but here we limit ourselves just to two:
$\e(t)$ and ${\ov\e}(t)$.
The local Susy variations associated to the two Susy charges of 
Eqs.(\ref{CPI28}) and (\ref{CPI29}) are:

\begin{equation}
\label{LS3}
	\left\{
	\begin{array}{l}
	\delta^{\s loc}_{\s{\QH}}=\delta^{\s loc}_{\s{Q}}-\beta\delta^{\s loc}_{\s{N}} \\
	{\delta}^{\s loc}_{\s\QBH}={\delta}^{\s loc}_{\ov{\s Q}}+
	\beta{\delta}^{\s loc}_{\ov{\s N}}. \\
	\end{array}
	\right.
\end{equation}

\noindent It is now straightforward to check that the local Susy variations 
of the Lagrangian $\LT$ of Eq.(\ref{CPI6}) give the following results:

\begin{equation}
\label{LS4}
	\delta^{\s loc}_{\s{\QH}}\widetilde{\mathcal L}=-i\dot{\e}\QH + (t.d.)
\end{equation}
\noindent 
and
\begin{equation}
\label{LS5}
	{\delta}^{\s loc}_{\s{\QBH}}\widetilde{\mathcal L}=-i\dot{\ov\e}\QBH + (t.d.).
\end{equation}

\noindent With $(t.d.)$ we indicate total derivative terms: they turn
into surface terms in the action and they disappear if we
require that  $\epsilon(t)$  and ${\ov\e}(t)$  be zero 
at the end points of integrations as we will do from now
on. To do things in a cleaner manner we should have actually
checked the invariance using the integrated charge as explained in
Appendix~\ref{app:intSusy}. Anyhow from Eqs.(\ref{LS4}) and
(\ref{LS5}) we see that the Lagrangian does not change by a total
derivative so the two local Susy transformations are not symmetries and we have to modify the 
Lagrangian to find another one which is invariant. If we find it, then it must also
be invariant\scite{Brin} under the composition of  two local 
Susy transformations which  we can prove
(see Appendix \ref{app:combSusy})  to be the sum of a local supersymmetry transformation 
plus a {\it local} time-translation generated by $\HT$. This last one
is not a symmetry
of $\LT$ and the Lagrangian would  change by a term proportional to $\HT$
multiplied by the time-derivative of the symmetry parameter, exactly as the
Noether theorem requires. The trick\scite{Brin} to get the
invariance is to add to $\LT$
some auxiliary fields multiplied by the charges under which $\LT$ is not
invariant. In our case the complete Lagrangian is:

\begin{equation}
\label{LS6}
	\widetilde{\mathcal L}_{\s Susy}\equiv\widetilde{\mathcal L}+
	\ov{\psi}\QH+\psi\QBH+g\widetilde{\mathcal H},	
\end{equation}

\noindent where $g(t),\psi(t),{\ov\psi}(t)$ are three new fields 
(the last two of Grassmannian nature)  whose variations
under the local Susy will be determined by the requirement that $\LT_{Susy}$
be invariant under the local Susy variations of Eq.(\ref{LS3}).
In detail we get:

\begin{equation}
\label{LS7}
	\delta_{\s{\QH}}\widetilde{\mathcal
	L}_{Susy}=-i\dot{\e}\QH+(\delta_{\s{\QH}}g)\widetilde{\mathcal
	H} + (\delta_{\s{\QH}}\ov{\psi})\QH + (\delta_{\s{\QH}}\psi)\QBH +
	\psi(2i\e\beta\widetilde{\mathcal H})
\end{equation}

\noindent and we see that the following transformations for the 
variables $g,\psi,{\ov\psi}$ would make $\LT_{Susy}$ invariant 
under the local transformation associated to $\QH$: 

\begin{equation}
\label{LS8}
	\left\{
	\begin{array}{l}
	\delta_{\s{\QH}}\ov{\psi}=i\dot{\e} \\
	\delta_{\s{\QH}}\psi=0 \\
	\delta_{\s{\QH}}g=+2i\e\beta\psi.
	\end{array}
	\right.
\end{equation}

\noindent For the variation under $\QBH$ we get:

\begin{equation}
\label{LS9}
	\delta_{\s{\QBH}}\widetilde{\mathcal L}_{Susy}=-i\dot{\ov\e}\QBH+({\delta}_{\s
	{\ov Q}_{H}}g)\widetilde{\mathcal H} + ({\delta}_{\s{\ov Q}_{H}}{\ov\psi})\QH +
	 ({\delta}_{\s{\ov Q}_{H}}\psi)\QBH +
	{\ov\psi}(2i{\ov\e}\beta\widetilde{\mathcal H})
\end{equation}

\noindent and we see that the following transformations for the 
variables $g,\psi,{\ov\psi}$ make $\LT_{Susy}$ invariant under the local
transformation associated to $\QBH$:

\begin{equation}
\label{LS10}
	\left\{
	\begin{array}{l}
	{\delta}_{\s{\QBH}}\ov{\psi}= 0\\
	{\delta}_{\s{\QBH}}\psi=i\dot{{\ov\e}} \\
	{\delta}_{\s{\QBH}}g=+2i{\ov\e}\beta\ov{\psi}.
	\end{array}
	\right.
\end{equation}

\noindent 
Last we should check how $\LT$ changes under a local time-reparametrization.
We have to do that because this reparametrization appears in the composition
of two local Susy transformations (Appendix \ref{app:combSusy}). 
The action of the local time reparametrization
on the variables \quattrova  is listed in formula (\ref{eq:C-nove}) of Appendix \ref{app:combSusy}.
Under those transformations we can easily prove that 

\begin{equation}
\label{LS11}
\delta \LT =-i\dot{\eta}\HT
\end{equation}

\noindent where $\eta (t)$ is the time-dependent parameter of the transformation. 
Let us now use this result in the variation of $\LT_{Susy}$ under
time reparametrization:

\begin{equation}
\label{LS12}
	\delta\LT_{Susy} = -i\dot{\eta}\HT  +\delta\ov{\psi}\QH +\delta\psi\QBH
	+\delta g\HT.
\end{equation}

\noindent We immediately notice that $\LTloc$ is invariant
under the local time-repara-\break metrization if we transform the variables
 $(\psi,\ov\psi,g)$ as follows 

\begin{equation}
\label{LS13}
	\left\{
	\begin{array}{l}
	\delta\psi = \delta\ov{\psi} = 0\\
	\delta g=i\dot{\eta}.
	\end{array}
	\right.
\end{equation}

\noindent 
So, putting together all the  three local symmetries, (\ref{LS8}),
(\ref{LS10}) and (\ref{LS13}), we can say that
$\LTloc$ (if we choose $\beta=1$ in Eqs.(\ref{LS3})-(\ref{LS10})) has the following local invariance:

\begin{equation}
\label{LS14}
	\left\{
	\begin{array}{l}
	\delta\psi=i\dot{{\ov\e}} \\
	\delta\ov{\psi}= i\dot{\e} \\
	\delta g= i\dot{\eta}+2i({\ov\e}\ov{\psi}+\e\psi).
	\end{array}
	\right.
\end{equation}

\noindent
It is not the first time that one-dimensional systems with local-Susy 
have been built. The first work was the classic one of Brink et al.\scite{Brin}.
Later on people\scite{Alv} played with the supersymmetric
Quantum Mechanical model (Susy-QM) of Witten\scite{Witten} turning 
its global Susy  into a local invariance. 
Regarding this model, the author of Ref.\cite{Alv} 
pretended to show that the locally Supersymmetric Quantum Mechanics 
was equivalent  to the  standard Susy QM with only global
invariance. The proof was based on the fact that, via the
analog of the transformations ({\ref{LS14}), it is possible
to bring the variables ($\psi,{\ov \psi}, g)$ to zero and so, looking 
at Eq.(\ref{LS6}), this would imply that we can turn  
$\LT_{Susy}$  into $\LT$.
This kind of reasoning is {\it misleading}. In fact, while it is easy to check
(see Appendix~\ref{app:gauge}) that it is possible to bring the $(\psi, \ov\psi, g)$ to zero
via the transformations (\ref{LS14}), it should
be remembered that the starting point was a gauge theory, $\LT_{Susy}$, 
and the value zero for
the variables $(\psi, \ov\psi, g)$ is equivalent to a particular choice
of gauge-fixing. Anyhow, the {\it physical} theory has to be gauge-fixing 
independent and this is achieved\scite{Teitel} by restricting the physical 
states via the
BRS charge associated to the local symmetries. So in the gauge-fixing where
the $(\psi, \ov\psi, g)$ are zero the locally-supersymmetric QM\scite{Alv}
has the same action as  the globally supersymmetric theory\scite{Witten} but
we have to restrict the states to the physical ones which are basically
those annihilated by the symmetry charges. So the two systems,
the one with global Susy and the one with local Susy, are not equivalent
even if they seem to be so in a particular gauge-fixing. Their Hilbert
spaces are different even if the dynamics, in a particular gauge-fixing,
is the same.  Moreover, even concerning the
degrees of freedom, we shall show that while
$\LT$ has $8n$ independent variables  the $\LTloc$ has $8n-6$. To do this analysis
we shall go through the business of studying the constraints associated
to the local symmetries of $\LTloc$.

The standard procedure is the one of Dirac~\cite{Vinc} which we will
follow here in detail. Looking at $\LTloc$ we see that the primary
constraints are:
\begin{equation}
\label{LS15}
	\left\{
	\begin{array}{l}
	\pipsi = 0 \\
	\pipsibar = 0 \\
	\pigi = 0
	\end{array}
	\right.
\end{equation}

\noindent
where $\pipsi$, $\pipsibar$ and $\pigi$ are the momenta associated\footnote{Remember
that here and in the sequel we shall use right derivatives to define the momenta of the
Grassmannian variables:
$\Pi_{\psi}=\displaystyle\frac{\overleftarrow{\p}\LT}{\p\dot{\psi}}$.}
to $\psi$, $\ov\psi$ and $g$. The {\it canonical} Hamiltonian\scite{Vinc} 
is then the following:

\begin{equation}
\label{LS16}
	\HT_{can.}=\HTloc=\HT-\psi\QBH-\bpsi\QH-g\HT
\end{equation}

\noindent
while the {\it primary}\scite{Vinc}
(or total)  Hamiltonian is:

\begin{equation}
\label{LS17}
	\HT_{\s
	P}=(1-g)\HT-\psi\QBH-\bpsi\QH+u_{1}\pipsi+u_{2}\pipsibar+u_{3}\pigi
\end{equation}

\noindent
and it is obtained by adding to $\HT_{can.}$ the primary
constraints (\ref{LS15}) via the Lagrange multipliers 
$u_{1},u_{2},u_{3}$. Next we have to impose that the primary constraints 
do not change under the time evolution, i.e:

\begin{equation}
\label{LS18}
	\begin{array}{lrc}
	[\pipsi,\HT_{\s P}] = 0, \;\; & [\pipsibar,\HT_{\s P}] = 0, \;\; &
	[\pigi,\HT_{\s P}] =0.
	\end{array}	
\end{equation}

\noindent
Here we have used the commutators instead of the Extended-Poisson-Brackets. We did that because
the two structures are isomorphic as explained in Section 1.2.
In particular the (graded) commutators we need in (\ref{LS18}) are

\begin{equation}
\label{eq:sessanta}
	\begin{array}{lrc}
	[\pipsi,\psi] = 1, \;\; & [\pipsibar,\bpsi] = 1, \;\; & [\pigi,g] =-i.
	\end{array}	
\end{equation}

\noindent
Using them we get from (\ref{LS18}) the following set of
{\it secondary}\scite{Vinc} constraints:
 
\begin{equation}
\label{LS20}
	\left\{
	\begin{array}{l}
	\QBH = 0 \\
	\QH = 0 \\
	\HT = 0.
	\end{array}
	\right.
\end{equation}

\noindent
At this point the careful reader could ask  which are
the operators  
generating the full set of transformations (\ref{LS14}),
especially the last one. It does not seem
that they are generated by the operators of Eqs. (\ref{LS20})
and ({\ref{LS15}). Actually the answer to this question
is rather subtle and tricky \scite{Teitel} and is given in full details
in Appendix \ref{app:constr}.

Having clarified this point, we can go on with our procedure.
We have now to require that also the secondary constraints
(\ref{LS20}) do not change
under time evolution using as operator of evolution always the primary
Hamiltonian $\HT_{\s P}$ as explained in Ref.\scite{Vinc}.
It is easy to realize that in our case we do not generate 
further constraints with this procedure and 
that, at the same time, we do not determine 
the Lagrange multipliers. The fact that the Lagrange multipliers are all 
left undetermined is a signal that the constraints are first 
class~\scite{Vinc}
as it is easy to check by doing the commutators among all the
six constraints (\ref{LS20}) and (\ref{LS15}).
Being them first class, one has to introduce six gauge-fixings
which will be used to determine the Lagrange multipliers\scite{Vinc}.

The gauge-fixings --- let us call them  $\chi_{i}$ --- must 
have a non-zero commutator with the
associated gauge generator. For the three constraints of
Eq.(\ref{LS15}) three suitable gauge-fixings can be:

\begin{equation}
\label{LS21}
	\begin{array}{lrc}
	\psi-\psi_{\s 0} = 0, \;\; & \bpsi-\bpsi_{\s 0} = 0, \;\; & g-g_{\s 0} =0,
	\end{array}	
\end{equation}

\noindent
where $\psi_{\scriptscriptstyle 0}$, ${\ov\psi}_{\scriptscriptstyle 0}$ 
and  $g_{\scriptscriptstyle 0}$ are three fixed functions. It is easy to check that 
each\break of them does not commute with
its associated generator. The gauge-fixing\break 
$\psi_{\scriptscriptstyle 0}={\ov\psi}_{\scriptscriptstyle 0}=g_{\scriptscriptstyle 0}=0$ 
is among the admissible ones, in the sense that there is a gauge 
transformation which
brings any configuration into this one as shown in Appendix \ref{app:gauge}.
In this gauge fixing 
we get that the $\gaug$ variables disappear from the $\LTloc$
and so $\LTloc$  apparently is reduced to $\LT$. Of course, as we said earlier,
this should not mislead us to think that the physics
of $\LTloc$ is the same as the one of $\LT$. In fact 
at the Hamiltonian level, even if $\HTloc$ is reduced to $\HT$
by the gauge-fixing, the Poisson Brackets for the two systems are different. 
For the one with local symmetries the Poisson Brackets are the Dirac ones
which, given two observables $O_{1}$ and $O_{2}$, are built as:

\begin{equation}
\label{LS22}
	\{O_{1},O_{2}\}_{\s DB} = \{O_{1},O_{2}\} -
	\{O_{1},G_{i}\}(C^{-1})^{ij}\{G_{j},O_{2}\},
\end{equation}

\noindent
We have indicated with $G_{i}$ any of the six first class
constraints of Eqs.(\ref{LS20}) and (\ref{LS15}),
and the matrix $C_{ij}$ has its elements built as $\{G_{i},\chi_{j}\}$
where $\chi_{i}$ are the six gauge-fixings  associated to
the constraints $G_{i}$. The brackets entering the expressions
on the RHS of (\ref{LS22})
are the standard Extended Poisson Brackets of Eq.({\ref{CPI9}).
If the dynamics is  the one of a system with global Susy only,
that is one whose Hamiltonian is really $\HT$ from the beginning,
then the Poisson Brackets among the  same two observables 
$O_{1}$ and $O_{2}$ would be $\{O_{1},O_{2}\}$ and it is clear that

\begin{equation}
\label{LS23}
	\{O_{1},O_{2}\} \neq \{O_{1},O_{2}\}_{\s DB}.
\end{equation}

\noindent
This explains why, even if the Hamiltonians of the two systems
(the one with local symmetries and the one with global ones) are the same
(in some gauge-fixings), the two dynamics are anyhow different because they
are ruled by different Poisson Brackets.

Also the counting of the degrees of freedom indicates that the
systems have different numbers of degrees of freedom. The one with global
symmetries, and Lagrangian $\LT$, has just the variables \quattrova  which 
are $8n$.
The one with local symmetries, $\LTloc$, has the variables \quattrova
 which are $8n$, plus the three gauge variables $\gaug$ and 
the relative momenta for a total of
6, minus the 6 constraints of Eqs.(\ref{LS15})(\ref{LS20}),
minus the 6 gauge fixings $\chi_{i}$, for a total of $8n-6$ variables.
This is the correct counting of variables as explained in 
Ref. \cite{Teitel}. So even from this we realize that the
two systems are different. 
As we said at the beginning, the classical path-integral is actually the counterpart
of the operatorial version of CM proposed by 
Koopman and von Neumann\scite{Koop}  and if we adopt this operatorial version
we should use the commutators derived in Eq.(\ref{CPI11}).
This operatorial version of course could be adopted also for the dynamics
with local symmetries associated to~$\HTloc$ . In the operatorial
formulation we have a Hilbert space but we know that, for a system
with local symmetries,
the Hilbert space is restricted to the {\it physical
states}. The selection of these states is done by a BRS-BFV charge\scite{Teitel}
associated to the gauge-symmetries of the system. This BRS-BFV charge 
of course  has nothing
to do with the $\Qb$ of Eq.(\ref{CPI21}). The construction of the 
BRS-BFV
charge for our $\LTloc$ goes as follows\scite{Teitel}. First we should introduce
a pair of {\it gauge} ghost-antighosts for each gauge generator. 
As our gauge generators are

\begin{equation}
\label{LS24}
	G_{i}=(\pipsibar, \pipsi, \pigi, \QH, \QBH, \HT)
\end{equation}

\noindent
the ghost-antighosts are twelve and can be indicated as:

\begin{equation}
\label{LS25}
	\begin{array}{l}
	\eta^{i} = (\eta_{\bpsi}, \eta_{\psi}, \eta_{g}, \eta_{\s H},
	\ov{\eta}_{\s H}, \widetilde{\eta}_{\s H})\\
	{\mathcal P}_{i} = ({\mathcal P}_{\bpsi}, {\mathcal P}_{\psi}, {\mathcal
	P}_{g}, {\mathcal P}_{\s H},
	\ov{\mathcal P}_{\s H}, \widetilde{\mathcal P}_{\s H}).
	\end{array}
\end{equation}

\noindent
The general BRS-BFV charge\scite{Teitel} is then\footnote{The graded commutators
among the ghosts of (\ref{LS25}) are $[\eta^{i},{\mathcal
P}_{j}]=\delta^{i}_{j}$}: 

\begin{equation}
\label{LS26}
	\Omega_{\s BRS} = \eta^{i}G_{i} -
	\frac{1}{2}(-)^{\e_{i}}\eta^{i}\eta^{j}C_{ji}^{k}{\mathcal P}_{k},
\end{equation}

\noindent
where the $\varepsilon_{\s i}$ is the Grassmannian grading of the
constraints $G_{i}$ and $C^{k}_{ij}$ are the structure constants
of the algebra of our constraints. In our case this algebra is:
\begin{equation}
\label{LS27}
	\begin{array}{l}
	\,[\QH,\QBH] = 2i\HT \\
	\,[\QH,\QH] = [\QBH,\QBH] = [\QH,\HT] = [\QBH,\HT] = 0 
	\end{array}
\end{equation}
\noindent
where we have put $\beta=1$ with respect to Eq.(\ref{CPI30}).

\noindent
It is now easy to work out the BRS-BFV charge for our  local
Susy invariance:

\begin{equation}
\label{LS28}
	\Omega_{\s 	BRS}^{(Susy)}=\eta_{\bpsi}\pipsibar+\eta_{\psi}\pipsi+\eta_{g}\pigi+\eta_{
	\s H}\QH+ \ov{\eta}_{\s H}\QBH+\widetilde{\eta}_{\s H}\HT-2i\eta_{\s
	H}\ov{\eta}_{\s H}
	\widetilde{\mathcal P}_{\s H}	
\end{equation}

\noindent
Note that it contains terms with three ghosts and so it is
hard to see how it works on the states. These terms with three
ghosts are there because the generators are not in involution.
As it is explained in Ref.\scite{Teitel} in case the constraints
$G_{i}$  are not in involution, one can build some new ones
$F_{i}=a^{j}_{i}G_{j}$ which are in involution. In our case the $F_{i}$
generators, replacing the $\QH$ and $\QBH$,
can be easily worked out and they  are:

\begin{equation}
\label{LS29}
	\left\{
	\begin{array}{l}
	Q_{\s A}\equiv(\QH + \psi\HT) \\
	Q_{\s B}\equiv(\QBH - 2i\pipsi) 
	\end{array}
	\right.
\end{equation}

\noindent
while the other $F_{i}$ are the same as the $G_{j}$. The associated 
BRS-BFV charge is then 

\begin{equation}
\label{LS30}
	\Omega^{\s (F)}_{\s
	BRS}=\eta_{\bpsi}\pipsibar+\eta_{\psi}\pipsi+\eta_{g}\pigi+\eta_{\s
	A}Q_{\s A}+
	\eta_{\s B}Q_{\s B}+\widetilde{\eta}_{\s H}\HT.		
\end{equation}
\noindent
We have called $\eta_{\s A}, \eta_{\s B}$ the BFV ghosts
associated to $Q_{\s A}, Q_{\s B}$. Note that this $\Omega^{\s (F)}_{\s BRS}$ does not
contain terms with three ghosts. The {\it physical states} are then defined
as

\begin{equation}
\label{LS31}
	\Omega^{\s (F)}_{\s BRS}\mid\mbox{phys}\rangle = 0
\end{equation}

\noindent
and, by following Ref.\scite{Teitel}, we can easily show that 
(\ref{LS31}) is equivalent to the following six constraints: 

\begin{equation}
\label{LS32}
	\begin{array}{ll}
	(1)\left\{
	\begin{array}{l}
	\pipsibar\mid\mbox{phys}\rangle = 0 \\
	\pipsi\mid\mbox{phys}\rangle  = 0 \\
	\pigi\mid\mbox{phys}\rangle  = 0
	\end{array}
	\right. 
	\;\;&\;\;
	(2)\left\{
	\begin{array}{l}
	Q_{\s A}\mid\mbox{phys}\rangle = 0 \\
	Q_{\s B}\mid\mbox{phys}\rangle = 0 \\
	\HT\mid\mbox{phys}\rangle = 0.
	\end{array}
	\right.
	\end{array}
\end{equation}

\noindent
The set (1) above means that the physical states must be independent
of the $\gaug$ that means independent of any choice of gauge fixing.
The set (2) instead (combined with some of the  conditions from the 
set (1)) is equivalent to the following conditions:

\begin{equation}
\label{LS33}
	\begin{array}{l}
	\QH\mid\mbox{phys}\rangle = 0 \\
	\QBH\mid\mbox{phys}\rangle = 0 \\
	\HT\mid\mbox{phys}\rangle = 0.
	\end{array}
\end{equation}

\noindent
We can summarize it
by saying that, even if in some gauge-fixing
the $\HTloc$ is the same as $\HT$, the dynamics of the first
is restricted to a subset (given by Eq.(\ref{LS33}))
of the full-Hilbert space while the dynamics of $\HT$ is not restricted.
This is what is not spelled out correctly in Ref.\scite{Alv}.
The author may have been brought to the wrong conclusions
not only because he  did not consider the correct Hilbert space
but also by the following fact that we like to draw to the attention
of the reader. The three quantities $\QH,\QBH,\HT$
entering the constraints (\ref{LS20})
are actually constants of motion in the space \quattrova. So fixing
them, as the constraints do, is basically fixing a set of initial 
conditions. The hypersurface defined by (\ref{LS20})
is the subspace of \quattrova  where the motion takes place and 
it takes place  with the same dynamics as in \quattrova. 
If the constraint surfaces were not made by constants of 
motion, then the dynamics would have to be  modified to force the
particle to move on them, but this is not the case here.

\section{New Local Susy Invariance and Equivariance }

\noindent
In Chapter 1 we have seen that, besides the Susy, there are other
global invariances of $\HT$. In the following lines we show what
happens  when we gauge separately the $\Qb$, $\QBb$, $N_{\s H}$, 
${\ov N}_{\s H}$ of Eqs.(\ref{CPI21}),(\ref{CPI22}),
(\ref{CPI26}) and (\ref{CPI27}). The local
variation of $\LT$ under these four combined gauge transformations is 

\begin{align}
\label{LS34}
\delta^{loc.}\LT & = [\e \Qb+{\ov\e}\QBb+\eta N_{\s H}+{\ov\eta}{\ov N}_{\s H},\LT]
\nonumber\\
& =  -i{\dot\e}\Qb-i{\dot{\ov\e}}\QBb-i{\dot\eta}N_{\s H}-i{\dot{\ov\eta}}{\ov
N}_{\s H},
\end{align}

\noindent
where $(\e(t),{\ov\e}(t),\eta(t),{\ov\eta}(t))$ are  four different
Grassmannian gauge
parameters. From equation (\ref{LS34}) one could be tempted to propose
the following as {\it extended} Lagrangian invariant under the local symmetries 
above:

\begin{equation}
\label{LS35}
\LT_{ext.}\equiv\LT+\alpha(t)\Qb+{\ov\alpha}(t)\QBb+\beta(t)N_{\s H}+{\ov\beta}(t)
{\ov N}_{\s H},
\end{equation}
\noindent
where $(\alpha(t),{\ov\alpha}(t),\beta(t),{\ov\beta}(t))$ are four Grassmannian
gauge-fields which we could transform in a proper way in order to 
make $\LT_{ext.}$ invariant.
This is actually impossible whatever transformation we envision
for the gauge-fields. In fact, as we did in the case of the Susy of 
the previous section, we should consider what happens when we compose two 
of the local symmetries of Eq.(\ref{LS34}). This information
is given by the following commutators\scite{Ennio}:

\begin{align}
\label{LS36}
&[\Qb,{\ov N}_{\s H}]=i\HT; &[\QBb,N_{\s H}]=-i\HT.
\end{align}

\noindent
This basically tells us that we should add to the Lagrangian $\LT_{ext.}$ of
Eq.(\ref{LS35}) an extra gauge field $g(t)$ and an extra gauge
generator\footnote{We will indicate with greek letters the gauge fields
associated to Grassmannian generators and with latin letters the one
associated to bosonic generators.} $\HT$:

\begin{equation}
\label{LS37}
\LT_{ext.}=\LT+\alpha(t)\Qb+{\ov\alpha}(t)\QBb+\beta(t)N_{\s H}+{\ov\beta}(t)
{\ov N}_{\s H}+g(t)\HT.
\end{equation}

\noindent
Doing now an extended gauge transformation like in (\ref{LS34})
we get:

\begin{align}
\label{LS38}
\delta_{loc.}\LT_{ext.}= & -  i{\dot\e}\Qb-i{\dot{\ov\e}}\QBb 
-i{\dot\eta} N_{\s H}- i{\dot{\ov\eta}}{\ov N}_{\s H}+(\delta\alpha)\Qb\nonumber\\
& +  i\alpha{\ov\eta}\HT+(\delta{\ov\alpha})\QBb-i{\ov\alpha}\eta\HT
+(\delta\beta) N_{\s H}\nonumber\\
& -  i\beta{\ov\e}\HT+(\delta{\ov\beta})~{\ov N}_{\s H}+i{\ov\beta}\e\HT +
(\delta g)\HT, 
\end{align}

\noindent
and from this it is easy to see that $\LT_{ext.}$ is invariant 
if the gauge-fields $(\alpha,{\ov\alpha}, \beta, {\ov\beta}, g)$ 
are transformed  as follows:

\begin{equation}
\label{LS39}
\left\{
\begin{array}{l}
\delta\alpha=i{\dot\e}\\
\delta{\ov\alpha}= i{\dot{\ov\e}}\\
\delta\beta=i{\dot\eta}\\
{\delta}{\ov\beta}=i{\dot{\ov\eta}}\\
\delta g = i{\ov\alpha}\eta-i\alpha{\ov\eta}+i\beta{\ov\e}-i{\ov\beta}\e~.
\end{array}
\right.
\end{equation}

\noindent From these transformations we notice that, for some choice of the gauge-fields
and of the gauge-transformations, we do not need to have the $g\HT$ in 
the $\LT_{ext.}$
The first choice is $\alpha={\ov\alpha}=\e={\ov\e}=0$
which, from Eq.(\ref{LS39}), implies that we can choose $g(t)=0$. The
$\LT_{ext.}$ is then 

\begin{equation}
\label{LS40}
\LT_{\s N}\equiv\LT+\beta(t)N_{\s H}+{\ov\beta}(t){\ov N}_{\s H}.
\end{equation}

\noindent
The second choice, which also implies that we can choose $g(t)=0$, is 
$\beta={\ov\beta}=\eta={\ov\eta}=0$
and this would lead to the following Lagrangian

\begin{equation}
\label{LS41}
\LT_{\s \Qb}\equiv\LT+\alpha(t)\Qb+{\ov\alpha}(t)\QBb.
\end{equation}

\noindent
We shall hang around  here for a moment  spending some time
on the Lagrangian $\LT_{\s \Qb}$ above
because it allows us to do some  crucial observations on the counting 
of the degrees of freedom. As we said in the previous
section the Lagrangian $\LT_{Susy}$ of Eq.(\ref{LS6}) has 
fewer degrees of freedom than the standard one $\LT$ of Eq.(\ref{CPI6}), 
and the same happens with ~$\LT_{\s \Qb}$. In fact in $\LT_{\s \Qb}$~ we have
two primary constraints: 

\begin{align}
\label{LS42}
&\Pi_{\s\alpha}=0; &\Pi_{\s {\ov\alpha}}=0;
\end{align}

\noindent
which generate two secondary ones:

\begin{align}
\label{LS43}
&\Qb=0; &\QBb=0.
\end{align}

\noindent
All these four constraints are first class so we need four gauge-fixings.
The total counting\scite{Teitel} is then $8n$ (original variables) $+ 4$ (gauge variables and momenta)
$-4$ (constraints) $-4$ (gauge-fixings) for a total of $8n-4$ phase-space variables. 
At this point one question that arises naturally is: ``{\it Is it possible to have a Lagrangian
with the local invariances generated by $\Qb$ and $\QBb$, but with the same number
of degrees of freedom as }$\LT$?". The answer is yes. In fact let us  start from the following
Lagrangian:

\begin{equation}
\label{LS44}
\LT_{\s \Qb}^{\prime}\equiv\LT+{\dot\alpha}(t)\Qb+{\dot{\ov\alpha}}(t)\QBb.
\end{equation}

\noindent
It is easy  to check that it is invariant under the following set of
local transformations generated by the $\Qb$ and $\QBb$:
\begin{align}
\label{LS45}
\delta({\s\ldots}) & = [\e(t)\Qb+{\ov\e}(t){\QBb},({\s\ldots})]\nonumber\\
\delta\alpha & = i{\e}\nonumber\\
\delta{\ov\alpha} & = i{\ov\e};
\end{align}

\noindent
where we have indicated with $({\s\ldots})$ any of the variables \quattrova.
Anyhow at the same time it is easy to check that the Lagrangian 
$\LT^{\prime}_{\s \Qb}$ of (\ref{LS44}) has only two primary constraints\footnote{Remember
that we use right derivatives in the definition of the momenta: $\Pi_{\alpha}\equiv\displaystyle
\frac{\overleftarrow{\p}\LT^{\prime}_{\Qb}}{\p\dot{\alpha}}$.} 
\begin{align}
\label{LS46}
\Pi_{\s\alpha}+{\Qb} & = 0 ;\nonumber\\
\Pi_{\s{\ov\alpha}} + {\QBb} & = 0;
\end{align}

\noindent
and no secondary ones. The above two constraints are first class so we need
just two gauge-fixings and not four as before. The counting of degrees of
freedom now goes as follows: $8n$ (original variables)$+4$(gauge variables and
momenta)$-2$ (constraints)$- 2$ (gauge fixings) for a total of $8n$ variables. So we see that the
system 
described by the Lagrangian
$\LT^{\prime}_{\s\Qb}$ of Eq.(\ref{LS44}) has the same number
of degrees of freedom as the original $\LT$. In Appendix \ref{app:Qb} we will show how
the constraints (\ref{LS46}) act in the Hilbert space of the system. The fact that
$\LT^{\prime}_{\s\Qb}$ and $\LT$ are somehow equivalent could also
be understood by doing an integration by parts of the terms 
of Eq.(\ref{LS44})  containing
${\dot\alpha}$ and ${\dot{\ov\alpha}}$. The integration by parts produces,
with respect to $\LT$, some terms which vanish  because of the conservation of
$\Qb$ and $\QBb$. Moreover we should mention
that it is not the first time  people thought of making the BRS-antiBRS
invariance local\scite{Baul}. We will come back in Ref.\scite{Hilb} to the issue of gauging the BRS
symmetry and all the $\mathfrak{isp}(2)$ charges of
Eqs.(\ref{CPI21})-(\ref{CPI29}).

Let us now return to Eq.(\ref{LS37}). The two choices
which led to Eqs.({\ref{LS40}) and ({\ref{LS41})
are not the only ones consistent with the transformations (\ref{LS39}).
Another choice is 

\begin{equation}
\label{LS47}
\begin{array}{ll}
	\left\{
	\begin{array}{l}
	\alpha=-{\ov\beta}\\
	{\ov\alpha}=\beta
	\end{array}
	\right. 
	&
	\left\{
	\begin{array}{l}
	\e=-{\ov \eta}\\
	{\ov\e}=\eta ~~.
	\end{array}
	\right. 
\end{array}
\end{equation}

\noindent
With this choice the Lagrangian that we get from (\ref{LS37})
is:

\begin{equation}
\label{LS48}
\LT_{eq}\equiv \LT +\alpha(t) Q_{\s (1)}+{\ov\alpha}(t)Q_{\s (2)}+g(t)\HT.
\end{equation}

\noindent
The suffix ``$eq$" on the Lagrangian stands for ``equivariant" and the 
reason will be clear in a while. The $Q_{\s (1)}, Q_{\s (2)}$ on the RHS of Eq.
$(\ref{LS48})$ are

\begin{equation}
\label{LS49}
	\left\{
	\begin{array}{l}
	Q_{\s (1)}\equiv\Qb-{\ov N}_{\s H}\\
	Q_{\s (2)}\equiv\QBb+N_{\s H}. 
	\end{array}
	\right.
\end{equation}
\noindent
Note that these charges, with respect to the $\QH$ and $\QBH$ 
of Eqs.(\ref{CPI28}) and (\ref{CPI29}), are somehow twisted in the sense that 
here we sum the $\Qb$ with ${\ov N}$
and not with $N$ and viceversa for the $\QBb$.
It is easy to check that:

\begin{align}
\label{LS50}
&Q^{2}_{\s (1)}=Q^{2}_{\s (2)}=-i\HT; &[Q_{\s (1)},Q_{\s (2)}]=0.
\end{align}

\noindent
So these two charges generate two supersymmetry transformations
which are anyhow different from those generated by the supersymmetry
generators of Eqs.(\ref{CPI28}) and (\ref{CPI29}). The Lagrangian of
Eq.(\ref{LS48}) has two {\it local} Susy invariances but different 
from the ones of $\LT_{Susy}$ (\ref{LS6}). 
In order to get the Lagrangian 
(\ref{LS6}) we should have made in Eq.(\ref{LS37})
and ({\ref{LS39}) the following choice: 

\begin{equation}
\label{LS51}
\begin{array}{ll}
	\left\{
	\begin{array}{l}
	\alpha=-\beta={\ov\psi}\\
	{\ov\alpha}={\ov\beta}=\psi
	\end{array}
	\right. 
	&
	\left\{
	\begin{array}{l}
	\e=-\eta\\
	{\ov\e}={\ov\eta}~.
	\end{array}
	\right. 
\end{array}
\end{equation}

\noindent
Going back to Eq.(\ref{LS48}), let us now restrict the Lagrangian
to the following one: 

\begin{equation}
\label{LS52}
\LT_{eq}=\LT +\alpha(t)Q_{\s (1)}+g(t)\HT,
\end{equation}
\noindent
which is locally invariant only under one Susy and the symmetry
transformations are:

\begin{equation}
\label{LS53}
\left\{
\begin{array}{l}
\delta ({\s\ldots})  =  [\e Q_{\s (1)}+\tau\HT, ({\s\ldots})]\\
\delta\alpha  =  i{\dot\e}\\
\delta g  =  2i\alpha\e+i{\dot\tau} ,
\end{array}
\right.
\end{equation}
\noindent
where $({\s\ldots})$ indicates any of the variables \quattrova
and $\e(t)$ and $\tau(t)$ are infinitesimal parameters.
Because of this gauge invariance, we have to handle the system
either via the Faddeev procedure\scite{Pavao} 
or the BFV method\scite{Teitel}. We will follow this last one.
The constraints (primary and secondary) derived from (\ref{LS52}) are:

\begin{equation}
\label{LS54}
\left\{
\begin{array}{l}
\Pi_{\s \alpha}=0\\
\Pi_{\s g}=0\\
Q_{\s (1)}=0\\
\HT=0
\end{array}
\right.
\end{equation}

\noindent
where $\Pi_{\s\alpha}$ and $\Pi_{\s g}$ are respectively the momenta
conjugate to the gauge fields $\alpha(t)$ and $ g(t)$. The BFV procedure,
as explained in the previous section, tells
us to add four new ghosts and their respective momenta to the system. We will
indicate them as follows:

\begin{equation}
\label{LS55}
\left\{
\begin{array}{l}
(C^{\s (1)}, C^{\s H}, {\ov C}_{\s (1)}, {\ov C}_{\s H})\\
({\ov{\cal P}}_{\s (1)}, {\ov{\cal P}}_{\s H},{\cal P}_{\s (1)},{\cal P}_{\s H})
\end{array}
\right.
\end{equation}
\noindent
We shall impose the following graded-commutators: 
	\begin{equation}
	\label{LS56}
      \left\{
	\begin{array}{l}
	\,[g,\Pi_{\s g}]=[C^{\s (1)},\ov{\mathcal P}_{\s (1)}] =
	\,[{\ov C}_{\s (1)},\mathcal{P}_{\s (1)}]=1 \\
	\,[\alpha,\Pi_{\s\alpha}]=[C^{\s H},\ov{\mathcal P}_{\s H}]=
	\,[{\ov C}_{\s H},{\mathcal P}_{\s H}]=1
	\end{array}
	\right.
	\end{equation}

\noindent
In the first line above the variables are all ``bosonic" while
in the second one are all Grassmannian.
Equipped with all these tools  we will now build the
BFV-BRS\scite{Teitel} charge associated to our constraints:

\begin{equation}
\label{LS57}
\Omega_{\s BRS}^{\s (Eq.)}\equiv C^{\s (1)}Q_{\s (1)}+C^{\s H}\HT+{\mathcal P}_{\s (1)}
\Pi_{\s\alpha}+{\mathcal P}_{\s (H)}\Pi_{\s g}+i (C^{\s (1)})^{2}
{\ov{\mathcal P}}_{\s H}
\end{equation}

\noindent
It is easy to check that $(\Omega_{\s BRS}^{\s (Eq.)})^{2}=0$ as a BRS charge
should be. The next step, analogous to what we did in Section \ref{sec:gaugeSusy}, is to
select as physical states those annihilated by the $\Omega_{\s BRS}^{\s (eq).}$
charge: 

\begin{equation}
\label{LS58}
	\Omega^{\s (Eq.)}_{\s BRS}\mid\mbox{phys}\rangle = 0
\end{equation}

\noindent
Because of the nilpotent character
of the $\Omega^{\s (Eq.)}_{\s BRS}$, we should remember that two 
physical states 
are equivalent if they differ by a BRS variation:

\begin{equation}
\label{LS59}
\mid\mbox{phys-2}\rangle=\mid\mbox{phys-1}\rangle+
\Omega^{\s (Eq.)}_{\s BRS}\mid\mbox{$\chi$}\rangle 
\end{equation}

\noindent
Performing the standard procedure\scite{Teitel} of abelianizing the constraints
(\ref{LS54}) and building the analog of the $\Omega_{\s BRS}^{(F)}$ 
of Eq.(\ref{LS30}), it is then easy to see that the physical state
condition $(\ref{LS58})$ is equivalent to the following four conditions:

\begin{equation}
\label{LS60}
\left\{
\begin{array}{l}
\HT\mid\mbox{phys}\rangle=0\\
Q_{\s (1)}\mid\mbox{phys}\rangle=0\\
\Pi_{\s \alpha}\mid\mbox{phys}\rangle=0\\
\Pi_{\s g}\mid\mbox{phys}\rangle=0
\end{array}
\right.
\end{equation}

\noindent
Let us now pause for a moment and, for completeness,
let us  briefly review the concept
of equivariant cohomology (for references see\scite{Berline}).
Let us indicate with $\psi$ and $\chi$ two {\it inhomogeneous} forms on
a symplectic space and with $V$ a vector field on the same  space.
One says that the form $\psi$ is equivariantly closed but not exact
with respect to the vector field $V$ if the following  conditions
are satisfied:

\begin{equation}
\label{LS61}
\left\{
\begin{array}{l}
{\cal L}_{\s V}\psi=0\\
{\cal L}_{\s V}\chi=0\\
(d-\iota_{\s V})\psi=0\\
\psi\neq(d-\iota_{\s V})\chi
\end{array}
\right.
\end{equation}
\noindent
The forms $\psi$ and $\chi$  have to be inhomogeneous
because, while the exterior derivative~$d$ increases the degree
of the form of one unit, the contraction with the vector
field $\iota_{\s V}$ decreases it of one unit, so the third
and fourth relations in Eq.(\ref{LS61}) would never have
a solution if $\psi$ and $\chi$ were homogeneous in the form degree. 
From now on let us use the notation: $d_{eq}\equiv (d-\iota_{\s V})$
and let us try to interpret the  relations contained in
Eq.({\ref{LS61}).
Restricting the forms to satisfy the first two relations contained in
({\ref{LS61}) and noting that $d_{eq}^{\;2}= -{\cal L}_{\s V}$,
we  have that on this restricted space $[d_{eq}]^{\;2}=0$, and so 
$d_{eq}$ acts
as an exterior derivative. If we  now consider the last two relations of 
(\ref{LS61}) it is then clear that they  define a cohomology 
problem for $d_{eq}$.

There are other more abstract definitions of
equivariant  cohomology\scite{Stora} based on the {\it basic} cohomology of the
Weil algebra associated to a Lie-algebra, but we will not dwell on it here.
Equivariant cohomology is a concept that entered also the famous localization 
formula of Duistermaat and Heckman\scite{Duit} thanks to the work of Atiyah and Bott,
and into Topological Field Theory\scite{Topol} thanks to the work of R. Stora
and collaborators.

Let us now go back to our Lagrangian $\LT_{eq}$ of Eq. (\ref{LS52})
whose physical state space is
restricted by the conditions (\ref{LS60}) because
of the gauge invariance given in (\ref{LS53}). It is easy to
realize that the first two physical state conditions of  
(\ref{LS60}) are equivalent to the  first and third conditions
of Eq.(\ref{LS61}) once the vector field $V$ is identified
with the Hamiltonian vector field $(dH)^{\sharp}$ (according to the notation
of Ref.\cite{Marsd}). In fact
let us remember the correspondence described 
in Section 1.3 between standard operations in differential geometry
and in our formalism and in particular formula (\ref{CPI46}).
This tells us that  the Lie-derivative ${\cal L}_{\s {dH}^{\sharp}}$ acts
on a form  as the commutator of $\HT$ with the same form written in terms of
$c^{a}$ variables. This commutator gives the same result as the action
of $\HT$  on functions of $\v^{a}$ 
and $c^{a}$ once $\HT$ is  written as a differential 
operator like in (\ref{CPI14})-(\ref{CPI15}). So 
Eq.(\ref{CPI46}) proves that the first condition in both Eqs.(\ref{LS61}) 
and (\ref{LS60}) is the same: 

\begin{equation}
\label{LS62}
{\cal L}_{\s ({dH}^{\sharp})}\psi=0\longrightarrow\HT\mid\mbox{phys}\rangle=0.
\end{equation}

\noindent
Next let us look at the second condition in (\ref{LS60})
and the third in (\ref{LS61}). From the form (\ref{LS49}) 
of $Q_{\s (1)}$ we see that its first term,
the $\Qb$,  via the correspondence given by Eq.(\ref{CPI38}), 
corresponds to the exterior derivative $d$ which is exactly the first term 
contained in the third relation of Eq.(\ref{LS61}). 
The second term in  $Q_{\s (1)}$ is the $\NH$ which is given in 
Eq.(\ref{CPI26}) and can be written as

\begin{equation}
\label{LS63}
{\ov N}_{\s H}=[\QBb, H]
\end{equation}

\noindent From the relation (\ref{CPI33}) we see that we can
interpret ${\ov N}_{\s H}$ as the Hamiltonian vector field built out of
the function $H$. Its action as a differential
operator on forms is then given by Eq.(\ref{CPI33}), that means it
acts  as the interior contraction, $\iota_{\s
({dH}^{\sharp})}$, of the
Hamiltonian vector field with forms. This basically proves
the correspondence between the second relation of (\ref{LS60})
and the third of (\ref{LS61}):

\begin{equation}
\label{LS64}
(d-\iota_{\s ({dH}^{\sharp})})\psi=0\longrightarrow Q_{\s (1)}\mid\mbox{phys}\rangle=0.
\end{equation}

\noindent
Note that this correspondence would not hold if we had gauged the $\QH$, like we
did in Section \ref{sec:gaugeSusy}. In fact the $\QH$, being made of $\Qb$ and $N_{H}$ and not
${\ov N}_{H}$, would not have had the meaning of equivariant exterior derivative.

Let us now conclude the proof that the conditions (\ref{LS60}) are really equivalent to
the equivariant
cohomology problem given by Eq.(\ref{LS61}). We have already explained in Eqs.(\ref{LS62})
and (\ref{LS64}) the correspondences\footnote{In the following it is understood that $V=(dH)^{\sharp}$.}:
	\[
	\begin{array}{c}
	{\cal L}_{\s V}\psi=0\longleftrightarrow\HT\mid\mbox{phys}\rangle=0 \\
	(d-\iota_{\s V})\psi=0\longleftrightarrow Q_{\s
	(1)}\mid\mbox{phys}\rangle=0. 
	\end{array}
	\]
\noindent We have not discussed yet the 2nd and 4th equations in (\ref{LS61}). These
conditions are equivalent to the following statement:
	\begin{equation}
	\label{LS65}
	\{\psi=(d-\iota_{\s V})\chi ~~~\mbox{with}~~~{\cal L}_{\s
	V}\chi=0\}~~\Longrightarrow ~~\{\psi\simeq 0\},
	\end{equation}
\noindent
where the symbol $\simeq$ means 
``cohomologically equivalent". Therefore, if we want to complete the proof of the
correspondence 
between the
$|\mbox{phys}\rangle$ states of (\ref{LS60}) and the
$\psi$ of (\ref{LS61}), we must show that:
	\begin{equation}
	\label{alfa}
	\{|\mbox{phys}\rangle = Q_{\scriptscriptstyle (1)} |\chi\rangle
	~~~\mbox{with}~~~\widetilde{\mathcal H}
	|\chi\rangle = 0\}~~\Longrightarrow ~~\{|\mbox{phys}\rangle \simeq 0\}.
	\end{equation}
\noindent 
Note that the state $|\chi\rangle$ is not 
required to be physical, but only to satisfy the LHS of Eq.(\ref{LS65}); this implies
that $|\chi\rangle$ 
in general does not satisfy the third and the fourth conditions of (\ref{LS60}). This
means that 
$|\chi\rangle$ in general can depend on $g$ and $\alpha$. The point is that this
dependence, 
due to the LHS of Eq.~(\ref{alfa}) and to the requirement that $|\mbox{phys}\rangle$ does not
depend on $g$ 
and $\alpha$, must have the following form, as proven in Appendix \ref{app:eq}:
	\begin{equation}
	\label{beta}
	|\chi\rangle = |\chi_{0}\rangle + |\zeta;\alpha,g\rangle , 
	\end{equation}
\noindent where $|\chi_{0}\rangle$ is independent of both $g$ and $\alpha$, and 
$|\zeta;\alpha,g\rangle \in \ker Q_{\scriptscriptstyle (1)}$. Moreover 
if $|\chi\rangle\in\ker\widetilde{\mathcal H}$ (as imposed by Eq.~(\ref{alfa})), also 
$|\chi_{0}\rangle\in\ker\widetilde{\mathcal H}$ as one can check by applying 
$ Q_{\scriptscriptstyle (1)}^2 = -i\widetilde{\mathcal H}$ to both members of 
Eq.~(\ref{beta}). We are now ready to show that states of the form (\ref{alfa}) are
cohomologically  equivalent to zero according to the cohomology defined by
$\Omega^{\scriptscriptstyle (Eq.)}_{BRS}$ of Eq.~(98). The proof goes as follows:
	\begin{equation}
	\label{gamma}
	\begin{array}{rl}
	\mid\mbox{phys}\rangle & =Q_{\s (1)}\mid\chi\rangle \\
	& = Q_{\s (1)}\mid\chi_{0}\rangle \\
	& = [C^{\s (1)}]^{-1}[C^{\s (1)}Q_{\s (1)}+C^{\s H}\HT+{\mathcal P}_{\s (1)}
	\Pi_{\s\alpha}+{\mathcal P}_{\s H}\Pi_{\s g}+i (C^{\s (1)})^{2}
	{\ov{\mathcal P}}_{\s H}]\mid\chi_{0}\rangle \\
	& = [C^{\s (1)}]^{-1}\Omega^{\s eq}_{\s BRS}\mid\chi_{0}\rangle \\
	& = \Omega^{\s (Eq.)}_{\s BRS}\mid\chi'\rangle;
	\end{array}
	\end{equation}

\noindent (where we have defined $\mid\chi'\rangle\equiv[C^{\s (1)}]^{-1}
\mid\chi_{0}\rangle$) and therefore $\mid\mbox{phys}\rangle$ is cohomologically
equivalent to zero. 
Note that the ghost $ C^{\s (1)}$ is bosonic in character and so we can 
build its inverse. In the third equality of (\ref{gamma}) we have used the fact that the
second, 
the third and the fourth terms give zero when applied to $|\chi_{0}\rangle$. 
The last term ($i (C^{\s (1)})^{2}
{\ov{\mathcal P}}_{\s H}$) also annihilates $|\chi_{0}\rangle$ by a similar
reasoning based on the 
fact that $\mid\mbox{phys}\rangle$ cannot contain any dependence on $C^{\s H}$.
This concludes our proof.

The next step would be to exploit the correspondence between Eqs.(\ref{LS60}) and Eqs.(\ref{LS61}) 
in the study of the topology of the space of classical trajectories. In fact, there is an important
property of the equivariant forms: if the action of the group $G$ (i.e. the group which defines the
equivariant forms) on the manifold $\cal M$ is free, that is:
	\begin{equation}
 	g\cdot x=x~~\Longleftrightarrow ~~ g=e ~~\forall~x, 
	\end{equation} 	
\noindent 
then we have the following isomorphism:
	\begin{equation}
	\label{isom}
	H_{\s G}({\mathcal M})\cong H({\mathcal M}/G),
	\end{equation}
\noindent 
where we have indicated with $H_{\s G}$ and $H$ the equivariant and de-Rham cohomologies respectively.
In our case, the RHS of the previous equation is the de-Rham cohomology of the quotient space
$\cal{M}/\HT$, which is precisely the space of the classical trajectories. This would mean that --- in
principle --- we could study the geometry of the space of the classical trajectories by simply analyzing
the
physics of the model (\ref{LS52}). Unfortunately the isomorphism (\ref{isom}) is not always true, because
in the case of the Hamiltonian evolution the action of $G$ on $\cal M$ is not always free\footnote{For
example
in the harmonic oscillator the hamiltonian $H=\frac{1}{2}(p^2+q^2)$ admits $p=q=0$ as a fixed point.}.   
We hope to shed some further light on this topics in future works.

Basically with our path-integral we have managed
to get the propagation of equivariantly non-trivial states by properly gauging
the Susy. This is what the Susy is telling us from a
geometrical point of view. It is not the first time that the equivariant
cohomology is
reduced to a sort of BRS formalism\scite{Niem}, but differently from these 
authors the BRS-BFV charge we obtained  is really linked to a local 
invariance problem associated to the Lagrangian (\ref{LS52}).

\section{Motion On Constant-Energy Surfaces.}
\noindent
In the previous sections we have gauged the Susy and we 
have ended up with a constrained motion on the hypersurfaces
given by Eq.(\ref{LS20}) or (\ref{LS54}). 
In this section we shall reverse the procedure. We want to constrain 
the motion on some particular hypersurfaces of phase-space
and see which local symmetries the associated Lagrangian will
exhibit. The hypersurfaces we
choose are those defined by fixed values of the constants of 
motion. We will explain later the reasons for this choice.

Let us start with the constant energy surface: $H(p,q)=E$.
The most natural thing to do is to add this constraint to the $\LT$ of
Eq.(\ref{CPI6}):

\begin{equation}
\label{eq:energia-uno}
\LT_{\s E}\equiv\LT+f(t)(H-E),
\end{equation}

\noindent where $f(t)$ is a gauge variable.
This Lagrangian has the following local invariance:

\begin{equation}
\label{eq:energia-due}
\left\{
\begin{array}{l}
\delta_{\s H}({\s\ldots})  =  [\tau(t) H,({\s\ldots})]\\
\delta_{\s H}f(t)  =  i{\dot\tau}(t),
\end{array}
\right.
\end{equation}

\noindent
where $\tau(t)$ is an infinitesimal bosonic parameter
and we have indicated
with $({\s\ldots})$ any of the variables \quattrova.
Anyhow this is not the whole story. In fact if we restrict the original 
phase-space to be a constant  energy surface, 
the forms $c^{a}$ themselves must be restricted to be
those living only on the energy surface, that means they must be 
``perpendicular" to the gradient of the Hamiltonian. 
This constraint is:

\begin{equation}
\label{eq:energia-tre}
c^{a}\partial_{a}H=0=N_{\s H}.
\end{equation}

\noindent
So basically we have to impose that the $N_{\s H}$-function of Eq.(\ref{CPI26})
be zero. This is a further constraint we should add to the $\LT_{\s E}$ of
Eq.(\ref{eq:energia-uno}).  One may think that an analogous restriction has 
to be done also for the vector fields considering that  forms and tensor
fields are paired  by the symplectic matrix as explained
in Section \ref{sec:sym}. If we accept
this we will have to add  the condition that the
${\ov N}_{\s H}$ of Eq.(\ref{CPI27}) be zero.

A manner to get all these constraints automatically, without
having to add them by hand, is beautifully achieved if we
request that the new Hamiltonian $\HT_{\s E}$, which describes the motion 
on the energy surface, be a Lie-derivative along a Hamiltonian vector field
like the original $\HT$ was. $\HT_{\s E}$ must be a Lie-derivative because,
after all, the motion is the same as before. This time the difference is 
that we fix a particular value of the energy and so we include this initial
condition directly into the Lagrangian. If $\HT_{\s E}$ is a Lie-derivative
along a Hamiltonian vector field then, from what we said in Chapter 1,
we gather that $\HT_{\s E}$ must be of
the following form:

\begin{equation}
\label{eq:energia-quattro}
\HT_{\s E}=[\Qb^{\s E}[\QBb^{\s E},({\s\ldots})]],
\end{equation}

\noindent
that means it must be the BRS variation of the antiBRS variation
of some function that we have indicated with $({\s\ldots})$. The BRS and antiBRS
charges in (\ref{eq:energia-quattro}) are not the ones relative to
$\HT$, that means those of Eqs.(\ref{CPI21})(\ref{CPI22}). For
this reason we have indicated them with  different symbols. 

If $\HT_{\s E}$ is of the form above then the associated Lagrangian must be
BRS invariant. Let us start from $\LT_{\s E}$ in Eq.(\ref{eq:energia-uno})
and see if it is BRS invariant at least under the old $\Qb$. It is easy to
do that calculation and we get:

\begin{equation}
\label{eq:energia-cinque}
[\Qb,\LT+f(H-E)]=f(t)N_{\s H}\neq 0.
\end{equation}

\noindent
So it is not BRS invariant. The way out 
is to add to $\LT_{\s E}$ the $N_{\s H}$ multiplied by a gauge field.
The new Lagrangian is:

\begin{equation}
\label{eq:energia-sei}
\LT^{\prime}_{\s E}\equiv\LT+f(t)(H-E)+i\alpha(t) N_{\s H},
\end{equation}

\noindent
where $\alpha(t)$ is the  Grassmannian gauge field. This Lagrangian is
BRS-invariant
provided that we define a proper BRS-variation also on the gauge-fields
$\alpha(t)$ and $f(t)$. These proper BRS transformations  are:

\begin{equation}
\label{eq:energia-sette}
\left\{
\begin{array}{l}
\delta ({\s\ldots})  =  [\e \Qb,({\s\ldots})] \\
\delta {\alpha}  =  i\e f \\
\delta f=0,
\end{array}
\right.
\end{equation}

\noindent
where we have indicated with $({\s\ldots})$ any of the \quattrova and with $\Qb$ the
old BRS charge of Eq.(\ref{CPI21}).

Next let us notice that if the form of $\HT_{\s E}$ is the one of Eq.(\ref{eq:energia-quattro}) 
then the associated Lagrangian has to be also antiBRS-invariant. Let us check if this happens with
the $\LT^{\prime}_{\s E}$ of Eq.(\ref{eq:energia-sei}):

\begin{equation}
\label{eq:energia-otto}
[\QBb,\LT^{\prime}_{\s E}]=f(t){\ov N}_{\s H}-\alpha(t)\HT.
\end{equation}

\noindent
So it is not antiBRS invariant and  the way out is again to add 
to $\LT^{\prime}_{\s E}$
the generators appearing on the RHS of (\ref{eq:energia-otto}).
The final Lagrangian is:

\begin{equation}
\label{eq:energia-nove}
\LT^{\prime\prime}_{\s E}\equiv\LT+f(t)(H-E)+i\alpha(t)N_{\s H}+
i{\ov \alpha}(t){\ov N}_{\s H}-g(t)\HT
\end{equation}

\noindent
where $(f,\alpha,{\ov\alpha},g)$ are gauge-fields. So we see that the
request that our $\HT_{\s E}$ be a Lie-derivative (\ref{eq:energia-quattro})
has automatically
produced the constraints $N_{\s H}=0$ and ${\ov N}_{\s H}=0$ that otherwise we
would have had to add by hand like we did in the reasoning leading to
Eq.(\ref{eq:energia-tre}).

The Lagrangian $\LT^{\prime\prime}_{\s E}$ is invariant under the 
following generalized BRS and antiBRS transformations:

\begin{equation}
\label{eq:energia-dieci}
\begin{array}{ll}
	\delta_{\s \Qb}\equiv
	\left\{
	\begin{array}{l}
	\delta({\s\ldots}) = [\e\Qb,({\s\ldots})]\\
	\delta \alpha = i\e f\\
	\delta {\ov\alpha} = 0\\
	\delta f = 0\\
	\delta g =\e{\ov\alpha}\\
	\end{array}
	\right. 
	&
	{\ov{\delta}}_{\s\QBb}\equiv
	\left\{
	\begin{array}{l}
	{\ov\delta}({\s\ldots}) = [{\ov\e}\QBb, ({\s\ldots})]\\
	{\ov\delta}\alpha  = 0\\
	{\ov\delta}{\ov\alpha}  = i{\ov\e} f\\
	{\ov\delta}f = 0\\
	{\ov\delta}g=-{\ov\e}{\alpha}.
	\end{array}
	\right. 
\end{array}
\end{equation}

\noindent
It is straightforward to build the BRS-antiBRS charges which produce the
variations indicated above. They are:

\begin{align}
\label{eq:energia-dodici}
&\Qb^{\s E}\equiv \Qb + i f\Pi_{\s \alpha}+i{\ov\alpha}\Pi_{\s g} \\
&\QBb^{\s E}\equiv \QBb + i f \Pi_{\s{\ov\alpha}}-i\alpha\Pi_{\s g}, \label{eq:energia-tredici}
\end{align}

\noindent
where $\Pi_{\s\alpha}$, $\Pi_{\s{\ov\alpha}}$, $\Pi_{\s g}$ are the
momenta conjugate to the variables $\alpha,{\ov\alpha}, g$ and their
graded commutators are:

\begin{equation}
\label{eq:energia-undici}
[\alpha,\Pi_{\s\alpha}] = [{\ov\alpha},\Pi_{\s{\ov\alpha}}]= i[\Pi_{\s g}, g ]
 = i [f, \Pi_{\s f}]=1.
\end{equation}

\noindent
The new BRS and antiBRS charges are nilpotent, as BRS charges should be, 
and anticommute among themselves

\begin{equation}
\label{eq:energia-quattordici}
({\Qb^{\s E}})^{2}=({\QBb^{\s E}})^{2}=[\Qb^{\s E},\QBb^{\s E}]=0.
\end{equation}

\noindent
Having obtained these charges it is then easy to prove that the 
$\HT_{\s E}^{\prime\prime}$ associated to the $\LT^{\prime\prime}_{\s E}$
of Eq.(\ref{eq:energia-nove}) has the form (\ref{eq:energia-quattro})
with the $({\s\ldots})$ in (\ref{eq:energia-nove}) given by\break  $-i(H+g(H-E))$, 
i.e:
 
\begin{equation}
\label{eq:energia-quindici}
\HT_{\s E}^{\prime\prime}=-i[\Qb^{\s E}[\QBb^{\s E},H+g(H-E)].
\end{equation}

\noindent
This shows, with respect to the $\HT$ of Eq.(\ref{CPI8}), 
that the 0-form out of which the Hamiltonian vector field is built
is not $H$ but $H+g(H-E)$. This is natural in the sense that this
0-form feels the constraint $H-E=0$.

\noindent
The symplectic structure behind our construction
can be made more manifest if we introduce the following notation:

\begin{equation}
\label{eq:energia-sedici}
	\left\{
	\begin{array}{l}
	\varphi^{\s A}\equiv (\varphi^{a}; g, \Pi_{\s f}) \\
	\lambda_{\s A}\equiv (\lambda_{a};\Pi_{\s g}, f)\\
	c^{\s A}\equiv(c^{a};{\ov\alpha}, \Pi_{\s \alpha})\\
	{\ov c}_{\s A}\equiv ({\ov c}_{a};\Pi_{\s {\ov\alpha}},\alpha) 
	\end{array},
	\right.
\end{equation}

\noindent
where the index in capital letter $({\s\ldots})^{\s A}$ runs from 1 to $2n+2$ while
the one in small letter $({\s\ldots})^{a}$ runs from 1 to $2n$ and it refers to
the usual variables \quattrova. Let us also  introduce an enlarged symplectic 
matrix:

\begin{equation}
\label{eq:energia-diciassette}
	\omega^{\s AB}=
	\left(
	\begin{array}{cc}
	\omega^{ab} & 0 \\
	0 &     \left(
		\begin{array}{cc}
		0 & 1 \\
		-1 & 0
		\end{array}
		\right)
	\end{array}
	\right)
\end{equation}

\noindent
and then, using the definitions (\ref{eq:energia-sedici}) and 
(\ref{eq:energia-diciassette}),
the BRS-antiBRS charges (\ref{eq:energia-dodici}) and (\ref{eq:energia-tredici})
can be written in the following compact form:

\begin{equation}
\label{eq:energia-diciotto}
	\left\{
	\begin{array}{l}
	\Qb^{\s E}=ic^{\s A}\lambda_{\s A} \\
	\QBb^{\s E}=i{\ov c}_{\s A}\omega^{\s AB}\lambda_{\s B}.
	\end{array}
	\right.
\end{equation}

\noindent
Note that this form resembles very much the one of the original BRS and antiBRS
charges
(\ref{CPI21})(\ref{CPI22}). It is also straightforward to prove that the
$\HT_{\s E}^{\prime\prime}$ has an N=2
supersymmetry like the old $\HT$. To build the Susy charges we should first
construct the $(N_{H},{\ov N}_{H})$ charges analogous to those in
Eqs.(\ref{CPI26}) and (\ref{CPI27}).
Replacing in (\ref{CPI27}) the symplectic matrix and 
the variables with those constructed respectively in (\ref{eq:energia-diciassette})
and (\ref{eq:energia-sedici}), and the 0-form $H$ with the 
0-form $H+g(H-E)$ entering the $\HT_{\s E}$, we get:

\begin{equation}
\label{eq:energia-diciannove}
\left\{
\begin{array}{l}
N^{\s E}_{\s H}= c^{\s A}\partial_{\s A}(H+g(H-E)) \\
{\ov N}^{\s E}_{\s H}={\ov c}_{\s A}\omega^{\s AB}\partial_{\s B}(H+g(H-E)).
\end{array}
\right.
\end{equation}
The supersymmetry charges analogous to those in Eqs.(\ref{CPI28}) and (\ref{CPI29}) 
are then 

\begin{equation}
\label{eq:energia-venti}
	\left\{
	\begin{array}{l}
	Q_{\s H}^{\s E}\equiv \Qb^{\s E}-\beta N_{\s H}^{\s E} \\
	\QBH^{\s E}\equiv {\overline Q}^{\s E}_{BRS}+\beta{\ov N}_{\s H}^{\s E},
	\end{array}
	\right.
\end{equation}

\noindent
where $\beta$ is a dimensional parameter like the one appearing in
(\ref{LS1}). It is then easy to check  that:

\begin{equation}
\label{eq:energia-ventuno}
[Q^{\s E}_{\s H}, \QBH^{\s E}]=2i\beta\HT_{\s E}^{\prime\prime}.
\end{equation}

\noindent Up to now we have found which are the global symmetries of our
Lagrangian (\ref{eq:energia-nove}), but let us not forget that the
goal of this section was to find out if, by imposing a constraint
from outside like the one of being on a constant energy surface,
we would get a Lagrangian with local symmetries. It is actually so
and a first hint was given by the local symmetry of 
Eq.(\ref{eq:energia-due}). The full set of local invariances 
of the Lagrangian $\LT_{\s E}^{\prime\prime}$ of Eq.(\ref{eq:energia-nove})
is:

\begin{equation}
\label{eq:energia-ventidue}
	\left\{
	\begin{array}{l}
	\delta({\s\ldots})=[\tau H+\bar\eta{\ov N}_{\s H}+\eta N_{\s H}+\epsilon
	\HT,({\s\ldots})] \\
	\delta f=i{\dot \tau}\\
	\delta\alpha={\dot \eta}\\
	\delta{\bar\alpha}={\dot{\bar\eta}}\\
	\delta g=-i{\dot\epsilon} ,
	\end{array}
	\right.
\end{equation}
\noindent
where $(\tau,\eta,{\bar\eta},\epsilon)$ are the local gauge-parameters
depending on $t$, and with $({\s\ldots})$ we have indicated the variables
\quattrova.

The above local symmetry is not a local supersymmetry as in the previous
sections but a different graded one whose generators are $(H,N,{\ov N},\HT)$.
While before, in section 2.1 and 2.2, the local symmetry was a clearly
recognizable one but the constraints --- being in the enlarged space
\quattrova --- were hard to visualize, here we have the inverse
situation: the constraint (the constant energy one) is easy to
visualize but not so much the local symmetries.

For a moment let us stop these formal considerations
and let us check that the Hamiltonian in Eq.(\ref{eq:energia-quindici})
is the correct one.
The procedure we have followed here of constraining the motion
on a constant energy surface can be applied also to any other
constant of motion $I(\varphi)$. The result would be the following
Hamiltonian:

\begin{equation}
\label{eq:energia-ventitre}
\HT^{\prime\prime}_{\s I}=\HT-f[I(\varphi)-k]-i\bar\alpha{\ov N}_{\s (I)}-
i\alpha N_{\s (I)}+g{\widetilde{\cal I}}
\end{equation}

\noindent where $k$ is a constant and

\begin{equation}
\label{eq:energia-ventiquattro}
\left\{
\begin{array}{l}
N_{\s (I)}=c^{a}\partial_{a}I \\
{\ov N}_{\s (I)}={\ov c}_{a}\omega^{ab}\partial_{b}(I)\\
{\widetilde{\cal I}}=-i[\Qb,[\QBb, I(\varphi)]].
\end{array}
\right.
\end{equation}

\noindent If we had an integrable system with $n$ constants of motion
$I_{\s i}$ in involution we would get as Hamiltonian the following expression:

\begin{equation}
\label{eq:energia-venticinque}
\HT^{\prime\prime}_{\s int.}=\HT-\sum_{i}\{f_{\s i}[I_{\s i}(\varphi)-k_{\s i}]
-i\bar\alpha_{\s i}{\ov N}_{\s (I)_{\s i}}-
i\alpha_{\s i}N_{\s (I)_{\s i}}+g{\widetilde{\cal I}}_{\s i}\}
\end{equation}

\noindent Let us now do a counting of the effective degrees of freedom of the
Hamiltonian $\HT^{\prime\prime}_{\s I}$ of Eq.(\ref{eq:energia-ventitre}).
We have $8n$ variables \quattrova, plus 4 gauge fields 
$(f,\alpha,{\bar\alpha},g)$,
plus 4 momenta associated to these gauge fields minus 4 primary constraints
(which are the gauge-momenta equal zero), minus 4 secondary constraints
($I(\v)-k=0$, $N_{\s (I)}=0$, ${\ov N}_{\s (I)}=0$, ${\widetilde{\cal I}}=0)$)
minus 8 gauge-fixings for a total of $8n-8$ independent phase-space variables.
For the Hamiltonian of an integrable system like
$\HT^{\prime\prime}_{\s int.}$ this counting would give $8n-8n=0$ as effective
number of phase-space variables describing the system. {\it This is absurd!}
This situation could already be seen in the one-dimensional harmonic oscillator
where $n=1$ and we have just one constant of motion (the energy). The 
number of variables of the associated $\HT^{\prime\prime}_{\s E}$ would be
$8n-8=8-8=0$. One could claim that our $\HT^{\prime\prime}_{\s int.}$,
 having zero degrees of
freedom, actually describes
a Topological-Field-Theory model, and maybe
it is so but for sure it does not describe the motion taking place on
the tori of an integrable system. On the tori we have the angles which
vary with time but here, having effectively zero phase-space variables,
we do not have any motion taking place at all. If it is a topological
theory  at most the 
$\HT^{\prime\prime}_{\s int.}$ can describe some static {\it geometric} feature
of our system. This in itself would be interesting and that is
why we have carried this construction so far.
We hope to come back to this issue in future papers but
for the moment we want to go back from where we started, that is
Eq.(\ref{eq:energia-uno}) and see which is the way to get a  Hamiltonian
describing really the motion on the constant energy surface.

What we basically want to get is a Hamiltonian whose counting of degrees
of freedom is correct. At the basic phase-space level labelled
by the variables $\varphi$ we have $2n$ variables minus 1 constraint
that is $H-E=0$ so the total number is $2n-1$. Going up to the
space \quattrova this number should be multiplied by 4 that is
$8n-4$. 

What went wrong in the construction of $\LT^{\prime\prime}_{\s E}$
of Eq.(\ref{eq:energia-nove})? One thing that we requested, but  which 
was not necessary, was that the vector fields obey a constraint ${\ov N}=0$
analogous to the one of the forms $N=0$. We made that request only in order 
to maintain the standard pairing between tensor fields and forms which
appear in any symplectic theory, but our theory is not a symplectic
one anymore because the basic space in $\varphi^{a}$ has odd dimension $2n-1$
and cannot be a symplectic space. So let us release the request of
having ${\ov N}=0$. We could have a weaker request by adding this
constraint via the derivative of a Lagrange multiplier (or gauge-field)
in the same manner  as we did in Eq.(\ref{LS44}). There we realized
that adding constraints in this manner does not decrease the
number of degrees of freedom.
By consistency then  also the $\HT$ constraint,
which appeared together with the ${\ov N}$ via 
the Eq.(\ref{eq:energia-otto}), should be added via the derivative
of its associated Lagrange multiplier. So in order to
describe the motion on constant energy surfaces, instead of
(\ref{eq:energia-nove}) the Lagrangian we propose is:

\begin{equation}
\label{eq:energia-ventisei}
L_{\s E}=\LT+f(H-E)+i\dot{\bar\alpha}{\ov N}+i\alpha N-{\dot g}\HT.
\end{equation}
\noindent
The constraints (primary and secondary) are:

\begin{equation}
\label{eq:energia-ventisette}
\left\{
\begin{array}{l}
\Pi_{\s f}=0~;~~~~H-E=0~; \\
\Pi_{\s {\alpha}}=0~;~~~~N=0~;\\
\Pi_{\s\ov\alpha}=-i{\ov N}~; \\
\Pi_{\s g}=-\HT.
\end{array}
\right.
\end{equation}
\noindent They are 6, all first class, and we need 6 gauge-fixings. So doing now
the counting of independent variables in phase-space
we have:  $8n$ variables \quattrova, plus $4+4$ gauge-fields and their momenta,
minus 6 constraints, minus 6 gauge fixings for a total of $8n-4$ which is
exactly the number we wanted!

Let us analyze the difference between the last constraint in
Eq.(\ref{eq:energia-ventisette})\break 
(that is $\Pi_{\s g}=-\HT$) and the one
associated to the Lagrangian $\LT^{\prime\prime}_{\s E}$ of Eq.
(\ref{eq:energia-nove}) (that is $\HT=0$). This last constraint seems to
totally freeze the motion while the one in Eq.(\ref{eq:energia-ventisette})
does not freeze it but just foliates the space of values of $\HT$.
Similar things can be said for the constraint ${\ov N}=0$ associated
to $\LT^{\prime\prime}_{\s E}$ and the one, $\Pi_{\s \alpha}=-i{\ov N}$,
associated to  $L_{\s E}$. This last one would not force the vector
fields in a configuration symplectically equivalent to the one of forms.

Let us now proceed to further analyze the Lagrangian $L_{\s E}$ 
of Eq.(\ref{eq:energia-ventisei}).
The associated Hamiltonian is: 
\begin{equation}
\label{eq:energia-ventotto}
H_{\s E}=\HT+\Pi_{\s\alpha}\dot{\alpha}+\Pi_{\s g}{\dot g}
+\Pi_{\s f}\dot{f}+\Pi_{\s\ov\alpha}\dot{\bar\alpha}
-f(H-E)-i{\alpha}N-i\dot{\bar\alpha}{\ov N}+\dot{g}\HT,
\end{equation}
\noindent
where we had to leave in some velocities because we could not perform
the Legendre transform. From the above Hamiltonian we can go to
the canonical one\scite{Vinc} by imposing the primary constraints.
The result is:
\begin{equation}
\label{eq:energia-ventinove}
H_{\s E}^{\s can.}\equiv\HT-f(H-E)-i{\alpha}N.
\end{equation}

\noindent
It is easy to prove that this $H_{\s E}^{\s can.}$ is a Lie-derivative
along a vector field but not along a Hamiltonian vector-field. To show that let
us first define the following new variables:

\begin{equation}
\label{eq:energia-trenta}
\left\{
\begin{array}{l}
\varphi^{\s A}=(\varphi^{a},\pi_{\s f})\\
\lambda_{\s A}=(\lambda_{a},f)\\
c^{\s A}=(c^{a},\Pi_{\s{\alpha}})\\
{\ov c}_{\s A}=({\ov c}_{a},{\alpha}).
\end{array}
\right.
\end{equation}

\noindent
In this enlarged phase-space the BRS charge (or exterior derivative) is

\begin{equation}
\label{eq:energia-trentuno}
Q_{\s BRS}^{\s can.}=\Qb+if\Pi_{\s{\alpha}},
\end{equation}

\noindent
and the analog of the Hamiltonian vector field ${\ov N}_{\s H}$ is

\begin{equation}
\label{eq:energia-trentadue}
{\ov N}_{\s H}^{\s can.}={\ov N}_{H}-{\ov\alpha}(H-E),
\end{equation}

\noindent 
which is not a Hamiltonian vector field anymore because it cannot
be written as the antiBRS variation of something as a Hamiltonian
vector field should be (see Eq.(\ref{CPI27})).

The proof that $H_{\s E}^{\s can.}$ of Eq.(\ref{eq:energia-ventinove})
is the Lie-derivative along the vector field ${\ov N}_{\s H}^{\s
can.}$ 
above comes from the fact that it can be written as the commutator 
of that vector field with the exterior derivative 
$Q_{\s BRS}^{\s can.}$  above:

\begin{equation}
\label{eq:energia-trentatre}
H_{\s E}^{\s can.}=-i[Q_{\s BRS}^{\s can.},{\ov N}_{\s H}^{\s can.}].
\end{equation}

\noindent
Proving this relation  is straightforward. One just needs to
use the standard commutators plus the following ones:

\begin{align}
\label{eq:energia-trentaquattro-1}
&[{\alpha},\Pi_{\s {\alpha}}]=1; &[f,\Pi_{\s f}]=-i.
\end{align}
\noindent
Eq.(\ref{eq:energia-trentatre}) implies that $H_{\s E}^{\s can.}$ of
Eq.({\ref{eq:energia-ventinove})
is invariant under the global BRS transformations generated by
the $Q_{\s BRS}^{\s can.}$ of Eq.(\ref{eq:energia-trentuno}). It is also easy to
see that the Lagrangian $L_{\s E}$ of Eq.(\ref{eq:energia-ventisei})
has the following local invariances different from those of
Eq.(\ref{eq:energia-ventidue}):

\begin{equation}
\label{eq:energia-trentaquattro}
	\left\{
	\begin{array}{l}
	\delta({\s\ldots})=[\tau H+{\bar\eta}{\ov
	N}_{H}+{\eta}N_{H}+\epsilon
	\HT,({\s\ldots})] \\
	\delta f=i{\dot \tau}\\
	\delta\alpha=\dot{\eta}\\
	\delta{\bar\alpha}={\bar\eta}\\
	\delta g=-i{\epsilon}. 
	\end{array}
	\right.
\end{equation}

\noindent
Again, as before, this is a local symmetry but not a local supersymmetry.

Regarding the supersymmetry we can find a global one under which our 
$H_{\s E}^{\s can.}$ of Eq.(\ref{eq:energia-ventinove}) is invariant.
It is the one generated by the following charge:

\begin{equation}
\label{eq:energia-trentacinque}
Q_{Susy}=Q_{\s BRS}^{\s can.}+{\ov N}_{\s H}^{\s can.},
\end{equation}

\noindent 
which is a Susy charge because it is  easy to prove that:

\begin{equation}
\label{eq:energia-trentasei}
[Q_{\s Susy}]^{2}=i H_{\s E}^{\s can.}.
\end{equation}

\noindent
Differently from the $\HT$ of our original system, we do not have
an N=2 supersymmetry like in Eq.(\ref{eq:energia-quattro}), but only an N=1
Susy. This is due to the loss of a symplectic structure on
the constant energy surface.

The reason to work out this supersymmetry is not just academical. In fact we
proved in Ref.\scite{Ergo} that there is a nice interplay between the
 loss of ergodicity of the system whose Hamiltonian
is $H$  and the spontaneous breaking of the Susy 
of $\HT$. We proved in particular that if the Susy of $\HT$ is unbroken 
then the system described by $H$ is in the ergodic phase and that if the system
is in the ordered phase (non-ergodic) then the Susy of $\HT$ must be broken.
We could not prove the inverse of these two statements that is that if the
system is in the ergodic phase then the Susy must be unbroken and that
if the Susy is broken then the system must be in an ordered or non-ergodic
regime. The reason we could not prove these inverse statements was that
the energy at which the motion took place had not been specified.
We have no time here to explain the detailed reasons why this lack of
specification could not allow us to do the inverse of that statement
and we advice the reader interested in understanding this point
to study in detail the full set of papers contained in Ref.\scite{Ergo}. 
Ergodicity\scite{Avez} is a concept
which is strongly energy dependent: a system can be ergodic at some energy
and not ergodic at other energies. So it was crucial to develop a formalism
giving us the motion on constant energy surfaces like we have done here.
The parameter $E$ entering our $H^{\s can.}_{\s E}$ is not a phase-space variable
and we can
consider it as a coupling constant. We know that  at some values of the 
coupling a symmetry can be broken while at others it can be restored. 
In  (\ref{eq:energia-ventisei}) the term containing the energy is like a 
tadpole term
because it is  proportional to a term linear in the field 
(the field in this case is
$f(t)$ while the coupling is $E$).

The attempt to have a formulation
of the CPI in which  $E$ enters explicitly was tried before\scite{Bill}
but along a different route. In that paper $E$ was not a coupling constant 
but a degree of freedom conjugate to time in a formulation of CM invariant 
under time-reparametrization. We think that, in order to understand the 
interplay {\it Susy/ergodicity}, it is better to treat $E$ as a coupling constant. 

The next step would be to check whether the Susy charge 
(\ref{eq:energia-trentacinque}) we have in $H_{\s E}^{\s can.}$
is that  for which the theorem\scite{Ergo}
mentioned above, regarding the interplay\break {\it Susy/ergodicity}, holds also 
in the inverse form.
If this were the case  then we would have a criterion to check if a system (at
some
energy) is ergodic or not using a universal symmetry like Susy.
Maybe even a sort of Witten index could be built which, by signaling if the Susy
is broken or not, could tell us if the system is ergodic or not.

All this project will be left to future works because there 
are several other difficulties that have to be cleared before
really embarking on a full understanding of the interplay 
between Susy and ergodicity. 
The main difficulty is  the presence of
zero and negative norm states which prevents the proper use of
something like a Witten index for the study of the above mentioned
interplay. 

At present we are just working on the issue about how to give the CPI a fully
satisfactory and consistent Hilbert space structure, with positive-definite
inner product and unitary evolution (i.e. the choice of the scalar product
must imply that $\HT$ is hermitian). The results of this analysis will be
contained in a forthcoming paper \cite{Hilb}.
\chapter*{3. CPI and $\kappa$-symmetry}
\addcontentsline{toc}{chapter}{\numberline{3}CPI and $\kappa$-symmetry}
\setcounter{chapter}{3}
\setcounter{section}{0}
\markboth{3. CPI and $\kappa$-symmetry}{}

\vspace{1cm}
\noindent
In the previous chapter we have analyzed some of the universal symmetries of the CPI Lagrangian 
(\ref{CPI6}). In particular we focused on the geometrical meaning of the classical
supersymmetry generated by the two charges $\QH$ and $\QBH$. In this chapter we 
switch our analysis to the two other fermionic charges $D_{\s H}$ and $\ov{D}_{\s H}$
introduced in Eqs.(\ref{CPIcov1}) and (\ref{CPIcov2}). These charges --- as we have already claimed in
Chapter 1 --- are strictly related to $\QH$ and $\QBH$ because in superspace they
become the covariant derivatives associated to these Susy charges.  

Following the lines of Ref.\cite{DG}, we make these two symmetries ($D_{\s H}$ and
$\ov{D}_{\s H}$) local and we note that the model we get exhibits a nonrelativistic 
local Susy which is very similar to the famous $\kappa$-symmetry introduced in the literature 
almost 20 years ago by Siegel \cite{Siegel}, who discovered it in the Lagrangian 
of the supersymmetric relativistic particle without mass. This model was analyzed also
in many following papers \cite{DeAz}\cite{Moshe},
where it was clarified that the most remarkable feature of this $\kappa$-symmetry
is that it gives rise to some problems in separating $1^{\text{st}}$-class from
$2^{\text{nd}}$-class constraints, and therefore in quantizing the model \cite{Sorokin}\cite{Kallosh}.
In our nonrelativistic framework, the main difference with respect to the relativistic
model above is that no difficulty arises in the separation of $1^{\text{st}}$-class from 
$2^{\text{nd}}$-class constraints because in our case no $2^{\text{nd}}$-class constraint survives 
after imposing the invariance under local reparametrizations of time \cite{chiccode}. 
\section{The $\kappa$-symmetry}
\noindent The model studied by Siegel \cite{Siegel} for the massless relativistic
superparticle is characterized by the following ($1^{\text st}$ order) action:
	\begin{equation}
	\label{1-1}
	S=\int d\tau\left\{p_{\mu}\left[\dot{x}^{\mu}-\frac{i}{2}\left(\zb
	\gamma^{\mu}\dot{\z}-\dot{\zb}\gamma^{\mu}\z\right)\right]-\frac{1}{2}\l
	p^2\right\},
	\end{equation}
\noindent where $x^{\mu}$ are $n$-dimensional space-time coordinates, $\z^{\,\s a}$ and $\zb_{\:\s a}$
are
Dirac spinors and $\l$ is a Lagrange multiplier 
introduced to realize the $p^2=0$ constraint. This action is invariant under the following
transformations:
	\begin{align}
	& 
	\begin{array}{ll}
	\text{\bf{$\tau$-reparametrization} (local)}\vspace{.1cm} \\
	\delta x^{\mu}=\epsilon\dot{x}^{\mu} & \delta p_{\mu}=\epsilon\dot{p}_{\mu} \\
	\delta \z=\epsilon\dot{\z} & \delta \zb=\epsilon\dot{\zb} \\
	\delta \l=\dot{(\epsilon\l)} &\vspace{.2cm}
	\end{array} \label{1-2}
	\\
	&
	\begin{array}{ll}
	\text{\bf{Supersymmetry} (global)}\vspace{.1cm} \\
	\delta x^{\mu}=\displaystyle\frac{i}{2} 
	\left(\ov{\ve}\gamma^{\mu}\z-\zb\gamma^{\mu}\ve\right) 
	& \delta p_{\mu}=0 \\
	\delta \z=\varepsilon & \delta \zb=\ov{\varepsilon} \\
	\delta \l=0 & \vspace{.2cm}
	\end{array}\label{1-3}
	\\
	&
	\begin{array}{ll}
	\text{\bf{$\kappa$-symmetry} (local)}\vspace{.1cm} \\
	\delta x^{\mu}=\displaystyle\frac{i}{2} 
	\left(\zb\gamma^{\mu}\pslsh\k-\kb\pslsh\gamma^{\mu}\z\right) 
	& \delta p_{\mu}=0 \\
	\delta \z=\pslsh\k & \delta \zb=\kb\pslsh \\
	\delta \l=2i(\dot{\zb}\k-\kb\dot{\z}) &
	\end{array}\label{1-4}
	\end{align}
\noindent 
where in (\ref{1-2}) the dot means derivation with respect to $\tau$ and $\pslsh$ is 
obviously $p_{\mu}\gamma^{\mu}$. As specified above, $\epsilon$ and $\kappa,\ov{\kappa}$ are local
parameters (the first is a commuting scalar, the others are anticommuting spinors) while $\ve$
and $\ov{\ve}$ are two global (i.e. they do not depend on the base space $\tau$) 
spinorial parameters. We are particularly interested in the structure of
the third symmetry, which has been deeply analyzed in the literature. Here we want to give a
pedagogical description of the structure of the transformation in phase space, and we want
to highlight the role of the various operators and various commutation structures (Dirac
Brackets) involved. This will turn out to be useful when we will analyze the analog of
the $\kappa$-symmetry in Classical Mechanics.    

First of all we notice that the first and third symmetries above are strictly related. In fact, if we
release the $p^2=0$ constraint introducing a mass $m$ in (\ref{1-1}) we get
 	\begin{equation}
	\label{1-5}
	S_{m}=\int d\tau\left\{p_{\mu}\left[\dot{x}^{\mu}-\frac{i}{2}\left(\zb
	\gamma^{\mu}\dot{\z}-\dot{\zb}\gamma^{\mu}\z\right)\right]-\frac{1}{2}\l
	(p^2-m^2)\right\}.
	\end{equation} 	
\noindent 
$S_m$ is still invariant under (\ref{1-3}) but the other two invariances
are lost. This is easy to see in
phase space if we apply the Dirac procedure to the actions (\ref{1-1}) and (\ref{1-5}).
Consider first the massive model. The constraints are the following:
	\begin{align}
	\text{First Class\hspace{.5cm}} 
	& 
	\begin{cases}
	\Pi_{\s \l} = 0 & (a) \\
	p^2-m^2=0 & (b)
	\end{cases} 
	\vspace{.5cm} \label{1-6}\\
	\text{Second Class\hspace{.3cm}}
	& 
	\begin{cases}
	\Pi_{\s p}^{\mu} = 0 & (a)\\
	(\Pi_{\s x})_{\mu}-p_{\mu} = 0 & (b) \vspace{.1cm}\\ 
	D^{\s a}\equiv(\Pi_{\s\zb})^{\s a}+\displaystyle\frac{i}{2}(\pslsh\,\z)^{\s a} = 0 &(c)
	\vspace{.15cm}\\
	\ov{D}_{\s a}\equiv(\Pi_{\s\z})_{\s a} +\displaystyle\frac{i}{2}(\zb\pslsh)_{\s a} = 0 & (d), 
	\end{cases}\label{1-7}
	\end{align} 
\noindent
where $\Pi_{(\ldots)}$ are the momenta conjugated\footnote{Here and in the sequel we choose right
derivatives for Grassmannian variables: $\Pi_{\z}:=\frac{\overleftarrow{\p}L}{\p\z}$.} to the variables indicated as $({\s\ldots})$, which
satisfy the following (graded) Poisson Brackets\footnote{In the sequel we shall omit the subcripts $+$
and $-$.}:
	\begin{equation}
	\begin{split}
	& \big[\l,\Pi_{\l}\big]_{-}=1; \hspace{2cm} \big[x^{\mu},p_{\nu}\big]_{-}=\delta^{\mu}_{\nu};
	\\
	& \big[\z^{\,\s a},(\Pi_{\z})_{\s b}\big]_{+}=\delta^{\s a}_{\s b}; \hspace{1.2cm}  
	\big[\zb_{\:\s a},(\Pi_{\zb})^{\s b}\big]_{+}=\delta_{\s a}^{\s b}.
	\end{split}
	\end{equation}
\noindent
The first thing to do is to construct the Dirac Brackets associated to the 
$2^{\text{nd}}$-class constraints. If we define the matrix 
	\begin{equation}
	\label{delta}
	\Delta_{ij}=[\phi_i,\phi_j]_{\s PB}
	\end{equation}
where $\phi_k$ are the second class constraints, we have that the Dirac Brackets between
two generic variables $A,B$ of phase space are defined as:
	\begin{equation}
	\label{1-8}
	[A,B]_{\s DB}=[A,B]_{\s PB}-[A,\phi_i]_{\s PB} 
	(\Delta^{-1})^{ij}[\phi_j,B]_{\s PB}.
	\end{equation}
\noindent
It is not difficult even if rather long (the details are confined in 
Appendix \ref{app:kappa}) to find out that the Dirac Brackets at hand are:
	\begin{gather}
	[x^{\mu},p_{\nu}]_{\s DB}=\delta^{\mu}_{\nu}; \label{1-13} \vspace*{0.3cm} \\
	[\z^{\,\s a} ,(\Pi_{\z})_{\s b}]_{\s DB}=\displaystyle\frac{1}{2}\delta^{\s a}_{\s b};
	 ~~~~[\zb_{\s a},(\Pi_{\zb})^{\s b}]_{\s DB}=\displaystyle\frac{1}{2}\delta_{\s a}^{\s
	 b}; \vspace*{0.3cm}\\
	[\z^{\,\s a},\zb_{\:\s b}]_{\s DB}=i(\pslsh^{\:-1})^{\s a}_{\s b}; \vspace*{0.3cm}\\
	[x^{\mu},\zb_{\:\s b}]_{\s DB}=-\displaystyle\frac{(\zb\gamma^{\mu}\pslsh^{\:-1})_{\s
	b}}{2}; \\
	[\z^{\,\s a},x^{\mu}]_{\s DB}=\displaystyle\frac{(\pslsh^{\:-1}\gamma^{\mu}\z)^{\s
	a}}{2}; \\
	[x^{\mu},(\Pi_{\z})_{\s b}]_{\s DB}=-\displaystyle\frac{i}{4}(\zb\gamma^{\mu})_{\s
	b}\vspace*{0.3cm}\\
	[x^{\mu},(\Pi_{\zb})^{\s a}]_{\s DB}=-\displaystyle\frac{i}{4}(\gamma^{\mu}\z)^{\s
	a}\label{1-17}
	\end{gather}	

\noindent
Once we have built the correct structure in phase space, it is not difficult to
realize that the generators of the {\it global} supersymmetry are the following operators:
	\begin{align}
	\label{1-18}
	Q=\pslsh\,\z; \\
	\ov{Q}=\zb\pslsh; \label{1-18b}
	\end{align}
\noindent 
which reproduce precisely the transformations (\ref{1-3}) if we define:
	\begin{equation}
	\label{1-18c}
	\delta({\s\ldots})\equiv\big[({\s\ldots}),i\ov{\varepsilon}Q-i\ov{Q}\varepsilon\big]_{\s DB}.
	\end{equation}
\noindent
Note that the minus sign in the RHS of the previous equation is chosen because of the anticommuting
character of the parameter $\varepsilon$. Moreover we have:
	\begin{equation}
	\label{1-19}
	\big[ Q,\ov{Q}\big]_{DB}=i\pslsh,
	\end{equation}
\noindent
which confirms that $Q$ and $\ov{Q}$ are two supersymmetry charges. 
Notice that we can induce the same Susy-transformations through the following operators:
	\begin{align}
	\label{1-20}
	Q^{\prime}=i\Pi_{\zb}+\frac{1}{2}\pslsh\z; \\
	\ov{Q}^{\prime}=i\Pi_{\z}+\frac{1}{2}\zb\pslsh; \label{1-21}
	\end{align}
\noindent
which is obvious because $Q\approx Q^{\prime}$ and $\ov{Q}\approx \ov{Q}^{\prime}$.

Let us now switch to the massless case (\ref{1-1}). The main difference is that we cannot 
repeat all the steps of the previous analysis. In fact the new constraint $p^2=0$ implies that
the matrix $\Delta$ of Eq.(\ref{delta}) is no longer invertible. This is due to the fact
that $\det\Delta\propto\det(\pslsh)\propto (p^{\mu}p_{\mu})^2\approx 0$. Thus 
the construction of the Dirac Brackets is not as simple as in the massive case. In fact half of 
the constraints in Eq.(\ref{1-7}) are now $1^{\text{st}}$-class while the other half remains 
$2^{\text{nd}}$-class and the separation of the two sets is not quite easy: see for example
Refs.\cite{Kallosh}. Nevertheless we can list the generators 
of the $\kappa$-transformations of Eq.(\ref{1-4}):
	\begin{align}
	K=i\pslsh D =i\pslsh\Pi_{\zb}-\frac{1}{2}\pslsh^{\;2}\z; \label{1-22}\\
	\ov{K}=i\ov{D}\pslsh=i\Pi_{\z}\pslsh-\frac{1}{2}\zb\pslsh^{\;2} \label{1-23}
	\end{align}
\noindent
($K$ and $\ov{K}$ generate the transformation (\ref{1-4}) through commutators like those in (\ref{1-18c})). 
Obviously we should remember that $(K,\ov{K})$ are not a set of independent constraints because
of the reason claimed above ($\pslsh$ is not invertible on the shell of the contraints).
Note that we can write down the form of the generators $K$, $\ov{K}$ even if we do not know exactly 
the form of the Dirac Brackets in this particular case. 
We can do that because the $K,\ov{K}$ constraints commute (weakly) 
with all the constraints in (\ref{1-7}{\it c}) and (\ref{1-7}{\it d}) and therefore we have 
$[K,({\s\ldots})]_{DB}\approx [K,({\s\ldots})]_{PB}$ (and the same holds for $\ov{K}$) whatever are the 
surviving $2^{\text{nd}}$-class constraints determining the Dirac Brackets at hand. 
	
\section{$\kappa$-symmetry and CPI}

\noindent
In Chapter 1 we showed that the formalism of the Classical Path Integral
exhibits a universal {\it global} Supersymmetry.
However, differently from the model of Siegel, it does not possess any local invariance.
If we want to build up a nonrelativistic analog of the model introduced in Section 1,
we firstly must inject the local $t$-reparametrization invariance into the Lagrangian
(\ref{CPI6}) by adding the corresponding constraint via a Lagrange multiplier $g$:
 	\begin{equation}
	\label{3-1}
	\LT_1\equiv\LT + g\HT.
	\end{equation}
\noindent
In fact it is easy to see that the previous Lagrangian is {\it locally} invariant under 
   	\begin{equation}
	\label{3-2}
	\begin{cases}
	\delta({\s\ldots})=\big[({\s\ldots}),\e(t)\HT\big] \\
	\delta g=-i\dot{\e}(t); 
	\end{cases}
	\end{equation}
\noindent 
where $({\s\ldots})$ denotes one of the variables $(\v^a,\l_b,c^a,\bc_b)$.
Moreover it is easy to check that it remains {\it globally} invariant under the $N=2$
classical Susy of Eqs.(\ref{CPI28}) and (\ref{CPI29}).
Nevertheless, in this simple model no local Susy is still present. 
If we want to complete the analogy, we must
add (following the lines of Ref.\cite{DG}) two further constraints 
to the Lagrangian (\ref{3-1}) and we get: 
 	\begin{equation} 
	\label{3-16}
	\LT_2\equiv\LT + \xi D_{\s H} + \ov{\xi}\ov{D}_{\s H} + g\HT.
	\end{equation} 
\noindent
In the previous equation $D_{\s H}$ and $\ov{D}_{\s H} $ are the operators introduced 
in Eqs.(\ref{CPIcov1}) and (\ref{CPIcov2}). 
We want to analyze this model following the same steps we used in Section 1 for the Lagrangian
(\ref{1-1}).   

First of all we remember again that, in our non-relativistic case, the analog of the ``$p^2=0$"
constraint 
is represented by the last term $g\HT$, which produces the constraint $\HT=0$. 
Thus, as we did in Eq.(\ref{1-5}), we start our analysis by releasing this constraint in the following
way:  
 	\begin{equation} 
	\label{3-17}
	\LT^{\prime}_2\equiv\LT + \xi D_{\s H} + \ov{\xi}\ov{D}_{\s H} + g(\HT-\widetilde{E}),
	\end{equation} 
\noindent 
which is the analog of Eq.(\ref{1-5}). It should be remembered that $\widetilde{E}$ is not the
energy of the system, but just a parameter related to the invariance under local time
reparametrization:
if $\widetilde{E}=0$ this symmetry is present, while if $\widetilde{E}\neq 0$ this symmetry is lost.

One can immediately work out the constraints:
	\begin{align}
	\text{First Class\hspace{.5cm}} 
	& 
	\begin{cases}
	\Pi_{\s \xi} =\Pi_{\s \ov{\xi}}=\Pi_{\s g}=0; \\
	\HT-\widetilde{E}=0;
	\end{cases} 
	\vspace{.5cm} \label{3-18}\\
	\text{Second Class\hspace{.3cm}}
	& 
	\begin{cases}
	D_{\s H}= 0; \\
	\ov{D}_{\s H} = 0, 
	\end{cases}\label{3-19}
	\end{align} 
\noindent
Now we can compare the previous constraints with those in Eqs.(\ref{1-6}) and (\ref{1-7}). Concerning
the
$1^{\text{st}}$-class constraints, we notice that $\HT-\widetilde{E}=0$ is the classical analog of the
relativistic mass-shell constraint $p^{\mu}p_{\mu}-m^2=0$. This implies that $\Pi_{\s g}=0$ plays the
same
role as $\Pi_{\s\l}=0$ in the relativistic case, while the remaining two constraints ($\Pi_{\s\z}=0$
and
$\Pi_{\s\zb}=0$) have no analog in the relativistic case. Consider now the $2^{\text{nd}}$-class
constraints.
The first thing to point out is that $D_{\s H}= 0$ and $\ov{D}_{\s H} = 0$ are precisely the classical
analogs of $D^{\s a}=0$ and $\ov{D}_{\s b}=0$ in the relativistic case. We can say that because $D_{\s
H}$ and
$\ov{D}_{\s H}$ are related to the classical Susy charges $\QH$ and $\QBH$ in the same way in which
$D^{\s a}$
and $\ov{D}_{\s b}$ are related to the relativistic Susy charges $Q^{\s a}$ and $\ov{Q}_{\s b}$.
In fact it is easy to see that in the relativistic framework $D^{\s a}$ and $\ov{D}_{\s b}$ commute
with 
$Q^{\s a}$ and $\ov{Q}_{\s b}$ and $[D^{\s a},\ov{D}_{\s b}]=[Q^{\s a},\ov{Q}_{\s b}]=i\pslsh~^{\s a}_{\s
b}$
in the same way in which, in the nonrelativistic context, $D_{\s H}$ and $\ov{D}_{\s H}$ commute with 
$Q_{\s H}$ and $\ov{Q}_{\s H}$ and $[D_{\s H},\ov{D}_{\s H}]=[Q_{\s H},\ov{Q}_{\s H}]=2i\beta\HT$.    
This is actually the kernel of the analogy. 
We start from a model which possesses a universal Susy generated by $\QH$
and $\QBH$ and we want to check whether it is possible to implement a classical analog of the
relativistic
$\kappa$-symmetry of Siegel. Since in the relativistic case the $2^{\text{nd}}$-class constraints are
$D^{\s a}=0$ and $\ov{D}_{\s b}=0$, we modify the CPI-Lagrangian (\ref{CPI6}) in such a way that the resulting
extension provides as $2^{\text{nd}}$-class constraints the classical analogs of $D^{\s a}$ and
$\ov{D}_{\s
b}$, that is $D_{\s H}$ and $\ov{D}_{\s H}$. This is precisely the model (\ref{3-17}).

If we go on with the same steps as in Section 1 we find that 
the matrix\break $\Delta_{ij}=[\phi_i,\phi_j]$ has the form:
	\begin{equation} 
	\label{3-20a}
	\Delta =
	\begin{pmatrix}
	0 & 2i\beta\HT \\
	2i\beta\HT & 0
	\end{pmatrix}
	\Longrightarrow
	\Delta^{-1}=
	\begin{pmatrix}
	0 & (2i\beta\HT)^{-1} \\
	(2i\beta\HT)^{-1} & 0
	\end{pmatrix}
	\end{equation} 
\noindent  
and consequently the Dirac Brackets deriving from (\ref{3-19}) have the form:
 	\begin{equation} 
	\label{3-20b}
	\big[A,B\big]_{\s DB}=\big[A,B\big]- \big[A,\ov{D}_{\s H}\big](2i\beta\HT)^{-1}
	\big[D_{\s H},B\big]-\big[A,D_{\s H}\big](2i\beta\HT)^{-1}
	\big[\ov{D}_{\s H},B\big].
	\end{equation} 
\noindent
Now that we have the correct structure of our phase space we can proceed with the analogy with the
relativistic case. First of all we can prove that the two supersymmetry charges $\QH$ and $\QBH$
introduced in 
Eqs.(\ref{CPI28})(\ref{CPI29}) become weakly equal to the $\Qb$ and $\QBb$ charges:
 	\begin{align} 
	& \QH\approx 2\Qb=2ic^a\l_a; \label{3-21} \\
	& \QBH\approx 2\QBb=2i\bc_a\w^{ab}\l_b; \label{3-22} 
	\end{align}
\noindent and consequently:	
	\begin{align}
	& \big[\Qb,\QBb\big]_{\s DB}=\displaystyle\frac{1}{4}\big[\QH,\QBH\big]_{\s DB}
	=\frac{i\beta}{2}\HT. \label{3-23}
	\end{align}
\noindent
This shows that $\QH$ and $\QBH$ are 
the analogs\footnote{This is not in contradiction with what we said few lines above, that 
is that $\QH$ and $\QBH$ are the nonrelativistic analogs of $Q^{\s a}$ and $\ov{Q}_{\s b}$. In fact 
it should be remembered that on the shell of the contraints we have $Q\approx Q^{\prime}$, 
$\ov{Q}\approx \ov{Q}^{\prime}$ (in the relativistic case) and $\QH\approx 2\Qb$,
$\QBH\approx 2\QBb$ (in the nonrelativistic case).} 
of the charges  $Q^{\prime}$ and $\ov{Q}^{\prime}$ of Eqs.(\ref{1-20}) and (\ref{1-21}) 
while the $\Qb$ and $\QBb$ charges are analogous 
to the $Q$ and $\ov{Q}$ charges of Eqs.(\ref{1-18}) and (\ref{1-18b}). 

Consider now the case in which $\widetilde{E}=0$. We reduce to the Lagrangian (\ref{3-16}) 
and we see that something happens which is
similar to the mechanism of $\kappa$-symmetry discussed in Section 1. In fact in that case we 
saw that half of the $2^{\text{nd}}$-class constraints became $1^{\text{st}}$-class.
Here, on the other side, we notice that 
both the  $2^{\text{nd}}$-class constraints $D_{\s H}=\ov{D}_{\s H}=0$ become $1^{\text{st}}$-class. In
other words all the constraints in the model (\ref{3-16}) are gauge constraints and contribute to
restrict the space of the physical states. Therefore we see that in our nonrelativistic framework
there is no difficulty in separating $1^{\text{st}}$-class from $2^{\text{nd}}$-class constraints (like
in the relativistic case). This is simply due to the fact that no  $2^{\text{nd}}$-class constraint
remains after imposing the constraint $\HT=0$ (which is the classical analog of $p_{\mu}p^{\mu}=0$).
Proceeding with the analogy it is very easy to 
construct the classical analogs of $K$ and $\ov{K}$ of Eqs.(\ref{1-22})(\ref{1-23}), that is 
the generators of the nonrelativistic $\kappa$-symmetry. They are simply ($NR$ stands for
``Non Relativistic"):
  	\begin{align}
	& K_{\s NR}=\HT D_{\s H};\\
	& \ov{K}_{\s NR}=\HT\ov{D}_{\s H};	
	\end{align} 
\noindent 
and the local transformations (under which the Lagrangian (\ref{3-16}) is invariant) generated by
$K_{\s NR}$ and 
$\ov{K}_{\s NR}$ are:
   	\begin{equation}
	\begin{cases}
	\delta({\s\ldots})=\big[({\s\ldots}),\varkappa(t)K_{\s NR}+\ov{\varkappa}(t)\ov{K}_{\s NR}\big]
	\\
	\delta\xi=-2i\dot{\varkappa}\HT \\
	\delta\ov{\xi}=-2i\dot{\ov{\varkappa}}\HT  \\
	\delta g=-2i\beta(\ov{\xi}\varkappa+
	\xi\ov{\varkappa})\HT. 
	\end{cases}
	\end{equation}
\noindent
It is interesting to determine the physical states
selected by the theory defined by Eq.(\ref{3-16}). Since all the constraints are now
$1^{\text{st}}$-class, 
we must impose them on the states as follows:
  	\begin{align}
	&\Pi_{\s \xi} \,\rho(\v,c;t) =\Pi_{\s \ov{\xi}}\,\rho(\v,c;t)=\Pi_{\s g}\,\rho(\v,c;t)=0; \\
	&D_{\s H}\,\rho(\v,c;t)= 0; \\
	&\ov{D}_{\s H} \,\rho(\v,c;t)= 0; \\
	&\HT\,\rho(\v,c;t)=0;
	\end{align} 
\noindent
and it is not difficult to obtain that the resulting (normalizable\footnote{Also a state of the form
$\rho(\v,c)\propto \exp[\beta H(\v)]c^1c^2\ldots c^{2n}$ would be admissible, but it is not
normalizable in
$\v$.}) physical states have the following shape:
 	\begin{equation} 
	\label{3-28}
	\rho(\v,c)\propto \exp[-\beta H(\v)].
	\end{equation} 
\noindent
This is precisely the {\it Gibbs} distribution characterizing the {\it canonical} ensemble, provided we
interpret the $\beta$ constant of Eqs.(\ref{CPI28})(\ref{CPI29}) as $(k_{\s B}T)^{-1}$, where
$T$ plays the role of the temperature at which the system is in equilibrium.
In fact we should remember that up to now the dimensional parameter $\beta$ introduced in 
Eqs.(\ref{CPI28}) and (\ref{CPI29}) has not been restricted by any constraint. It is a 
completely free parameter with a dimension of $(\text{\it Energy})^{-1}$ which characterizes
the particular $N=2$ classical supersymmetry. 
The canonical Gibbs state made its appearance earlier in the context of the CPI and precisely in
the first of Refs.\cite{Ergo}.
There it was shown that, in the pure CPI model (\ref{CPI6}), the zero eigenstates of $\HT$ which are
also Susy-invariant are precisely the canonical Gibbs states. In our model instead we have obtained the
Gibbs states as the entire set of physical states associated to the gauge theory described by 
the Lagrangian (\ref{3-16}).

However the model (\ref{3-16}), though interesting for the peculiar physical subspace it determines, 
is not the nonrelativistic Lagrangian which is closest to the Siegel model. We mean that one should
remember
that the Lagrangian (\ref{3-16}) gives rise to a canonical Hamiltonian of the form:
 	\begin{equation} 
	\label{3-29}
	\HT_{2}\equiv\HT -\xi D_{\s H} -\ov{\xi}\ov{D}_{\s H} - g\HT,
	\end{equation} 
\noindent
but on the other hand we have already checked that the two couples of operators $(\QH,\QBH)$ and 
$(D_{\s H},\ov{D}_{\s H})$ close on $\HT$ and not on $\HT_{2}$. Therefore, if we want to 
construct a more precise nonrelativistic analog of the model of Siegel, we should consider a 
slightly modified version of the Lagrangian (\ref{3-16}) which is:
 	\begin{equation} 
	\label{3-30}
	\LT_3\equiv\LT + \dot{\xi} D_{\s H} + \dot{\ov{\xi}}\ov{D}_{\s H} + g\HT.
	\end{equation} 
\noindent
One can easily check that the Lagrangian (\ref{3-30}) yields, a part from a factor $(1-g)$, the same
Hamiltonian as the 
CPI. Therefore we can proceed following the same steps as before: we turn the $\HT=0$ constraint into 
$\HT-\widetilde{E}=0$
 	\begin{equation} 
	\label{3-31}
	\LT^{\prime}_3\equiv\LT + \dot{\xi} D_{\s H} + \dot{\ov{\xi}}\ov{D}_{\s H} + g(\HT-\widetilde{E})
	\end{equation} 
\noindent
and we find out that the new constraints are:     
	\begin{align}
	&\text{$1^{\text{st}}$-Class\hspace{.25cm}}  
	\begin{cases}
	\Pi_{\s g}=0; \\
	\HT-\widetilde{E}=0;
	\end{cases} 
	&\text{$2^{\text{nd}}$-Class\hspace{.25cm}} 
	\begin{cases}
	\Pi_{\s\xi}+D_{\s H}\equiv D^{\prime}_{\s H}= 0; \\
	\Pi_{\s\ov{\xi}}+\ov{D}_{\s H}\equiv D^{\prime}_{\s H}= 0.
	\end{cases}\label{3-32}
	\end{align} 
\noindent
Then, it is easy to check that we can repeat all the considerations we did below Eq.(\ref{3-19}), if
we replace $D_{\s H}$ and $\ov{D}_{\s H}$ with $D^{\prime}_{\s H}$ and $\ov{D}^{\prime}_{\s H}$. As a
second
remark, we notice that the two constraints $\Pi_{\s\xi}=\Pi_{\s\ov{\xi}}=0$, which had no analog in the
relativistic context, have now disappeared. Moreover, because $\big[D^{\prime}_{\s
H},\ov{D}^{\prime}_{\s H}\big]=
\big[D_{\s H},\ov{D}_{\s H}\big]=2i\beta\HT$, we have also that the Dirac Brackets remain the same as
those in
Eq.(\ref{3-20b}), which lead to Eqs.(\ref{3-21})-(\ref{3-23}). Again, when we put $\widetilde{E}=0$, we
obtain
that the two $2^{\text{nd}}$-class constraints $D^{\prime}_{\s H}=\ov{D}^{\prime}_{\s H}=0$ become both
$1^{\text{st}}$-class, differently from the relativistic case. However, the two models described by the
two Lagrangians (\ref{3-16}) and (\ref{3-30})
are not equivalent. There are basically two differences. The first is the new 
form of the nonrelativistic $\kappa$-symmetry which now reads:   
   	\begin{align}
	&\begin{cases}
	\delta({\s\ldots})=\HT\big[({\s\ldots}),\varkappa(t)D_{\s H}+\ov{\varkappa}(t)\ov{D}_{\s H}\big] 
	\approx
	\big[({\s\ldots}),\varkappa(t)K^{\prime}_{\s
	NR}+\ov{\varkappa}(t)\ov{K}^{\prime}_{\s NR}\big] \\
	\delta\xi=\big[\xi,\varkappa(t)K^{\prime}_{\s NR}+\ov{\varkappa}(t)\ov{K}^{\prime}_{\s NR}\big]
	=-i\varkappa\HT \\
	\delta\ov{\xi}=\big[\ov{\xi},\varkappa(t)K^{\prime}_{\s NR}+\ov{\varkappa}(t)\ov{K}^{\prime}_{\s
	NR}\big]
	=-i\ov{\varkappa}\HT  \\
	\delta g=2i\beta\HT(\dot{\ov{\xi}}\varkappa+
	\dot{\xi}\ov{\varkappa}). 
	\end{cases} 
	\end{align}
\noindent where ``$\approx$" is understood in the Dirac sense and
	\begin{align}
	&
	K^{\prime}_{\s NR}\equiv\HT D^{\prime}_{\s H}; \hspace{2cm} \ov{K}^{\prime}_{\s NR}\equiv\HT
	\ov{D}^{\prime}_{\s H}.
	\end{align}
\noindent
The second difference, which is the most important, is represented by the two physical spaces
associated to the two models 
(\ref{3-16}) and (\ref{3-30}). In fact we have already seen that the physical states associated to the 
first model are the Gibbs distributions $\rho(\v)\propto\exp(-\beta H(\v))$; on the other hand the
physical 
states determined by the Lagrangian (\ref{3-30}) must obey the following conditions:
  	\begin{align}
	&\Pi_{\s g}\,\rho(\v,c,\xi,\ov{\xi},g)=0~; 
	&D^{\prime}_{\s H}\,\rho(\v,c,\xi,\ov{\xi},g) =\big(-i\p_{\xi}+D_{\s H}\big)
	\rho(\v,c,\xi,\ov{\xi},g)=0~; \label{3-35}\\
	&\HT\,\rho(\v,c,\xi,\ov{\xi},g)=0~;
	&\ov{D}^{\prime}_{\s H}\,\rho(\v,c,\xi,\ov{\xi},g) =\big(-i\p_{\ov{\xi}}+
	\ov{D}_{\s H}\big)\rho(\v,c,\xi,\ov{\xi},g)=0~.  \label{3-38}
	\end{align} 
\noindent
It is not difficult to realize that the solution of Eqs.(\ref{3-35})-(\ref{3-38}) has the form:
	\begin{equation}
	\rho(\v,c,\xi,\ov{\xi},g)\propto\exp\big(-i\xi D_{\s H}-i\ov{\xi}\ov{D}_{\s
	H}\big)\tilde{\rho}(\v,c)~,\label{3-39}
	\end{equation}
\noindent where	
	\begin{equation}
	\HT\,\tilde{\rho}(\v,c)=0~, \label{3-40}
	\end{equation}
\noindent
which implies that $\tilde{\rho}(\v,c)$ is a function of constants of motion only. 
Therefore we can say that the physical states associated to the Lagrangian (\ref{3-30}) are isomorphic
to the functions $\tilde{\rho}(\v,c)$ which are annihilated by the Hamiltonian $\HT$ and are
consequently constants of motion. Obviously the Gibbs distributions are a subset of them. This
allows us to claim that the model (\ref{3-30}) is actually more general than 
that characterized by the Lagrangian (\ref{3-16}). More precisely the theory described by 
(\ref{3-30}) is equivalent to that characterized by
the Lagrangian (\ref{3-1}). In fact it is easy to see that the physical Hilbert space associated to the
latter is characterized by the distributions $\tilde{\rho}(\v,c,g)$ obeying to the constraints:
	\begin{align}
	\label{3-41}
	&\displaystyle\frac{\p}{\p g}\tilde{\rho}(\v,c,g)=0; 
	&\HT\,\tilde{\rho}(\v,c,g)=0; 
	\end{align}
\noindent
and the physical space is precisely the same as that in (\ref{3-40}), which is isomorphic to that
determined by Eqs.(\ref{3-35})-(\ref{3-38}).

\chapter*{4. The Rescaling of the Action}
\addcontentsline{toc}{chapter}{\numberline{4}The Rescaling of the Action}
\setcounter{chapter}{4}
\setcounter{section}{0}
\markboth{4. The Rescaling of the Action}{}

\vspace{1cm}
\noindent
In the previous chapters we studied the geometrical meaning of some of the 
symmetry charges listed in Section 1.3. Here we want to continue this project, and 
we focus on a very particular symmetry which we have not introduced yet.
The reason we introduce it only now is that this new transformation does not possess
a canonical generator like all the other symmetries introduced in Section 1.3. This is due to
the fact that the effect of the former is a rescaling of the action
$\widetilde{S}$ of the CPI, which is not a canonical transformation. Nevertheless, if we go to
the superspace $(t,\t,\tb)$, we see that we can find the expression of the generator, which
we shall call ${\mathscr Q}_{\s S}$, through which we can recontruct the transformation on all the
variables \quattrova.  

\section{The symmetry charge ${\mathscr{Q}}_{\s S}$}
First of all let us come back to the superspace representation of the charges
(\ref{CPI49})-(\ref{CPI54}):
	\begin{gather}
	{\mathscr Q}_{\s BRS}=-\displaystyle\frac{\p}{\p\t};~~~~
	{\ov{\mathscr Q}}_{\s BRS}=\displaystyle\frac{\p}{\p\tb}; \label{CPI49b}\\
	{\mathscr K}=\tb\displaystyle\frac{\p}{\p\t};~~~~
	{\ov{\mathscr K}}=\t\displaystyle\frac{\p}{\p\tb}; \\
	{\mathscr Q}_{g}=\tb\displaystyle\frac{\p}{\p\tb}-
	\t\displaystyle\frac{\p}{\p\t}; \\
	{\mathscr N}_{\s H}=\tb\displaystyle\frac{\p}{\p t};~~~~
	{\ov{\mathscr N}}_{\s H}=\t\displaystyle\frac{\p}{\p t}; \\
	{\mathscr Q}_{\s H}=-\displaystyle\frac{\p}{\p\t}-
	\tb\displaystyle\frac{\p}{\p t};~~~~
	{\ov{\mathscr Q}}_{\s H}=\displaystyle\frac{\p}{\p\tb}+
	\t\displaystyle\frac{\p}{\p t}; \\
	\widetilde{\mathscr H}=i\displaystyle\frac{\p}{\p t}. \label{CPI54b}
	\end{gather}
It is easy to see that in this framework all the representations of the various charges
commute with ${\mathscr H}$, which is the superspace representation of the Hamiltonian
$\HT$. This is obvious because they are symmetries of the CPI formalism and they commute with
$\HT$ already at the level of the phase space $\MT$, where they are expressed as functions of
\quattrova and the commutators are those defined in Eq.(\ref{CPI11}). However, at the level of
superspace, it is not difficult to check that also the following operator 

	\begin{equation}
	\label{CME91}
	{\mathscr{Q}}_{\s S}=\theta{\partial\over\partial\theta}+{\bar\theta}{\partial
	\over{\partial\bar\theta}}
	\end{equation}
\noindent
does commute with $\mathscr H$. In this sense we can say that (\ref{CME91}) generates a symmetry of
the system at hand. The main difference with respect to all the other operators listed in
Eqs.(\ref{CPI49b})-(\ref{CPI54b}) is that ${\mathscr{Q}}_{\s S}$ does not correspond to any
operator in $\MT$, and this is the reason why we did not introduce it together with all the
others. The fact that ${\mathscr{Q}}_{\s S}$ does not correspond to any operator in $\MT$ is
strictly bound to the fact that it does not generate a {\it canonical} transformation on the
variables \quattrova. 
In fact, remember that the representation of an operator $O$ in superspace is defined by:
	\begin{equation}
	\label{CME91b}
	\delta\Phi^{a}=[\e O, \Phi^a]\equiv -\e\mathscr{O}\Phi^{a},
	\end{equation}	
\noindent
where the brackets $[{\s\ldots},{\s\ldots}]$ are the (canonical) commutators introduced either in
Eq.(\ref{CPI9}) or in Eq.(\ref{CPI11}). Consequently, if we are given an operator $\mathscr{O}$ and we can
find its counterpart $O$, it also means that $\mathscr{O}$ generates a canonical transformation on
$\MT$. Actually this is not the case of $\mathscr{Q}_{\s S}$. In fact, 
let us substitute Eq.(\ref{CME91}) into Eq.(\ref{CME91b}) and see what variation it induces
on the variables \quattrova:

	\begin{equation}
	\label{CME92}
	\delta_{{\scriptscriptstyle \QS}}\Phi^{a} =-\varepsilon{\mathscr{Q}}_{\s S}\Phi^{a}=
	-\varepsilon(\theta c^{a}+{\bar\theta}\omega^{ab}{\bar c}_{b}+2i{\bar\theta}
	\theta\omega^{ab}\lambda_{b}).
	\end{equation}

\noindent Comparing the coefficients with the same power of $\theta$ and ${\bar\theta}$
on both sides of the equation above, we get 

	\begin{align}
	\label{CME93}
	&\delta_{{\scriptscriptstyle\QS}}\v^{a} = 0\\
	&\delta_{{\scriptscriptstyle\QS}}c^{a} = -\varepsilon c^{a}\\
	&\delta_{{\scriptscriptstyle\QS}}{\bar c}_{a} = -\varepsilon {\bar c}_{a}\\
	&\delta_{{\scriptscriptstyle\QS}}\lambda_{a} = -2\varepsilon \lambda_{a}.
	\end{align}

\noindent It is easy to check how the CPI-Lagrangian (see Eq.(\ref{CPI6})) changes
under the variations above:

	\begin{equation}
	\label{CME97}
	\delta_{\QS}{\LT}= -2\varepsilon\LT.
	\end{equation}

\noindent
We can conclude that the transformations induced by $\mathscr{Q}_{\s S}$
on the variables\break \quattrova are a symmetry of our system. In fact they just rescale
the overall Lagrangian $\LT$ and therefore they keep the equations of motion invariant
and these qualifies them as symmetry transformations. Of course
they are {\it non-canonical} symmetries, as we have already claimed, 
because the rescaling of the whole
Lagrangian is not a canonical transformation in $\MT$.

We could ask whether this is a symmetry also of
the Susy-QM model of Witten\scite{Witten} or at least of the
conformal-QM of Ref.\cite{FUB} (which will be introduced in the following chapter).
Interestingly enough, the answer is {\it No!}, The 
technical reason
being that in those QM models the analogs of the $\lambda_{a}$ variables
enter the Lagrangian with a quadratic term while in our $\LT$ they enter linearly.
There is also an important physical reason why that symmetry is not present in those
QM models while it is present in our CM one. 
The reason is that in QM one cannot rescale the action (as our symmetry
does) due to the presence of $\hbar$ setting a scale for it, while this
can be done in CM where no scale is set. The reader may object that our
transformation rescales the Lagrangian but not the action

	\begin{equation}\label{CME98}
	{\widetilde S}=\int \LT~dt 
	\end{equation}

\noindent because one could compensate the rescaling of the $\LT$ with a rescaling
of $t$. That is not so because our $\mathscr{Q}_{\s S}$ transforms only the Grassmannian
partners of time ($\theta,{\bar\theta})$ and not the time itself.

Therefore, this generalized symmetry seems to be very meaningful in the transition from Classical
to Quantum Mechanics: in CM it is a true symmetry, in QM it is lost.
We hope to shed some further light on this issue in the future. 

For the time being, let us continue our analysis about rescaling the action, but from a 
slightly different point of view. 
Up to now we have dealt with a noncanonical rescaling of the Lagrangian $\LT$. If we look at the explicit
form (\ref{CPI6}) of this Lagrangian, we notice that the transformation (\ref{CME97})
can be induced by a rescaling of $H(\v)$ (or equivalently of $L(q,\dot{q})$) at the level of
the ordinary phase space $\cal M$. Then, the question which spontaneously arises is whether we
can think of the $\mathscr{Q}_{\s S}$-transformation as the extension of another symmetry of CM, formulated
at the level of $\cal M$. This is the question we shall try to answer in the next sections. In other
words we want to check whether there exists a {\it universal} transformation (because the 
$\mathscr{Q}_{\s S}$-rescaling is also universal) in the space $(q,t)$ whose
effect is the rescaling of the action of any classical system.

\section{The MSA transformation}

\noindent
Consider a classical system defined by a set of coordinates (in the configurations
space) denoted by $\{q(t)\}$, where $q(t)$ is an $N$-dimensional vector and $2N$ is 
the dimension of the phase space. The action of the system is 
obviously:
	\begin{equation}
	\label{MSA1}
	S(t,t_{\s 0})_{[q(\tau),\dot{q}(\tau)]}=\int_{t_0}^t d\tau L(q(\tau),\dq;\tau);
	\end{equation}
\noindent
where the LHS means that $S$ is an ordinary function of the endpoints $(t,t_{\s 0})$ and a
functional of the trajectories $\{\q\}$. In the following lines we want to find out
a {\it universal} transformation of the time $t$ which has the effect of rescaling 
the action (\ref{MSA1}) by a factor $(1+\e)$ (we consider for simplicity an infinitesimal 
rescaling) regardless of the form of the Lagrangian $L(\q,\dq,\tau)$ out of which the action
in question is built. We make a further requirement and we assume that the configuration
$q(t)$ be a scalar under the transformation we are looking for. This means that we are in search for
a transformation 
	\begin{equation}
	\label{MSA2}
	t^{\prime}=f(t,t_{\s 0})_{[\q,\dq]}=t + \delta t(t,t_{\s 0})_{[\q,\dq]}
	\end{equation}
with the following properties:
	\begin{equation}
	\label{MSA3}
	\begin{cases}
	\displaystyle\int_{t^{\prime}_0}^{t^{\prime}}d\tau^{\prime}\,L^{\prime}(\qp,\dqp,\tau^{\prime})=(1+\e)
	\int_{t^{\prime}_0}^{t^{\prime}} d\tau^{\prime}\, L(q^{\prime}(\tau),\dqp;\tau^{\prime});\\
	\qp=\q.
	\end{cases}
	\end{equation}
\noindent From 
now on we shall use the following notation: we denote by $\delta q$ and $\ov{\delta} q$ the 
following quantities:
	\begin{equation}
	\label{MSA4}
	\begin{cases}
	\delta q (\tau)\equiv \qp-\q;  \\
	\ov{\delta} q(\tau)\equiv q^{\prime}(\tau)-\q;
	\end{cases}
	\end{equation}
\noindent
the relation between $\delta q$ and $\ov{\delta} q$ is given by:
	\begin{equation}
	\label{MSA5}
	\ov{\delta} q(\tau) = -\dot{q}^{\prime}(\tau)\delta\tau + \delta q(\tau)
	\end{equation}
\noindent
which, by use of Eq.(\ref{MSA3}), reduces to:
	\begin{equation}
	\label{MSA6}
	\ov{\delta} q(\tau) = -\dot{q}^{\prime}(\tau)\delta\tau. 
	\end{equation}
\noindent
Finally the following variations will be useful:
	\begin{equation}
	\label{MSA7}
	\begin{cases}
	\delta\dot{q}(\tau) = \displaystyle\frac{d}{d\tau^{\prime}}\qp -\displaystyle\frac{d}{d\tau}\q
	=
	-\dot{q}(\tau)\dot{(\delta\tau)}; \vspace{.2cm}\\
	\ov{\delta} \dot{q}(\tau)
	=\displaystyle\frac{d}{d\tau}q^{\prime}(\tau)-\displaystyle\frac{d}{d\tau}\q 
	=\displaystyle\frac{d}{d\tau}\ov{\delta}{q}(\tau). 
	\end{cases}
	\end{equation}
\noindent
Now we can proceed and try to work out the transformation (\ref{MSA2}). First of all we can rewrite
the first equation in (\ref{MSA3}) in the following way\footnote{Remember that we have chosen $\delta q(\tau)
=0$ (See Eq.(\ref{MSA3})).}:
	\begin{multline}
	\label{MSA8}
	\displaystyle\int_{t^{\prime}_0}^{t^{\prime}}d\tau^{\prime}\,L^{\prime}(\qp,\dqp,\tau^{\prime})=
	\displaystyle\int_{t^{\prime}_0}^{t^{\prime}}d\tau^{\prime}\,L^{\prime}(\q,\dq+\delta\dot{q}(\tau),
	\tau+\delta\tau)=\\
	=\displaystyle\int_{t^{\prime}_0}^{t^{\prime}}d\tau^{\prime}\left[L^{\prime}(\q,\dq,\tau)+\frac{\p
	L^{\prime}}
	{\p\dot{q}}\delta\dot{q}+\frac{\p L^{\prime}}{\p\tau}\delta\tau\right].
	\end{multline}
\noindent
But now we can use the definition of transformed action $S\rightarrow S^{\prime}$:
	\begin{equation}
	\label{MSA9}
	S^{\prime}=\displaystyle\int_{t^{\prime}_0}^{t^{\prime}}d\tau^{\prime}\,L^{\prime}(\qp,\dqp,\tau^{\prime})=
	\int_{t_0}^t d\tau L(q(\tau),\dq,\tau) = S
	\end{equation}
\noindent
and rewrite Eq.(\ref{MSA8}) as follows:
	\begin{equation}
	\label{MSA10}
	\displaystyle\int_{t_0}^{t}d\tau\,L(\q,\dq,\tau)=
	\displaystyle\int_{t^{\prime}_0}^{t^{\prime}}d\tau^{\prime}\left[L^{\prime}(\q,\dq,\tau)+\frac{\p
	L^{\prime}}
	{\p\dot{q}}\delta\dot{q}+\frac{\p L^{\prime}}{\p\tau}\delta\tau\right].
	\end{equation}
\vspace{.3cm}\noindent
According to Eq.(\ref{MSA3}) we have that $L^{\prime}=(1+\e)L$ and consequently we can write:
	\begin{equation}
	\label{MSA11}
	\displaystyle\int_{t_0}^{t}d\tau\,L(\q,\dq,\tau)=(1+\e)
	\displaystyle\int_{t^{\prime}_0}^{t^{\prime}}d\tau^{\prime}\left[L(\q,\dq,\tau)+\frac{\p L}
	{\p\dot{q}}\delta\dot{q}+\frac{\p L}{\p\tau}\delta\tau\right]
	\end{equation}
\vspace{.3cm}\noindent
which, since $\e$ is an infinitesimal parameter, implies:
	\begin{equation}
	\label{MSA12}
	\displaystyle\int_{t_0}^{t}d\tau\frac{\p\tau^{\prime}}{\p\tau}\left[L(\q,\dq,\tau)+\frac{\p L}
	{\p\dot{q}}\delta\dot{q}+\frac{\p L}{\p\tau}\delta\tau\right]=(1-\e)
	\displaystyle\int_{t_0}^{t}d\tau\,L(\q,\dq,\tau).
	\end{equation}
\vspace{.3cm}\noindent
But remember that:
	\begin{equation}
	\label{MSA13}
	\frac{d\tau^{\prime}}{d\tau}=1+\frac{d}{d\tau}\delta\tau,
	\end{equation}
\vspace{.3cm}\noindent
and if we plug this result into Eq.(\ref{MSA12}) and keep only $1^{\text{st}}$-order terms, we obtain:
	\begin{equation}
	\label{MSA14}
	\int_{t_0}^t d\tau\frac{\p L}{\p\tau}\,\delta\tau -\int_{t_0}^t d\tau\frac{\p L}{\p\dot{q}}\,
	\dot{q}\,\dot{(\delta\tau)} + \int_{t_0}^t d\tau\,\dot{(\delta\tau)}\,L = 
	-\e\int_{t_0}^t d\tau L.
	\end{equation}
\vspace{.3cm}\noindent
Suppose for simplicity that $\displaystyle\frac{\p L}{\p\tau}=0$; then we can rewrite the previous equation
as:
	\begin{equation}
	\label{MSA15}
	\int_{t_0}^t d\tau\frac{d}{d\tau}(\delta\tau)\,L - \int_{t_0}^t d\tau\frac{\p L}{\p\dot{q}}\,
	\dot{q}\:\frac{d}{d\tau}(\delta\tau) = -\e\int_{t_0}^t d\tau\, L.
	\vspace{.3cm}
	\end{equation}
\noindent
The previous equation must hold for every choice of $(t_{\s 0},t)$ and therefore we arrive at the following
expression:
	\begin{equation}
	\label{MSA16}
	\frac{d}{d t}(\delta t) = -\e\: \frac{L(q(t),\dot{q}(t))}{L(q(t),\dot{q}(t))-\dot{q}(t)\,
	\displaystyle\frac{\p L}{\p\dot{q}(t)}}\:.
	\end{equation}
\vspace{.3cm}\noindent
If we denote by $H(q,\dot{q})$ the so-called ``energy function" \cite{Goldstein}:
	\begin{equation}
	\label{MSA17}
	H(q(t),\dot{q}(t))\equiv-L(q(t),\dot{q}(t))+\dot{q}(t)\,
	\displaystyle\frac{\p L}{\p\dot{q}(t)},
	\end{equation}
\vspace{.3cm}\noindent
we can rewrite Eq.(\ref{MSA16}) as
	\begin{equation}
	\label{MSA18}
	\frac{d}{d t}(\delta t) = \e\: \frac{L(q(t),\dot{q}(t))}{H(q(t),\dot{q}(t))}
	\end{equation}
\vspace{.3cm}\noindent
which finally yields:
	\begin{equation}
	\label{MSA19}
	\boxed{
	\delta t = \e\:\int_0^t d\tau\frac{L(q(\tau),\dot{q}(\tau))}{H(q(\tau),\dot{q}(\tau))} }\:.
	\vspace*{.5cm}
	\end{equation}
\noindent
This is precisely the transformation we were looking for. For the moment we consider $\e$ as a
global parameter and this is the reason why we kept it outside the integral in (\ref{MSA19}).
As we said in the Introduction, the name MSA is the acronym of ``Mechanical Similarity
Anomaly". The name ``Mechanical Similarity" was used by Landau \cite{Landau} to denote a 
transformation  which --- though in a completely different context --- had the effect of
rescaling the classical action. The reason for the word ``Anomaly", on the other hand, will be
clear in the next Section.

\section{MSA as a ``standard" symmetry}

\noindent
In the previous section we have built a general time-transformation which induces a rescaling of the
overall action of {\it every} classical system, regardless of the form of the Lagrangian defining the
system. Obviously this transformation is very peculiar; for example it is a {\it path-dependent}
transformation, and we shall come back on that later. The reason why we think it is worth analyzing 
this ``strange" symmetry is that we feel that its behaviour can become very interesting when we pass to the
quantum domain. In particular our goal would be to implement this transformation in the Quantum Path Integral
via a procedure \`a la Fujikawa \cite{Fuji}. What we could get is that this {\it universal classical}
symmetry becomes {\it anomalous} at the quantum level (that is where the ``Anomaly" word in MSA comes from)
and in this case it could be very interesting to evaluate this anomaly.  
Therefore, if we want to proceed along these lines, the first step we should do is to
find what is the conserved current associated to this transformation. The point is that, as
it stands, this is not a ``standard" symmetry because it does not leave the action invariant, but rescales
it. In this section we shall try to give the MSA a formulation in which it appears as a ``standard"
symmetry. 
The proposal we make is to enlarge the configuration space with two further variables $(\gamma, S)$
and consider a new kind of {\it extended} Lagrangian $L_{\s ext}$ defined as follows:
	\begin{equation}
	\label{MSA20}
	L_{\s ext}(q,\dot{q},\gamma,S)\equiv L(q,\dot{q})+\gamma(L(q,\dot{q})-\dot{S}).
	\end{equation} 
\noindent
Next let us extend the transformation (\ref{MSA19}) to the new enlarged configuration space in the
following way:

	\begin{equation}
	\label{MSA21}
	\begin{cases}
	\delta t = \e\:\displaystyle\int_0^t
	d\tau\frac{L(q(\tau),\dot{q}(\tau))}{H(q(\tau),\dot{q}(\tau))}\\
	\delta q=0 \\
	\delta S=? \\
	\delta\gamma= ?
	\end{cases}
	\end{equation} 

\noindent
and we want to find out which are the variations $\delta S$ and $\delta\gamma$ turning the
transformation (\ref{MSA21}) into a symmetry of the action associated to the 
Lagrangian $L_{\s ext}$ in Eq.(\ref{MSA20}). Clearly, in order to be a symmetry of the latter, the
variations in (\ref{MSA21}) must obey the following relation:
	\begin{equation}
	\label{MSA22}
	L_{\s ext}\big(q+\delta
	q,\dot{q}+\delta\dot{q},\gamma+\delta\gamma,\dot{S}+\delta\dot{S}\big)
	\frac{dt^{\prime}}{dt}=L_{\s ext}\big(q,\dot{q},\gamma,\dot{S}\big),
	\end{equation} 
\noindent
which, after a little algebra, can be rewritten as follows:
	\begin{equation}
	\label{MSA23}
	-(1+\gamma)\,H(q,\dot{q})\dot{(\delta t)}+(1+\gamma)\frac{\p L}{\p\dot{q}}\dot{(\delta q)}
	+(1+\gamma)\frac{\p L}{\p q}(\delta q)-\gamma\dot{(\delta S)}+(L-\dot{S})\,\delta\gamma = 0.
	\end{equation} 
\noindent
If we plug Eq.(\ref{MSA21}) into Eq.(\ref{MSA23}) we obtain:
 	\begin{equation}
	\label{MSA24}
	-(1+\gamma)\,\e\,L(q,\dot{q})-\gamma\dot{(\delta S)}+(L-\dot{S})\,\delta\gamma=0,
	\end{equation}  
\noindent
and we have the two following possibilities.
\subsection{Case $\delta\gamma=0$}

\noindent
We require $\delta\gamma=0$ in Eq.(\ref{MSA24}) and we are left with:
 	\begin{equation}
	\label{MSA25}
	\gamma\dot{(\delta S)}=-(1+\gamma)\,\e\,L(q,\dot{q}).
	\end{equation} 
\noindent	
The most general solution of the previous equation is given by the following expression:
  	\begin{equation}
	\label{MSA26}
	\delta S = -\frac{\e(\gamma+1)}{\gamma}\int_0^t d\tau\,L(\q,\dq).
	\end{equation}
\noindent
This in turn implies that the complete transformation is (remember Eq.(\ref{MSA21})):

	\begin{equation}
	\label{MSA27}
	\begin{cases}
	\delta t = \e\:\displaystyle\int_0^t
	d\tau\frac{L(q(\tau),\dot{q}(\tau))}{H(q(\tau),\dot{q}(\tau))}\\
	\delta q=0 \\
	\delta S=-\displaystyle\frac{\e(\gamma+1)}{\gamma}\int_0^t d\tau\,L(\q,\dq) \\
	\delta\gamma=0.
	\end{cases}
	\end{equation} 	   

\noindent
The next step is to find out what is the current associated to the symmetry transformation
(\ref{MSA27}). It is easy to see that this current is:
  	\begin{equation}
	\label{MSA28}
	\begin{split}
	Q_1 &= \frac{\p L_{\s ext}}{\p\dot{q}}\ov{\delta}q+\frac{\p L_{\s
	ext}}{\p\dot{S}}\ov{\delta}S+
	L_{\s ext}\delta t \\
	&=(1+\gamma)\,\delta t \,\displaystyle\left(-\frac{\p L}{\p\dot{q}}\dot{q}+L\right)-
	\gamma\,\delta S \\
	&=\e\,(1+\gamma)\displaystyle\left(-H\int_0^t d\tau\frac{L}{H}+\int_0^t d\tau\,L\right),
	\end{split}
	\end{equation}
\noindent
where in the last step we used the results in Eq.(\ref{MSA27}). Now, Noether's theorem states that
the current $Q_1$ is conserved on the shell of the equations of motion. Anyway, it is not difficult 
to check that $Q_1$ vanishes on the shell of the equations of motion ($H$ is a constant on this shell
and can be passed inside the integral), in such a way that Noether's theorem is satisfied in a
trivial way. More interesting is the following possibility. 
\subsection{Case $\delta\gamma\neq 0$}

\noindent
Differently from the previous case we make the choice $\delta\gamma=\e\,(1+\gamma)$ and from
Eq.(\ref{MSA24}) we obtain:
 	\begin{equation}
	\label{MSA29}
	\gamma\dot{(\delta S)}+\e\,\dot{S}\,(1+\gamma)=0;
	\end{equation} 
\noindent
this in turn implies that:
 	\begin{equation}
	\label{MSA30}
	\delta S=-\int_0^t d\tau\,\frac{\e\,(1+\gamma)}{\gamma}\,\dot{S} +const.
	\end{equation} 
\noindent
After an integration by parts we can rewrite the previous formula as:
 	\begin{equation}
	\label{MSA31}
	\delta S=-\frac{\e\,(1+\gamma)}{\gamma}\,S(t)+\e\int_0^t d\tau\,
	\left[\frac{d}{d\tau}\left(\frac{(1+\gamma)}{\gamma}\right)\right]\,S(\tau) +const;
	\end{equation} 
\noindent
We can now summarize the complete transformation in the following way:

	\begin{equation}
	\label{MSA32}
	\begin{cases}
	\delta t = \e\displaystyle\int_0^t
	d\tau\frac{L(q(\tau),\dot{q}(\tau))}{H(q(\tau),\dot{q}(\tau))};\\
	\delta q=0; \\
	\delta S=-\displaystyle\frac{\e\,(1+\gamma)}{\gamma}\,S(t)+\e\int_0^t d\tau\,
	\left[\frac{d}{d\tau}\left(\frac{(1+\gamma)}{\gamma}\right)\right]\,S(\tau) +const; \\
	\delta\gamma=\e\,(1+\gamma).
	\end{cases}
	\end{equation} 

\noindent
The conserved current associated to (\ref{MSA32}) is easily found:
  	\begin{equation}
	\label{MSA33}
	\begin{split}
	Q_2 &= \frac{\p L_{\s ext}}{\p\dot{q}}\ov{\delta}q+\frac{\p L_{\s
	ext}}{\p\dot{S}}\ov{\delta}S+
	L_{\s ext}\delta t \\
	&=\e\,(1+\gamma)\displaystyle\left(S-H\int_0^t d\tau\frac{L}{H}\right)
	-\e\,\gamma\int_0^t d\tau\,\dot{S}\frac{d}{d\tau}\left(\frac{1+\gamma}{\gamma}\right) +
	const;
	\end{split}
	\end{equation}
\noindent
This current, on the shell of the equations of motion, takes the form:
 	\begin{equation}
 	\label{MSA34}
	Q_2 = \e\,(1+\gamma)\displaystyle\left(S-\int_0^t d\tau\,L\right)+ const;
	\end{equation}	
\noindent
and Noether's theorem becomes:
 	\begin{equation}
 	\label{MSA35}
	\frac{d Q_2}{d t}=0 ~ \Longrightarrow ~ \frac{d S}{d t}=L.
	\end{equation}	
\noindent
Let us stop here for a while and open a brief parenthesis. Everybody knows that the Hamilton-Jacobi
equation of Classical Mechanics is:
  	\begin{equation}
 	\label{MSA36}
	\frac{\p S(q,t)}{\p t} + H\left(q,\frac{\p S(q,t)}{\p t}\right)=0,
	\end{equation}
\noindent
where $S(q,t)$ is called the {\it Hamilton generating function}. It is easy to check that the
previous equation is equivalent to the following set of equations:
   	\begin{equation}
 	\label{MSA37}
	\begin{cases}
	\displaystyle\frac{d S(q,t)}{d t} = L(q,\dot{q}),\vspace{.2cm} \\
	\displaystyle\frac{\p S(q,t)}{\p q} = \frac{\p L(q,\dot{q})}{\p \dot{q}};
	\end{cases}
	\end{equation}
\noindent
in fact, the first equation in (\ref{MSA37}) can be rewritten as
  	\begin{equation}
 	\label{MSA38}
	\frac{\p S(q,t)}{\p t} + \frac{\p S(q,t)}{\p q}\,\dot{q}-L(q,\dot{q})=0,
	\end{equation} 
\noindent
and if we substitute the second equation of (\ref{MSA37}) into (\ref{MSA38}), we finally obtain the
Hamilton-Jacobi equation (\ref{MSA36}). Now we see that the Noether theorem stated in
Eq.(\ref{MSA35}) corresponds to the first relation in (\ref{MSA37}), and therefore we are close to
our goal which was to find out the conserved current associated to the MSA
symmetry. Unfortunately it is not so simple to complete the project and obtain also the second
relation (the missing one in Eq.(\ref{MSA35})) of  (\ref{MSA37}). The reasons is basically twofold. Firstly it is not so
clear how to implement on $S(t)$, in a consistent way, a dependence on $q$ at the level of
Eq.(\ref{MSA35}). The second motivation is conceptually deeper and therefore it is described with
full details in the next section.  
\section{MSA as an anholonomic transformation}

\noindent
In the previous sections we have already claimed many times that the MSA-transformation is not 
a standard symmetry of Classical Systems for several reasons. One of them (maybe the most
important) is that this transformation is path-dependent. In particular it 
transforms the differentials $dq$ in such a way that they
cannot be integrated to a diffeomorphism, as we shall show in this section. 
These kinds of transformations are known in the literature as ``anholonomic transformations" 
\cite{Kleinert} and it is useful to spend few lines in describing what are their most important
features.

The most general definition of symmetry is a transformation which leaves invariant the equations
of motion of a dynamical system. In most cases, this requirement is equivalent to leaving
invariant the action of the system at hand, but there are cases in which this is no longer true
(the anholonomic transformations are one example). This problem never arises when we deal with
diffeomorphisms: for instance consider a dynamical system with Lagrangian $L(q,\dot{q},t)$ and
suppose to make the following transformation:
  	\begin{equation}
 	\label{MSA39}
	q \longrightarrow q^{\prime}(q,t);
	\end{equation} 
\noindent
the transformed Lagrangian is then given by:
  	\begin{equation}
 	\label{MSA40}
	L^{\prime}(q^{\prime},\dot{q}^{\prime};t)=L(q,\dot{q};t).
	\end{equation} 	
\noindent
What we are going to show is that if a function $\tilde{q}(t)$ is a solution of Lagrange
equations for $L$, then the function 
    	\begin{equation}
 	\label{MSA41}
	\tilde{q}^{\prime}(t)=q^{\prime}(\tilde{q},t)
	\end{equation} 
\noindent
is a solution of the Lagrange equations associated to $L^{\prime}$. In fact the transformed
Lagrange derivative\footnote{Given a function $f(q,\dot{q})$, the Lagrange derivative of
$f(q,\dot{q})$ is defined as 
\[ \frac{d}{dt}\left(\frac{\p f(q,\dot{q})}
{\p\dot{q}}\right)-\frac{\p f(q,\dot{q})}{\p q}. \]}
is:
  	\begin{equation}
 	\label{MSA42}
	\left[\frac{d}{dt}\left(\frac{\p L^{\prime}(q^{\prime},\dot{q}^{\prime};t)}
	{\p\dot{q}^{\prime}}\right)-\frac{\p L^{\prime}(q^{\prime},\dot{q}^{\prime};t)}{\p q^{\prime}}
	\right]_{\tilde{q}^{\prime}(t)},
	\end{equation} 		
\noindent
and if we use Eq.(\ref{MSA40}) in Eq.(\ref{MSA42}), we can rewrite it in the following way:
    	\begin{equation}
 	\label{MSA43}
	\left[\frac{d}{dt}\left(\frac{\p L}{\p\dot{q}}\right)
	\,\frac{\p\dot{q}}{\p\dot{q}^{\prime}}+\frac{\p L}{\p\dot{q}}\,
	\frac{d}{dt}\left(\frac{\p\dot{q}}{\p\dot{q}^{\prime}}\right)-\frac{\p L}{\p q}\,\frac{\p q}
	{\p q^{\prime}}-\frac{\p L}{\p\dot{q}}\,\frac{\p\dot{q}}{\p{q}^{\prime}}
	\right]_{\tilde{q}(t)} .
	\end{equation}
\noindent
Now, inverting Eq.(\ref{MSA39}), it is not difficult to show that:
    	\begin{equation}
 	\label{MSA44}
	\frac{\p\dot{q}}{\p\dot{q}^{\prime}}=\frac{\p q}{\p q^{\prime}},
	\end{equation}   
\noindent
and if we substitute Eq.(\ref{MSA44}) into Eq.(\ref{MSA43}) we obtain:
    	\begin{multline}
 	\label{MSA45}
	\left[\frac{d}{dt}\left(\frac{\p L^{\prime}}
	{\p\dot{q}^{\prime}}\right)-\frac{\p L^{\prime}}{\p q^{\prime}}
	\right]_{\tilde{q}^{\prime}(t)}=\\
	\left[\frac{d}{dt}\left(\frac{\p L}{\p\dot{q}}\right)-\frac{\p L}{\p
	q}\right]_{\tilde{q}(t)}\frac{\p q}{\p q^{\prime}}+
	\frac{\p L}{\p\dot{q}}\,\left[
	\frac{d}{dt}\left(\frac{\p{q}}{\p{q}^{\prime}}\right)
	-\frac{\p\dot{q}}{\p{q}^{\prime}}\right]_{\tilde{q}(t)}.
	\end{multline}
\noindent
But now, by hypothesis, we know that for the unprimed system the Lagrange equations hold:
  	\begin{equation}
 	\label{MSA46}
	\left[\frac{d}{dt}\left(\frac{\p L}
	{\p\dot{q}}\right)-\frac{\p L}{\p q}
	\right]_{\tilde{q}(t)}=0,
	\end{equation} 	
\noindent
and therefore the RHS of Eq.(\ref{MSA45}) reduces to the second term which is zero because:
    	\begin{equation}
 	\label{MSA47}
	\frac{d}{dt}\left(\frac{\p q}{\p q^{\prime}}\right)-\frac{\p\dot{q}}{\p{q}^{\prime}}
	=\frac{\p^2 q}{\p q^{\prime 2}}\,\dot{q}^{\prime}+\frac{\p^2 q}{\p t\p q^{\prime}}-
	\frac{\p}{\p q^{\prime}}\left[\frac{\p q}{\p q^{\prime}}\dot{q}^{\prime}+\frac{\p q}
	{\p t}\right]=0,
	\end{equation}  	 	
\noindent
where in the last step we used the fact that $\displaystyle\frac{\p\dot{q}^{\prime}}{\p
q^{\prime}}=0$. Finally we can write:
  	\begin{equation}
 	\label{MSA48}
	\left[\frac{d}{dt}\left(\frac{\p L^{\prime}}
	{\p\dot{q}^{\prime}}\right)-\frac{\p L^{\prime}}{\p
	q^{\prime}}
	\right]_{\tilde{q}^{\prime}(t)}=0
	\end{equation} 	
\noindent
which is precisely what we wanted to prove. Note that we could repeat all the steps above (longer
calculations are involved) if we considered, instead of Eq.(\ref{MSA39}), a more general
transformation like the following:
  	\begin{equation}
 	\label{MSA49}
 	\begin{cases}
	q^{\prime}=q^{\prime}(q,t); \\
	t^{\prime}=t^{\prime}(t);
	\end{cases}
	\end{equation} 
\noindent
but this is not necessary for our purposes and we shall not do it here. 
This for the diffeomorphisms. On the other hand,
a general anholonomic transformation $x^{\mu}\rightarrow y^{\rho}$ is
defined as:
   	\begin{equation}
 	\label{MSA50}
	\begin{cases}
	dx^{\mu}=e^{\mu}_{\rho}(y)\,dy^{\rho} \vspace{.2cm}\\
	\displaystyle\frac{\p e^{\mu}_{\rho}(y)}{\p y^{\si}}\neq\frac{\p e^{\mu}_{\si}(y)}{\p
	y^{\rho}} 
	\end{cases}
	\end{equation} 
\noindent
where the second condition corresponds to the failure of the Schwarz integrability condition. We can
rewrite the first of Eqs.(\ref{MSA50}) by introducing a new parameter $\tau$:
  	\begin{equation}
 	\label{MSA51}
	x^{\mu}(\tau)-x^{\mu}(0)=\int_0^{\tau}d\eta\:e^{\mu}_{\rho}(y)\frac{dy^{\rho}}{d\eta}.
	\end{equation} 
\noindent
If the Schwarz integrability condition were satisfied, Eq.(\ref{MSA51}) could be integrated (at
least in principle\footnote{The fact that it can be integrated does not mean that the 
result has necessarily an analytic expression.}) to obtain $x(\si)=x(q(\si))$ and we would get a 
diffeomorphism; in general this
is not possible. This means that an anholonomic transformation can relate a system which obeys
the Lagrange equations to a system where these equations are no longer satisfied.
This introduces us to the main point of this section. Up to now we have always called the
MSA-transformation a symmetry (even if particular) because its effect is the universal rescaling of
the classical action. Now we can ask: is this sufficient to call it a symmetry, or is there anything
else which must be taken into account? Actually we shall show that the MSA-transformation is
anholonomic and the rescaling of the action, in this case, does not imply the invariance of the
equations of motion. 
To show that, we could proceed in two ways. One possibility is to rephrase the MSA transformation
in a formalism in which the time $t$ is at the same level of $q$ (Maupertuis formalism) and prove
the failure of the Schwarz integrability condition (\ref{MSA50}). The other possibility is to 
prove directly that the MSA transformation does not leave invariant the equations of motion of the
system at hand. In the sequel we shall follow this strategy.

First of all, let us
recast the MSA transformation in a form similar to (\ref{MSA50}). Remember that we have:
   	\begin{equation}
 	\label{MSA52}
	\begin{cases}
	\delta t=\e\,\displaystyle\int_0^t d\tau\,\frac{L(q,\dot{q})}{H(q,\dot{q})} \vspace{.2cm}\\
	\delta q = 0;
	\end{cases}
	\end{equation} 
\noindent
and from the first equation we get:
   	\begin{equation}
 	\label{MSA53}
	t^{\prime}(t)-t=\e\,\int_0^t d\tau\,\frac{L(q,\dot{q})}{H(q,\dot{q})} \vspace{.1cm}
	\end{equation} 
\noindent
which implies:
   	\begin{equation}
 	\label{MSA54}
	dt^{\prime}= dt+\e\, dt \, R(q,\dot{q})\vspace{.1cm} 
	\end{equation} 
\noindent
where we have denoted by $R(q,\dot{q})$ the following ratio:
   	\begin{equation}
 	\label{MSA55}
	R(q,\dot{q})=\frac{L(q,\dot{q})}{H(q,\dot{q})}=\frac{m\dot{q}^2-2V(q)}{m\dot{q}^2+2V(q)}.
	\vspace{.1cm}
	\end{equation} 
\noindent From Eq.(\ref{MSA54}) we also obtain:
   	\begin{equation}
 	\label{MSA56}
	\frac{dt^{\prime}}{dt}=1+\e\,R(q,\dot{q}), \vspace{.1cm}
	\end{equation}
\noindent
which allows us to write:	 		
   	\begin{equation}
 	\label{MSA57}
	\dot{q}^{\prime}=\frac{dq^{\prime}(t^{\prime})}{dt^{\prime}}=
	\frac{dq^{\prime}(t^{\prime})}{dt}\frac{dt}{dt^{\prime}}=
	\frac{dq(t)}{dt}\frac{dt}{dt^{\prime}}=\dot{q}\big[1-\e\,R(q,\dot{q})\big], \vspace{.1cm}
	\end{equation}
\noindent
which in turn implies:
   	\begin{equation}
 	\label{MSA58}
	\ddot{q}^{\prime}=\frac{d}{dt^{\prime}}\big\{ \dot{q}\big[1-\e\,R(q,\dot{q})\big]\big\}=
	\ddot{q}\big[1-2\e\,R(q,\dot{q})\big]-\e\,\dot{q}\left(\frac{\p R}{\p q}\dot{q}
	+\frac{\p R}{\p\dot{q}}\ddot{q}\right). \vspace{.1cm}
	\end{equation}
\noindent
Now we have all the ingredients to check whether the equations of motion of the old (the unprimed)
system are really invariant, i.e. if they have the same form as the new (primed) ones.
Let us start from these last equations which are simply:
   	\begin{equation}
 	\label{MSA59}
	m\ddot{q}^{\prime}=-\frac{\p V(q^{\prime})}{\p q^{\prime}}, \vspace{.1cm}
	\end{equation}
\noindent
and substitute Eq.(\ref{MSA58}) into Eq.(\ref{MSA59}); we get:
   	\begin{multline}
 	\label{MSA60}
	m\ddot{q}^{\prime}=m\left\{
	\ddot{q}\big[1-2\e\,R(q,\dot{q})\big]-\e\,\dot{q}\left(\frac{\p R}{\p q}\dot{q}
	+\frac{\p R}{\p\dot{q}}\ddot{q}\right)\right\}=-\frac{\p V(q^{\prime})}{\p q^{\prime}}=
	-\frac{\p V(q)}{\p q}. \vspace{.1cm}
	\end{multline}
\noindent
Next, if we use $\displaystyle\frac{\p V}{\p q}=-m\ddot{q}$ in Eq.(\ref{MSA60}) we obtain:
   	\begin{equation}
 	\label{MSA61}
	2\ddot{q}\,R + \dot{q}^2\frac{\p R}{\p q} + \dot{q}\ddot{q}\frac{\p R}
	{\p\dot{q}}=0; \vspace{.1cm}
	\end{equation}	
\noindent
and we must check whether this last equation holds on the shell of the old equations of motion.
To do that we need the two following expressions:
	\begin{align}
	&\frac{\p R}{\p q}=-\frac{4m\dot{q}^2\frac{\p V}{\p q}}{(m\dot{q}^2+2V)^2};\\
	&\frac{\p R}{\p\dot{q}}=\frac{8m\dot{q}\,V}{(m\dot{q}^2+2V)^2};
	\end{align}
\noindent
which, when substituted in Eq.(\ref{MSA61}), yield:
   	\begin{equation}
 	\label{MSA64}
	2m^2\ddot{q}\dot{q}^4 - 8\ddot{q}\,V^2(q) - 4m\dot{q}^4\frac{\p V(q)}{\p q} +
	8m\dot{q}^2\ddot{q}\,V(q)=0.
	\vspace{.1cm}
	\end{equation}	
\noindent
Now, if we use again the unprimed equations of motion $\displaystyle\frac{\p V}{\p q}=-m\ddot{q}$ we
can simplify Eq.(\ref{MSA64}) and reduce it to the following form:
   	\begin{equation}
 	\label{MSA65}
	3m^2\dot{q}^4 + 4m\dot{q}^2\,V(q) - 4 V^2(q) =0;
	\vspace{.1cm}
	\end{equation}		
\noindent
but this equation is not satisfied for every classical system. 
Therefore we have proven that in general the equations of motion of a classical system are not
invariant under the MSA transformation. 
This means that in this case the rescaling of the action is not a sufficient condition for leaving
invariant the equations of motion (the Lagrange equations), which is the main property of a symmetry
transformation. Therefore we can conclude that the $\mathscr{Q}_{\s S}$-transformation which we have
introduced at the beginning of this chapter cannot be interpreted as a generalization to the space
$\MT$ of a universal symmetry of Classical Mechanics at the level of $(q,t)$. 

\chapter*{5. The Conformal Extension of the CPI}
\addcontentsline{toc}{chapter}{\numberline{5}The Conformal Extension of the CPI}
\setcounter{chapter}{5}
\setcounter{section}{0}
\markboth{5. The Conformal Extension of The CPI}{}

\noindent
In the previous chapter we have introduced two transformations: the first one (generated by ${\mathscr Q}_{\s S}$) 
rescales the action of the CPI while the second (the MSA) rescales the standard action of Classical Mechanics. 
We have seen that in both cases the formalism is not easy to manage because on one hand ${\mathscr Q}_{\s S}$
induces a noncanonical transformation on \quattrova while, on the other hand, the MSA is even anholonomic and
path dependent. 
In this chapter we want to analyze another rescaling of time which is 
much easier to handle because of its canonical structure.
This transformation is known as {\it superconformal transformation} and is the result of the composition of a
conformal transformation together with a Susy. 
In the following sections we shall present a model which exhibits a kind of superconformal 
invariance derived from the classical Susy of the CPI. Our hope is that this particular model 
may be a playground to tackle the more general problems of the transformations introduced in the previous chapter.   

\section{Conformal Mechanics}
\label{sec:CME}
\noindent 
In this section we briefly describe the model which was originally presented in
\cite{DFF}. It is a one dimensional model described by a configuration $q(t)$ where
$t$ is the base space which in our context will be always understood as the time. The
action is:
	\begin{equation}
	\label{CME1}
	S=\int dt\, \frac{1}{2}\left[{\dot q}^{2}-{g\over q^{2}}\right],
	\end{equation}
\noindent 
where $g$ is a dimensional constant. It is not difficult to check that this
Lagrangian is invariant under the following transformations:
	\begin{align}
	\label{CME2}
	&t^{\prime} = {{\alpha t+\beta}\over {\gamma t + \delta}};\\
	\label{CME3}
	&q^{\prime}(t^{\prime}) = {q(t)\over (\gamma t +\delta)}; \\
	\vspace{-.2cm}\nonumber \\
	&(\mbox{with } ~~\alpha\delta-\beta\gamma=1) \nonumber;
	\end{align}
\noindent
which are the conformal transformations in 0+1 dimensions\footnote{``0+1" refers to
the spatial+temporal dimensions of the base space. In our case we have 0 spatial
dimensions and 1 (i.e. the time $t$) temporal dimension.}. We have confined in Appendix \ref{app:Conf} a
brief review of the Conformal Group and its representations; for our purposes here,
it is enough to say that a generic transformation (\ref{CME2}) is a composition of
the following three transformations:
	\begin{align}
	\label{CME5}
	& t^{\prime} = \alpha^{2}t & &\mbox{\it dilations},\\
	\label{CME6}
	& t^{\prime} = t+\beta  & &\mbox{\it time-translations},\\
	\label{CME7}
	& t^{\prime} = {t\over {\gamma t+1}} & &\mbox{\it
	special-conformal~transformations.}
	\end{align}     
\noindent
We can apply Noether's theorem to the following transformations and what we get is
that the conserved charges are the following:
	\begin{align}
	\label{CME8}
	H & = \frac{1}{2}\left(p^{2}+{g\over q^{2}}\right);\\
	\label{CME9}
	D & = tH-{1\over 4}(qp+pq);\\
	\label{CME10}
	K & = t^{2}H-\frac{1}{2}t(qp+pq)+\frac{1}{2}q^{2}.
	\end{align}
\noindent
Using the quantum commutator $[q,p]=i$, we see that the three charges above obey
the following algebra:
	\begin{align}
	\label{CME11}
	&[H,D] = iH;\\
	\label{CME12}
	&[K,D] = -iK;\\
	\label{CME13}
	&[H,K] = 2iD.
	\end{align}
\noindent 
The reader should not be surprised by the fact that $D$ and $K$ do not commute
with $H$, even if they are constants of motion. In fact they depend explicitly on
time and therefore Noether's theorem --- which claims:
	\begin{align}
	\label{CME14}
	&0=\frac{d}{dt}D= \frac{\p}{\p t}D + i[H,D]; \\
	&0=\frac{d}{dt}K= \frac{\p}{\p t}K + i[H,K]; \label{CME15}
	\end{align}
\noindent
does not imply that $[H,D]=0$ or $[H,K]=0$.
Anyway, since the algebra above does not involve the time $t$ explicitly, the same
commutation relations are satisfied also by the same operators evaluated at time
$t=0$, namely:
	\begin{align}
	\label{CME16}
	&H_{0}=\frac{1}{2}\left[p^{2}+{g\over q^{2}}\right],\\
	\label{CME17}
	&D_{0} =  -{1\over 4}\left[qp+pq\right], \\
	\label{CME18} 	
	&K_{0} = \frac{1}{2}q^{2},
	\end{align}
\noindent 
which will turn useful in the following. The algebra (\ref{CME11})-(\ref{CME13})
is that of the group $SO(2,1)$, which is isomorphic (see Appendix \ref{app:Conf} for the details)
to the conformal group in 0+1 dimensions.
\section{Conformal Mechanics and General relativity}

\noindent
Before going to the supersymmetric extensions of Conformal Mechanics, we give
a brief account of the reason why this model has gained again much attention in
the context of black holes and General Relativity \cite{KAL}\cite{MAL}. 

First of all we consider the metric of an {\it extreme} Reissner-Nordstr{\o}m black hole: for 
the reader who is not familiar with it we give a brief account of this topic in Appendix \ref{app:Reiss}. 
For our purposes here, it is enough to write down the form of the metric:
	\begin{equation}
	\label{CME19}
	ds^2=-\left(1-\frac{M}{r}\right)^2 dt^2 + \left(1-\frac{M}{r}\right)^{-2}dr^2 +
	r^2d\Omega^2,
	\end{equation}
\noindent
where:
	\begin{equation}
	\label{CME20}
	d\Omega^2=\sin ^2\t\, d\v^2 +d\t^2
	\end{equation}
\noindent 
is the usual expression of the solid angle. As a first step we change the coordinates by introducing a new variable 
$\rho\equiv r-M$: the metric (\ref{CME19}) takes the form:
	\begin{equation}
	\label{CME21}
	ds^2=-\left(1+\frac{M}{\rho}\right)^{-2} dt^2 + \left(1+\frac{M}{\rho}\right)^{2} 
	\big[d\rho^2 + \rho^2 d\Omega^2 \big].
	\end{equation}
\noindent
The effect of this change of variable is simple: we have shifted the singularity from $r_o=M$ to $\rho_o=0$. 
Therefore, in the near-horizon region ($\rho\cong 0$), we can approximate the previous expression with the following:
	\begin{equation}
	\label{CME22}
	ds^2=-\left(\frac{\rho}{M}\right)^{2} dt^2 + \left(\frac{M}{\rho}\right)^{2} d\rho^2 + M^2 d\Omega^2.
	\end{equation}
\noindent
If we finally introduce two further new variables $(z,\tau)$ defined as:
	\begin{equation}
	\label{CME23}
	\begin{cases}
	\rho=: M\text{e}^{-z}\cos{\tau} \\
	t=: M\text{e}^{z}\tan{\tau},
	\end{cases}
	\end{equation}
\noindent
after a little algebra we find that (\ref{CME22}) becomes:
	\begin{equation}
	\label{CME24}
	ds^2=M^2\big[-d\tau^2 + \cos^2\tau dz^2 + d\Omega^2\big],
	\end{equation}
\noindent 
which goes under the name of Bertotti-Robinson metric. The first two terms of the RHS are invariant 
under $SO(2,1)$ while $d\Omega^2$ is clearly invariant under $SO(3)$. The metric exhibits an 
overall $SO(2,1)\times SO(3)$ invariance, which is commonly called in the literature $AdS_2\times S_2$ (see 
Appendix \ref{app:Desitt} for the details). Therefore we already have a hint about how the conformal group enters the game, 
because --- as we have seen previously --- the conformal mechanics in $0+1$ dimensions is invariant under 
transformations of the group $SO(2,1)$. These subject has recently obtained a big interest because of the 
famous Maldacena's conjecture \cite{MAL}, which claims that the large $N$ limit of a conformally invariant 
theory in $d$-dimensions (the boundary) is equivalent to supergravity on $d+1$-dimensional $AdS$ space (the bulk) 
times a compact manifold (which in the maximally supersymmetric case is a sphere).

Let us now proceed with the description of the relation between the Conformal Mechanics (\ref{CME1}) and the 
Reissner-Nordstr{\o}m metric (\ref{CME22}). Our next step is to determine the Hamiltonian which governs the 
dynamics of a particle in the near-horizon region (i.e. $\rho\cong 0$) in which the approximation (\ref{CME22}) 
holds. First of all we make the last change of variables by introducing a new parameter $R$ defined as follows:
	\begin{equation}
	\label{CME25}
	\frac{\rho}{M}=:\left(\frac{2M}{R}\right)^2
	\end{equation}
\noindent
which implies:
	\begin{equation}
	\label{CME26}
	ds^2=-\left(\frac{2M}{R}\right)^{4} dt^2 + \left(\frac{2M}{R}\right)^{2} dR^2 + M^2 	d\Omega^2.
	\end{equation}
\noindent From 
the previous equation it is easy to see that the metric coefficients $g_{\mu\nu}$ do not depend on $t$ and 
therefore (see Appendix \ref{app:Killing}) $p_t$ is a constant of the motion, which is just the Hamiltonian 
we are looking for. We can calculate it explicitly in the following way (other methods are possible, of course). 
First of all we exploit the fact that we have a constraint on the momenta:
	\begin{equation}
	\label{CME27}
	(p_{\mu}-qA_{\mu}) g^{\mu\nu} (p_{\nu}-qA_{\nu})+m^2=0
	\end{equation}
\noindent
where $q$ and $m$ are the charge and the mass of the particle moving in the near-horizon region, and 
$A_{\mu}$ is the electromagnetic potential generated by the black hole. Since the latter is a {\it static} 
charge sitting at $r=0$ we have that the only non vanishing component of $A_{\mu}$ is:
	\begin{equation}
	\label{CME28}
	A_{0}(\rho)=-\frac{Q}{r(\rho)}=-\frac{1}{1+\frac{\rho}{M}}.
	\end{equation}
\noindent
In the near-horizon region ($\rho\cong 0$) we can rewrite it as:
	\begin{equation}
	\label{CME29}
	A_{0}(\rho)=-\left(1-\frac{\rho}{M}\right)\cong\left(\frac{2M}{R}\right)^{2},	\end{equation}
\noindent
where in the last step we have used the fact that an electromagnetic potential does not feel (at least 
at the classical level) additive constants. Now, if we plug (\ref{CME29}) into (\ref{CME27}) and we 
solve for $p_t$, after a little algebra we find:
	\begin{equation}
	\label{CME30}
	H\equiv-p_t= \left(\frac{2M}{R}\right)^{2}\left\{\left[m^2+\frac{1}{4M^2}\left(R^2p_{\s R}
	^2+4L^2\right)\right]^{\frac{1}{2}}-q\right\},	
	\end{equation}
\noindent
where
	\begin{equation}
	\label{CME31}
	L^2\equiv p_{\t}^2+\frac{p_{\v}^2}{\sin^2\t}
	\end{equation}
\noindent
is clearly the angular momentum which is another constant of motion both at the classical and the quantum 
level (where becomes $l(l+1)\hslash^2$). Finally we can make more manifest the conformal symmetry if we 
rewrite the Hamiltonian (\ref{CME30}) as:
	\begin{equation}
	\label{CME32}
	H= \frac{p_{\s R}^2}{2f}+\frac{g}{2fR^2},
	\end{equation}
\noindent
where $f$ and $g$ are respectively:
	\begin{align}
	& f\equiv \frac{1}{2}\left[\left(m^2+\frac{R^2 p_{\s R}^2+4L^2}{4M^2}\right)^{\frac{1}{2}}
	+q\right] \label{CME33} \\
	& g= 4M^2(m^2-q^2)+4L^2.   \label{CME33b}
	\end{align}
\noindent
Then, apart from the factor $f$, we can recognize the Hamiltonian (\ref{CME8}) of the Conformal Mechanics.
\section{Superconformal Mechanics}

\noindent 
In this section we briefly review the first supersymmetric extension of Conformal Mechanics,
which was provided in Ref.\cite{FUB}.
This extension of model (\ref{CME1}) is tailored on the celebrated Supersymmetric Quantum Mechanics invented
by Witten \cite{Witten} and subsequently developed by others \cite{Cooper}\cite{Salom}.
The Hamiltonian is:
	\begin{equation}
	\label{CME34}
	\HS =\frac{1}{2}\left(p^{2}+{g\over q^{2}}+{\sqrt {g}\over
	q^{2}}[\psi^{\dag},\psi]_{\scriptscriptstyle -}\right)
	\end{equation}
\noindent
where $\psi,\psi^{\dag}$ are Grassmannian variables whose anticommutator
is $[\psi,\psi^{\dag}]_{\scriptscriptstyle +}=1$. As one can notice, in $\HS$
there is a bosonic piece which is the conformal Hamiltonian 
(\ref{CME8}), plus a Grassmannian part. Note that the equations
of motion of ``$q$" have an extra piece with respect to the equations
of motion of the old Conformal Mechanics \cite{DFF}.

To make contact with Supersymmetric Quantum Mechanics 
let us notice that $\HS$ can be written as:
	\begin{equation}
	\label{CME35}
	\HS =\frac{1}{2}\left(p^{2}+\left({dW\over dq}\right)^{2}-[\psi^{\dag},\psi
	]_{\scriptscriptstyle -}{d^{2}W\over dq^{2}}\right)
	\end{equation}
\noindent 
(which is the Hamiltonian of Susy-QM) where the superpotential $W$ turns out to be:
	\begin{equation}
	\label{CME36}
	W(q)=\sqrt{g}\log q.
	\end{equation}
\noindent
The two Supersymmetry charges associated to the Hamiltonian (\ref{CME34}) are easily found:
	\begin{align}
	\label{CME37}
	&Q=\psi^{\dag}\left(-ip+{dW\over dq}\right),\\
	\label{eq:ventitre}
	&Q^{\dag}=\psi\left(ip+{dW\over dq}\right),
	\end{align}
\noindent 
whose commutator closes on the Hamiltonian:
	\begin{equation}
	\label{CME38}
	\big[ Q,Q^{\dag} \big]_{\s +} = 2\HS.
	\end{equation}
\noindent 
It is interesting to see what we obtain when we combine a supersymmetric
transformation with a conformal one generated by the $(H,K,D)$
elements of the $SO(2,1)$ algebra (\ref{CME11})-(\ref{CME13}).
We get what is called a {\it superconformal} transformation. In order to understand 
this better let us list the following eight operators:

\begin{center}
{\bf TABLE 1} 
\end{center}
\begin{small}
\[
\begin{array}{|rcl|}
\hline && \\
H & = &\displaystyle\frac{1}{2}\Bigl[p^{2}+{{g+2\sqrt{g}B}\over q^{2}}\Bigr];\\
D & = &\displaystyle -{[q,p]_{\scriptscriptstyle +}\over 4};\\
K & = &\displaystyle{q^{2}\over 2};\\
B & = &\displaystyle{[\psi^{\dag},\psi]_{\scriptscriptstyle -}\over 2};\\
Q & = &\displaystyle \psi^{\dag}\Bigl[-ip+{\sqrt{g}\over q}\Bigr];\\
Q^{\dag} & = &\displaystyle \psi\Bigl[ip+{\sqrt{g}\over q}\Bigr];\\
S & = &\displaystyle \psi^{\dag}q;\\
S^{\dag} & = &\displaystyle \psi q.\\
&& \\
\hline
\end{array}
\]
\end{small}
\vspace{.5cm}

\noindent
The algebra of these operators is closed and given in {\bf TABLE 2}:

\begin{center}
{\bf TABLE 2}
\end{center}
\[
\begin{array}{|lll|}
\hline & & \\

[H,D]=iH; \hspace{3cm}&[K,D]=-iK; \hspace{3cm} &[H,K]=2iD; \\

[Q,H]=0; &[Q^{\dag},H]=0; &[Q,D]={i\over 2}Q; \\

[Q^{\dag},K]=S^{\dag}; &[Q,K]=-S; &[Q^{\dag},D]={i\over 2}Q^{\dag}; \\

[S,K]=0; &[S^{\dag},K]=0; &[S,D]=-{i\over 2}S; \\

[S^{\dag},D]=-{i\over 2}S^{\dag}; &[S,H]=-Q; &[S^{\dag},H]=Q^{\dag}; \\

[Q,Q^{\dag}]=2H; &[S,S^{\dag}]=2K; & \\

[B,S]=S; &[B,S^{\dag}]=-S^{\dag}; & \\
 
[Q,S^{\dag}]=\sqrt{g}-B+2iD; &[B,Q]=Q; &[B,Q^{\dag}]=-Q^{\dag}; \\

& & \\
\hline
\end{array} 
\]

\vskip 1cm
\noindent all other commutators are zero or derivable from these by Hermitian conjugation.
The square-brackets $[({\s\ldots}),({\s\ldots})]$ in the algebra above 
are  {\it graded}-commutators and from now on we shall
omit the subindex $+$ or $-$ as we did before. They are commutators
or anticommutators according to the Grassmannian nature of the operators 
entering the brackets.

As is well known, a {\it superconformal} transformation is a combination
of a supersymmetry transformation and a conformal one. We see from the
algebra above that the commutators of the supersymmetry 
generators $ (Q,Q^{\dag}) $ with the three conformal generators 
$ (H,K,D) $ generate
two new operators  $(S,S^{\dag})$. Including these operators we generate an algebra
which is closed  provided that we introduce the operator $B$ of {\bf TABLE 1}. This is the last 
operator we need. 
\section{A New Supersymmetric Extension of Conformal Mechanics}

\noindent
In the previous Section we have described the supersymmetric extension of Conformal Mechanics
which is based on the Susy-QM model of Witten. In this Section we want to obtain a new
supersymmetric extension of the action (\ref{CME1}) by use of another strategy. 
In fact, as we have seen in Section 1.3, the Hamiltonian $\HT$ of the Classical Path Integral is
automatically supersymmetric, regardless of the particular shape of $H(\v)$ from which it is built. Therefore, if we
choose as $H(\v)$ the conformal Hamiltonian of Eq.(\ref{CME8}) we will obtain a model of
Superconformal Mechanics which is basically different from that tailored on the Susy-QM
model. Thus, as a first step, we insert
the $H$ of Eq.(\ref{CME8}) into the $\HT$ of Eq.(\ref{CPI8}).
The result is:
	\begin{equation}
	\label{CME39}
	\HT=\lambda_q p+\lambda_p\frac{g}{q^3}
	+i\bar{c}_qc^p-3i\bar{c}_pc^q\frac{g}{q^4}
	\end{equation} 
\noindent where the indices $({\s\ldots})^{q}$ and $({\s\ldots})^{p}$ on the variables $(\lambda, c, {\bar c})$
replace
the indices $({\s\ldots})^{a}$ which appeared in the general formalism. 
In fact, as the system
is one-dimensional, the index ``$({\s\ldots})^{a}$" can only indicate the variables ``$(p,q)$" and that
is why we use $(p,q)$ as index. The two supersymmetric 
charges of Eqs.(\ref{CPI28}) and (\ref{CPI29}) are in this case
	\begin{align}
	\label{CME40}
	\QH &=\Qb+\beta\left(\frac{g}{q^3}c^q-pc^p\right) \\
	\label{CME41}
	\QBH &=\QBb+\beta\left(\frac{g}{q^3}\bar{c}_p+pc_q\right). 
	\end{align}
\noindent It was one of the central points of the original paper\scite{DFF} on Conformal
Mechanics that the Hamiltonians of the system could be, beside $H_{0}$
of Eq.(\ref{CME16}), also $D_{0}$ or $K_{0}$ of Eqs.(\ref{CME17})(\ref{CME18}) 
or any linear combination of them. In the same manner as we built the
Lie-derivative~$\HT$~associated to $H_{0}$, we can also
build the Lie-derivatives associated to the flow generated by $D_{0}$
and $K_{0}$. We just have to insert\footnote{We will neglect ordering
problem in the expression of $D_{0}$ because we are doing
a classical theory. The sub-index ``$({\s\ldots})_{0}$" that we will put on $\DT$ and $\KT$
below is to indicate that they were built from $D_{0}$ and $K_{0}$.}  $D_{0}$ or $K_{0}$
in place of $H$ as superpotential in the $\HT$ of Eq.(\ref{CPI8}). If we denote the
associated Lie-derivatives by $\DT_0$ and 
$\KT_0$, what we get is:
	\begin{align}
	\label{CME42}
	&\DT_0=\frac{1}{2}[\lambda_pp-\lambda_qq+i(\bar{c}_pc^p-\bar{c}_qc^q)]\\
	\label{CME43} 
	&\KT_0=-\lambda_pq-i\bar{c}_pc^q 
	\end{align}
\noindent The construction is best illustrated in Figure 1.

\begin{figure}
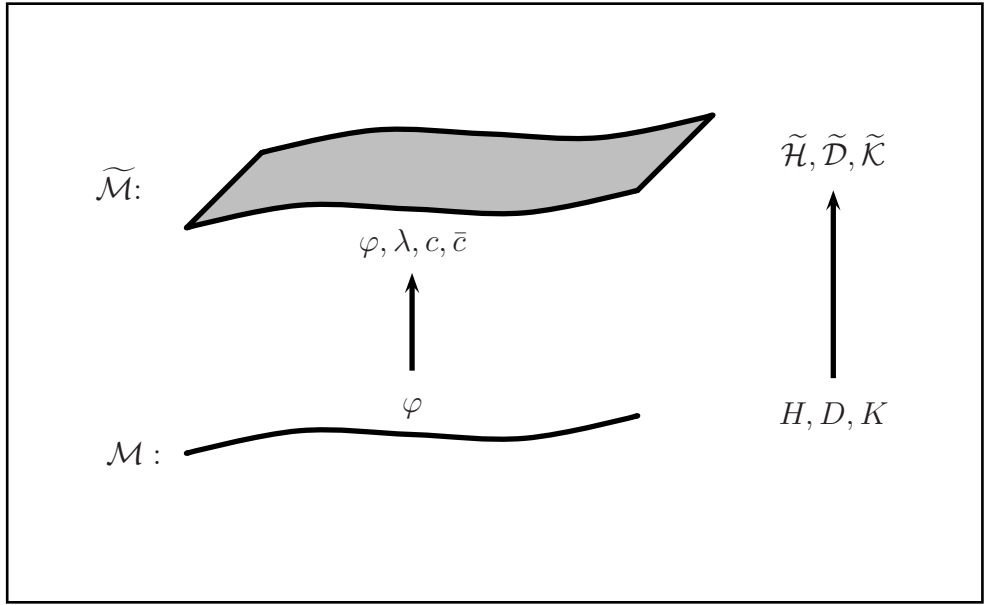

\pspicture(-1.25,0)(12.75,9)
\psset{linewidth=2pt}
\psframe[linewidth=1pt](0,0)(13,8)
\pscurve[]{c-c}(2.4,2)(3.9,2.3)(5.4,2.25)(6.9,2.2)(8.4,2.5)
\pscustom[fillstyle=solid,fillcolor=lightgray]{
\pscurve[]{c-c}(2.4,5)(3.9,5.3)(5.4,5.25)(6.9,5.2)(8.4,5.5)
\psline[](8.4,5.5)(9.4,6.5)
\pscurve[liftpen=2]{c-c}(9.4,6.5)(7.9,6.2)(6.4,6.25)(4.9,6.3)(3.4,6)
\psline[](3.4,6)(2.4,5)}

\rput(5.4,4.8){$\v,\lambda,c,\bar{c}$}
\rput(5.4,2.6){$\v$}
\psline[]{->}(5.4,3.1)(5.4,4.4)

\rput(11,6){$\HT,\DT,\KT$}
\rput(11,2.5){$H,D,K$}
\rput(1.7,2){$\mathcal M :$}
\rput(1.5,5.6){$\MT$:}
\psline[]{->}(11,3)(11,5.5)
\endpspicture
\caption{\small The correspondence between $\mathcal M$ and $Super \mathcal M$.}
\end{figure}

As it is easy to prove, both $\DT_0$ and $\KT_0$ are supersymmetric. It is 
possible in fact to introduce the following charges:
	\begin{align}
	\label{CME44}
	&\QD=\Qb+\gamma(qc^p+pc^q);\\
	\label{CME45}
	&\QBD=\QBb+\gamma(p\bar{c}_p-q\bar{c}_q);\\
	\label{CME46}
	&\QK=\Qb-\alpha qc^q; \\
	\label{CME47}
	&\QBK=\QBb-\alpha q\bar{c}_p;
	\end{align}
($\gamma$ and $\alpha$ play the same role as $\beta$ for $H$) which close on $\DT_0$ and
$\KT_0$:
	\begin{align}
	\label{CME48}
	&[\QD,\QBD]=4i\gamma\DT_0\\
	\label{CME49}
	&[\QK,\QBK]=2i\alpha\KT_0.
	\end{align}
\noindent One further point to notice is that 
the conformal algebra of Eqs.(\ref{CME11})-(\ref{CME13}) 
is now realized, via the commutators 
(\ref{CPI11}) of our formalism, by the $(\HT,\DT_0,\KT_0)$
and not by the old functions~$(H,D_0,K_0)$. In fact, via  these new commutators,
we get:
	\begin{align}
	\label{CME50}
	&[\HT,\DT_0]=i\HT; &&[\KT_0,\DT_0]=-i\KT_0; &&[\HT,\KT_0]=2i\DT_0;\\
	\label{CME51}
	&[H,D_0]=0; && [K_0,D_0]=0; && [H,K_0]=0.
	\end{align}
\noindent The next thing to find out, assuming $\HT$ as basic Hamiltonian
and $\QH$ and $\QBH$ as supersymmetry charges, is to perform
the commutators between supersymmetries and conformal operators so to 
get the superconformal generators. It is easy to work this out and we
get:
	\begin{align}
	\label{CME52}
	&[\QH,\DT_0]=i(\QH-\Qb); & &[\QBH,\DT_0]=i(\QBH-\QBb); \\
	\label{CME53}
	&[\QH,\KT_0]=\frac{i\beta}{\gamma}(\QD-\Qb); & &[\QBH,\KT_0]=\frac{i\beta}{\gamma}(\QBD-\QBb).
	\end{align}

\noindent From what we have above we realize immediately the role of the $\QD$ and $\QBD$:
besides being the ``square roots" of $\DT_0$ they are also (combined with the
$\Qb$ and $\QBb$) the generators of the superconformal transformations.
It is also a simple calculation to evaluate the commutators
between the various ``supercharges" $\QH$,$\QBD$, $\QBK$, $\QBH$, $\QD$, $\QK$:

	\begin{align}
	\label{CME54}
	&[\QH,\QBD]=i\beta\HT+2i\gamma\DT_0-2\beta\gamma H; & 
	&[\QBH,\QD]=i\beta\HT+2i\gamma\DT_0+2\beta\gamma H; \\
	\label{CME55}
	&[\QK,\QBD]=i\alpha\KT_0+2i\gamma\DT_0+2\alpha\gamma K_0; & 
	&[\QBK,\QD]=i\alpha\KT_0+2i\gamma\DT_0-2\alpha\gamma K_0; \\
	\label{CME56}
	&[\QH,\QBK]=i\beta\HT+\;i\alpha\KT_0-2\alpha\beta D_0; & 
	&[\QBH,\QK]=i\beta\HT+i\alpha\KT_0+2\alpha\beta D_0.
	\end{align}
\noindent From the RHS of these expressions we see that one needs also the
old functions $(H,D_0,K_0)$ in order to close the algebra.

The complete set of operators which close the algebra is listed in the
following table:

\begin{center}
{\bf TABLE 3}
\end{center}

\[
\begin{array}{|lcl|}
\hline & & \\
\HT=\displaystyle\lambda_q p+\lambda_p\frac{g}{q^3}+i\bar{c}_q
c^p-3i\bar{c}_pc^q\frac{g}{q^4}; &
&H=\displaystyle\frac{1}{2}\left(p^2+\frac{g}{q^2}\right);\\
\KT_0=\displaystyle-\lambda_pq-i\bar{c}_pc^q; && K_0=\frac{1}{2}q^2;\\
\DT_0=\displaystyle\frac{1}{2}[\lambda_pp-\lambda_qq+i(\bar{c}_pc^p-\bar{c}_qc^q)];&&D_0=\displaystyle
-\frac{1}{2}qp;\\
\Qb= \displaystyle i(\lambda_qc^q+\lambda_pc^p);&&\QBb= \displaystyle 
i(\lambda_p\bar{c}_q-\lambda_q\bar{c}_p);\\
\QH=\displaystyle\Qb+\beta\left(\frac{g}{q^3}c^q-pc^p\right);
&&\QBH=\displaystyle\QBb+\beta\left(\frac{g}{q^3}\bar{c}_p+p\bar{c}_q\right); \\
\QK=\displaystyle\Qb-\alpha qc^q; &&\QBK=\displaystyle\QBb-\alpha q\bar{c}_p;\\
\QD=\displaystyle\Qb+\gamma(qc^p+pc^q);&&\QBD=\displaystyle\QBb+\gamma(p\bar{c}_p-q\bar{c}_q).\\
&&\\
\hline
\end{array}
\]

\noindent The complete algebra among these generators is given in {\bf TABLE 4}, where the missing commutators are
zero and the $Q_{(\ldots)}$ appearing in the table can be
any of the following operators: $\Qb$,$\QH$,$\QD$,$\QK$ (the same holds for
$\overline{Q}_{(\ldots)}$). 
Obviously all commutators are understood between quantities calculated at the same time.

We notice that for our supersymmetric extension we need 14 charges
(see {\bf TABLE~3}) 
to close the algebra, while in the extension of
Ref.\scite{FUB} one needs only 8 charges (see {\bf TABLE 1}).
This is so not only because ours is an $N=2$ supersymmetry (while
the one of\scite{FUB} is an $N=1$) but also because of the totally
different character of the model.
\newpage
\begin{center}
{\bf TABLE 4}
\end{center}
\begin{small}
\[
\hspace{-1.1cm}
\begin{array}{|lll|}
\hline 
& & \\

[\HT,\DT_0]=i\HT; \hspace{3cm}&[\KT,\DT_0]=-i\KT_0; \hspace{3cm}  &[\HT,\KT_0]=2i\DT_0; \\

[\QH,\HT]=0; &[\QBH,\HT]=0; &[\QH,\QBH]=2i\beta\HT; \\

[\QH,\DT_0]=i(\QH-\Qb); &[\QBH,\DT_0]=i(\QBH-\QBb); & \\

[\QH,\KT_0]=i\beta\gamma^{-1}(\QD-\Qb); &[\QBH,\KT_0]=i\beta\gamma^{-1}(\QBD-\QBb); & \\

[\Qb,\HT]=[\QBb,\HT]=0; & [\Qb,\KT_0]=[\QBb,\KT_0]=0; & [\Qb,\DT_0]=[\QBb,\DT_0]=0; \\

[\QD,\HT]=-2i\gamma\beta^{-1}(\QH-\Qb); &[\QBD,\HT]=-2i\gamma\beta^{-1}(\QBH-\QBb); & \\

[\QD,\KT_0]=2i\gamma\alpha^{-1}(\QK-\Qb); &[\QBD,\KT_0]=2i\gamma\alpha^{-1}(\QBK-\QBb); & \\

[\QD,\DT_0]=0; &[\QBD,\DT_0]=0; &[\QD,\QBD]=4i\gamma\DT_0; \\

[\QK,\HT]=-i\alpha\gamma^{-1}(\QD-\Qb); &[\QBK,\HT]=-i\alpha\gamma^{-1}(\QBD-\QBb); & \\

[\QK,\DT_0]=-i(\QK-\Qb); &[\QBK,\DT_0]=-i(\QBK-\QBb); & \\

[\QK,\KT_0]=0; &[\QBK,\KT_0]=0; &[\QK,\QBK]=2i\alpha\KT_0; \\

[\QH,\QBD]=i\beta\HT+2i\gamma\DT_0-2\beta\gamma H; & 
[\QBH,\QD]=i\beta\HT+2i\gamma\DT_0+2\beta\gamma H; & \\

[\QK,\QBD]=i\alpha\KT_0+2i\gamma\DT_0+2\alpha\gamma K; &
[\QBK,\QD]=i\alpha\KT_0+2i\gamma\DT_0-2\alpha\gamma K; & \\

[\QH,\QBK]=i\beta\HT+i\alpha\KT_0-2\alpha\beta D; &
[\QBH,\QK]=i\beta\HT+i\alpha\KT_0+2\alpha\beta D; 
& \\

[\QH,\QBb]=[\QBH,\Qb]=i\beta\HT; & [\QK,\QBb]=[\QBK,\Qb]=i\alpha\KT_0; & \\

[\QD,\QBb]=[\QBD,\Qb]=2i\gamma\DT_0; & & \\

[Q_{\scriptscriptstyle(\ldots)},H]=\beta^{-1}(\Qb-\QH); &
[\overline{Q}_{\scriptscriptstyle(\ldots)},H]= 
\beta^{-1}(\QBb-\QBH); & \\

[Q_{\scriptscriptstyle(\ldots)},D_0]=(2\gamma)^{-1}(\Qb-\QD); & 
[\overline{Q}_{\scriptscriptstyle(\ldots)},D_0]=(2\gamma)^{-1}(\QBb-\QBD); & \\

[Q_{\scriptscriptstyle(\ldots)},K_0]=\alpha^{-1}(\Qb-\QK); & 
[\overline{Q}_{\scriptscriptstyle(\ldots)},K_0]=\alpha^{-1}(\QBb-\QBK); & \\

[\HT,H]=0; & [\KT_0,K_0]=0; & [\DT_0,D_0]=0; \\

[\HT,K_0]=[H,\KT_0]=2iD_0; &
[\HT,D_0]=[H,\DT_0]=iH; & [\DT_0,K_0]=[D_0,\KT_0]=iK_0. \\

& & \\
\hline
\end{array}
\]
\end{small}


\section{Study of the two Superconformal Algebras}

\noindent
A Lie superalgebra\scite{KAC} is an algebra made of even $E_{n}$ and
odd $O_{\alpha}$ generators whose graded commutators look like:
	\begin{align}
	\label{CME57}
	&[E_{m},E_{n}]=F^{p}_{mn}E_{p};\\
	\label{CME58}
	&[E_{m},O_{\alpha}]=G^{\beta}_{m\alpha}O_{\beta};\\
	\label{CME59}
	&[O_{\alpha},O_{\beta}]=C^{m}_{\alpha\beta}E_{m};
	\end{align} 
\noindent and where the structure constants $F^{p}_{mn},G^{\beta}_{m,\alpha},
C^{m}_{\alpha,\beta}$ satisfy generalized Jacobi identities.

One can interpret the relation (\ref{CME58}) by saying that
the even part of the algebra has a representation on the odd part. 
This is clear if we consider the odd part as a vector space and that
the even part acts on this vector space via the graded
commutators. The structure constants $F^{\beta}_{m\alpha}$ are then
the matrix elements which characterize the representations.

For superconformal algebras the usual folklore says that the even part of the
algebra has his conformal subalgebra represented spinorially
on the odd part. The reasoning roughly
goes as follows: the odd part of the algebra must contain the supersymmetry
generators which transform as spinors under the Lorentz group which is
a subgroup of the conformal algebra. So it is impossible that  the whole
conformal algebra  is represented non-spinorially on the odd part.

Actually this line of reasoning is true in a relativistic context in which 
the supersymmetry is a true relativistic supersymmetry and the charges must 
carry a spinor index due to their nature. Instead, in our non-relativistic
point particle case, the charges  do not carry any space-time index 
and so we do not have as a consequence that necessarily the even part of the
algebra is represented spinorially on the odd part. It can happen but it can
also not happen. In this respect we will analyze the superalgebras
of the two supersymmetric extensions of conformal mechanics seen here, the
one of\scite{FUB} and ours.

Let us start from the one of Ref.\scite{FUB}  which is given in {\bf TABLE 1}.
The conformal subalgebra ${\mathcal G}_{0}$ of the even part  can be 
organized in an $SO(2,1)$ form as follows:
	\[
	{\mathcal G}_{0}: \left\{ 
	\begin{array}{l}
	B_{1}=\displaystyle{1\over 2}\left[ {K\over a}-aH \right] \\
	B_{2}= D \\
	J_{3}=\displaystyle{1\over 2}\left[ {K\over a}+a H \right]
	\end{array}
	\right.
	\]
\noindent where $a$ is the same parameter introduced in \cite{DFF} with a physical
dimension of time.

The odd part ${\mathcal G}_{1}$ is:
	\[
	{\mathcal G}_{1}: \left\{ \begin{array}{l}
	Q\\
	Q^{\dag}\\
	S\\
	S^{\dag}
	\end{array}
	\right.
	\]
\noindent It is easy to work out, using the results of {\bf TABLE 2}, the action
of ${\mathcal G}_{0}$  on ${\mathcal G}_{1}$. The result
is summarized in the following table:
\newpage
\begin{small}
\begin{center}
{\bf TABLE 5}
\end{center}
\[
\begin{array}{|lll|}
\hline & & \\

[B_{1},Q]=\displaystyle{1\over 2a}S; \hspace{2cm}&[B_{2},Q]=-\displaystyle{i\over 2}Q;
\hspace{2cm} &
\displaystyle[J_{3},Q]={1\over 2a}S; \\

[B_{1},Q^{\dag}]=\displaystyle-{1\over 2a}S^{\dag}; \hspace{2cm}&[B_{2},Q^{\dag}]=
\displaystyle-{i\over 2}Q^{\dag}; \hspace{2cm} & [J_{3},Q^{\dag}]=\displaystyle-{1\over
2a}S^{\dag};\\

[B_{1},S]=-\displaystyle{a\over 2}Q; \hspace{2cm}&[B_{2},S]=\displaystyle{i\over 2}S;
\hspace{2cm} &
[J_{3},S]=\displaystyle{a\over 2}Q; \\

[B_{1},S^{\dag}]=\displaystyle{a\over 2}Q^{\dag};
\hspace{2cm}&[B_{2},S^{\dag}]=\displaystyle{i\over 2}
S^{\dag}; \hspace{2cm} & [J_{3},S^{\dag}]=\displaystyle-{a\over 2}Q^{\dag}. \\

&& \\
\hline
\end{array}
\]
\end{small}

\noindent As we said before, in order to act with the even part of the
algebra on the odd part, we have to consider the odd part as a vector
space. Let us then introduce the following ``vectors":
	\begin{align}
	\label{CME60}
	|q\rangle&\equiv Q+Q^{\dag}\\
	\label{CME61}
	|p\rangle & \equiv S-S^{\dag}\\
	\label{CME62}
	|r\rangle & \equiv Q-Q^{\dag}\\
	\label{CME63}
	|s\rangle & \equiv  S+S^{\dag};
	\end{align}
\noindent 
they label a 4-dimensional vector space.
On these vectors we act via the commutators, for example:
	\begin{equation}
	\label{CME64}
	B_{1}|q\rangle\equiv [B_{1},Q+Q^{\dag}]
	\end{equation}
\noindent 
It is then immediate  to realize from {\bf TABLE 5} that the 2-dim. space with
basis ~$(|q\rangle,|p\rangle)$ forms a closed space under the action
of the even part of the algebra so it carries a 2-dim. representation
and the same holds for the space $(|r\rangle,|s\rangle)$. We can immediately check
which kind of representation this is: Let us take the Casimir operator
of the algebra ${\mathcal G}_{0}$ which is~
${\mathcal C}=B_{1}^{2}+B_{2}^{2}-J_{3}^{2}$
and apply it to a state of one of the two 2-dim. representations:
	\begin{align}
	\label{CME65}
	{\mathcal C}|q\rangle & = [B_{1},[B_{1},Q+Q^{\dag}]]+[B_{2},[B_{2},Q+Q^{\dag}]]-
	[J_{3},[J_{3},Q+Q^{\dag}]]\nonumber \\
	& = -{3\over 4}(Q+Q^{\dag})\nonumber\\
	& = -{3\over 4}|q\rangle.
	\end{align}
\noindent 
This factor ${3\over 4}=-{1\over 2}({1\over 2}+1)$ indicates that the
$(|q\rangle,|p\rangle)$ space carries a spinorial representation. It is possible to prove the
same for the other space.
\newpage
Let us now turn the same crank for our supersymmetric extension
of conformal mechanics. Looking at the {\bf TABLE 3} of our operators,
we can organize the even part ${\mathcal G}_{0}$, as follows:

\begin{small}
\begin{center}
{\bf TABLE 6 (${\mathcal G}_{0}$)}
\end{center}
\[
\begin{array}{|lcl|}
\hline 

&&\\
B_{1}=\displaystyle{1\over 2}\left( {\KT\over a}-a\HT\right); && P_{1}=2D;\\

B_{2}=\displaystyle\DT; && P_{2}=\displaystyle aH-{K\over a};\\

J_{3}=\displaystyle{1\over 2}\left( {\KT\over a}+a\HT\right); && P_{0}=\displaystyle
aH+{K\over a}.\\
&&\\
\hline
\end{array}
\]
\end{small}
\noindent The LHS is the usual $SO(2,1)$ while the RHS is formed by three translations
because they commute among themselves. So the overall algebra is the Euclidean
group $E(2,1)$.

The odd part of our superalgebra is made of 8 operators (see {\bf TABLE 3})
which are:

\begin{center}
{\bf TABLE 7 (${\mathcal G}_{1}$)}
\end{center}
\[
\begin{array}{|lcl|}
\hline & & \\

\QH; && \QBH;\\

\QK; && \QBK;\\

\QD; && \QBD;\\

\Qb; && \QBb.\\
& & \\
\hline
\end{array}
\]

\noindent As we did before in {\bf TABLE 5} for the model of\scite{FUB},
we will now evaluate for our model the action of ${\mathcal G}_{0}$ 
on ${\mathcal G}_{1}$. The result is summarized in the next table,
where for simplicity we have made the choice $\displaystyle
a=\sqrt{\frac{\beta}{\alpha}}$ 
and $\displaystyle\eta\equiv\frac{\gamma}{\sqrt{\alpha\beta}}$.
\newpage
\begin{small}
\begin{center}
{\bf TABLE 8}
\end{center}
\[
\begin{array}{|ll|}
\hline
& \\

[B_1,\QH]=\displaystyle\frac{i}{2\eta}(\Qb-\QD); &
[B_1,\QBH]=\displaystyle\frac{i}{2\eta}(\QBb-\QBD);\\

[B_1,\QK]=\displaystyle\frac{i}{2\eta}(\Qb-\QD); &
[B_1,\QBK]=\displaystyle\frac{i}{2\eta}(\QBb-\QBD);\\

[B_1,\QD]=-i\eta(\QH+\QK-2\Qb); &
[B_1,\QBD]=-i\eta(\QBH+\QBK-2\QBb);\\

[B_1,\Qb]=0; & [B_1,\QBb]=0;\\

[B_2,\QH]=i(\Qb-\QH); & [B_2,\QBH]=i(\QBb-\QBH);\\

[B_2,\QK]=i(\QK-\Qb); & [B_2,\QBK]=i(\QBK-\QBb);\\

[B_2,\QD]=0; & [B_2,\QBD]=0;\\

[B_2,\Qb]=0; & [B_2,\QBb]=0;\\

[J_3,\QH]=\displaystyle\frac{i}{2\eta}(\Qb-\QD); &
[J_3,\QBH]=\displaystyle\frac{i}{2\eta}(\QBb-\QBD);\\

[J_3,\QK]=\displaystyle -\frac{i}{2\eta}(\Qb-\QD); & [J_3,\QBK]=\displaystyle
-\frac{i}{2\eta}(\QBb-\QBD);\\

[J_3,\QD]=i\eta(\QH-\QK); & [J_3,\QBH]=i\eta(\QBH-\QBK);\\

[J_3,\Qb]=0; & [J_3,\QBb]=0;\\

[P_1,Q_{\scriptscriptstyle(\ldots)}]=\gamma^{-1}(\QD-\Qb); & 
[P_1,\overline{Q}_{\scriptscriptstyle(\ldots)}]=-\gamma^{-1}(\QBD-\QBb);\\

[P_2,Q_{\scriptscriptstyle(\ldots)}]=\gamma^{-1}\eta(\QH-\QK); & 
[P_2,\overline{Q}_{\scriptscriptstyle(\ldots)}]=-\gamma^{-1}\eta(\QBH-\QBK);\\

[P_0,Q_{\scriptscriptstyle(\ldots)}]=\gamma^{-1}\eta(\QH+\QK-2\Qb); & 
[P_0,\overline{Q}_{\scriptscriptstyle(\ldots)}]=-\gamma^{-1}\eta(\QBH+\QBK-2\QBb).\\

& \\

\hline
\end{array}
\]
\end{small}

\noindent
As we have to represent the conformal subalgebra of ${\mathcal G}_{0}$
(see {\bf TABLE 6}) on the vector space ${\mathcal G}_{1}$ of {\bf TABLE 7} 
it is easy to realize from {\bf TABLE 8} that the following three vectors

	\begin{equation}
	\left\{
	\begin{array}{l}
	|q_{\scriptscriptstyle H}\rangle = (\QH-\Qb)-(\QBH-\QBb)\\
	|q_{\scriptscriptstyle K}\rangle = (\QK-\Qb)-(\QBK-\QBb)\\
	|q_{\scriptscriptstyle D}\rangle = \eta^{-1}[(\QD-\Qb)-(\QBD-\QBb)]
	\end{array}
	\right.
	\end{equation}

\noindent
make an irreducible representation of the conformal subalgebra. In fact, using\break
{\bf TABLE~8}, we get:
	
	\begin{equation}
	\left\{ 
	\begin{array}{l}
	B_{1}|q_{\scriptscriptstyle H}\rangle =-{i\over 2}
	|q_{\scriptscriptstyle D}\rangle\\
	B_{2}|q_{\scriptscriptstyle H}\rangle=-i|q_{\scriptscriptstyle H}\rangle\\
	J_{3}|q_{\scriptscriptstyle H}\rangle=-{i\over 2}
	|q_{\scriptscriptstyle D}\rangle\\
	B_{1}|q_{\scriptscriptstyle K}\rangle=-{i\over 2}
	|q_{\scriptscriptstyle D}\rangle\\
	B_{2}|q_{\scriptscriptstyle K}\rangle=i|q_{\scriptscriptstyle K}\rangle\\
	J_{3}|q_{\scriptscriptstyle K}\rangle={i\over 2}|q_{\scriptscriptstyle
	D}\rangle\\
	B_{1}|q_{\scriptscriptstyle D}\rangle= -i(|q_{\scriptscriptstyle H}\rangle
	+|q_{\scriptscriptstyle K}\rangle)\\
	B_{2}|q_{\scriptscriptstyle D}\rangle=0\\
	J_{3}|q_{\scriptscriptstyle D}\rangle=i(|q_{\scriptscriptstyle H}\rangle-
	|q_{\scriptscriptstyle K}\rangle).
	\end{array}
	\right.
	\end{equation}
\noindent Having three vectors in this representation we presume it is an `integer'
spin representation, but to be sure let us apply the Casimir operator
on a vector. The Casimir is given, as before, by: 
${\mathcal C}=B_{1}^{2}+B_{2}^{2}-J_{3}^{2}$ but we must remember to use as 
$B_1$, $B_2$ and $J_3$ the operators contained
in {\bf TABLE~6}. It is then easy to check that
	\begin{equation}
	\label{CME68}
	{\mathcal C}|q_{\scriptscriptstyle H}\rangle=-2|q_{\scriptscriptstyle H}\rangle.
	\end{equation}
The same we get for the other two vectors $|q_{\scriptscriptstyle K}\rangle,
|q_{\scriptscriptstyle D}\rangle$, so the eigenvalue in the equation above is $-2=-1(1+1)$ 
and this indicates that those vectors make a ``spin 1" representation.

In the same way as before it is easy to prove that the following three vectors:
	\begin{equation}
	\left\{
	\begin{array}{l}
	|\widetilde{q_{\scriptscriptstyle H}}\rangle= (\QH-\Qb)+(\QBH-\QBb)\\
	|\widetilde{q_{\scriptscriptstyle K}}\rangle= (\QK-\Qb)+(\QBK-\QBb)\\
	|\widetilde{q_{\scriptscriptstyle D}}\rangle=(\QD-\Qb)+(\QBD-\QBb)
	\end{array}
	\right.
	\end{equation}
\noindent
make another irreducible representation with ``spin 1".

Of course, as the vector space ${\mathcal G}_{1}$ of {\bf TABLE 7} is
8-dimensional and up to now we have used  only 6 vectors  to build the
two integer representations, we expect that there must be
some other representations which can be built using the two remaining vectors. In fact it is so.
We can build the following two other vectors:
	\begin{align}
	\label{CME70}
	|q_{\scriptscriptstyle BRS}\rangle=\Qb-\QBb\\
	\label{CME71}
	|\widetilde{q_{\scriptscriptstyle BRS}}\rangle=\Qb+\QBb
	\end{align}
\noindent
and it is easy to see that each of them carries a representation of spin zero:
	\begin{equation}
	\label{CME72}
	{\mathcal C}|q_{\scriptscriptstyle BRS}\rangle={\mathcal C}|\widetilde{q_{\scriptscriptstyle
	BRS}}\rangle=0.
	\end{equation} 
So we can conclude that our vector space ${\mathcal G}_{1}$
carries a reducible representation of the conformal algebra
made of two spin one and two spin zero representations.

We wanted to do this analysis in order to underline a further difference
between our supersymmetric extension and the one of\scite{FUB}
whose odd part ${\mathcal G}_{1}$, as we showed before, carries
two ``spin $\frac{1}{2}$" representations.
\section{Superspace Formulation of the Model}

\noindent
In this Section we use the techniques introduced in Section (\ref{sec:superspace}) and we give a
superspace formulation of our new Superconformal Mechanics. 
Proceeding in the same way as we did for the $\QH$ and $\QBH$ charges,
it is a long but easy procedure to give a superspace
representation also for the charges $\QD,\QBD,\QK,\QBK$ of\break
Eqs.(\ref{CME44})-(\ref{CME47}). This long derivation 
is contained in Appendix \ref{app:super1} and the result is:
	\begin{align}
	\label{CME73}
	\mathscr{Q}_{\s K} &= -{\partial\over\partial\theta}-\alpha~\omega^{ad}K_{db}~{\bar\theta}\\
	\label{CME74}
	\ov{\mathscr{Q}}_{\s K} &= {\partial\over\partial{\bar\theta}}+\alpha~\omega^{ad}K_{db}~\theta\\
	\label{CME75}
	\mathscr{Q}_{\s D} & = -{\partial\over\partial\theta}-2\gamma~\omega^{ad}D_{db}~{\bar\theta}\\
	\label{CME76}
	\ov{\mathscr{Q}}_{\s D} & =  {\partial\over\partial{\bar\theta}}+2\gamma~\omega^{ad}D_{db}~\theta
	\end{align}
\noindent
where the matrices $K_{db}$ and $D_{db}$ are:
	\begin{equation}
	\label{CME77}
	K_{db}  = \left(
	\begin{array}{lr}
	1 & 0\\
	0 & 0
	\end{array}
	\right);~~~~~~~
	D_{db}  = 
	-{1\over 2}\left(\begin{array}{lr}
	0 & 1\\
	1 & 0
	\end{array}
	\right)
	\end{equation}
(the repeated indices in Eqs.(\ref{CME73})-(\ref{CME76})
are summed). The matrices $K_{db}$ and $D_{db}$ are $2\times 2$ just because
the symplectic matrix itself $\omega^{ab}$ is $2\times 2$ in our case. The
conformal mechanics system in fact has just a pair of phase-space
variables $(p,q)$ and the index in the $\v^{a}$-phase space variables
can take only 2 values to indicate either $q$ or $p$ (see the Eqs. of motion
(\ref{CPI1})).

\indent From the expressions of $\QCK,\QCBK,\QCD,\QCBD$ above, we see that they have
two free indices. This implies that those
operators ``turn" the various superfields in the sense that they turn a
$\Phi^{q}$~into combinations of $\Phi^{q}$ and $\Phi^{p}$ and viceversa.
This is something the other charges did not do.

Reached this point, we should  stop and think a little
bit about this superspace representation. We gave the superspace
representation of the various charges
$(\QD,\QBD,\QK,\QBK)$ of Eqs.(\ref{CME44})-(\ref{CME47}) which were linked to the
$\DT_0,\KT_0$ of\break Eqs.(\ref{CME42})(\ref{CME43}). But these last quantities were built
using the $D_{0}$ and $K_{0}$ that is the $D$ and $K$ at $t=0$.
If we had used, in building the $\DT_0, \KT_0$, the $D$ and $K$ at $t\neq 0$
of Eqs.(\ref{CME9}) and (\ref{CME10}), we would have obtained a $\DT$ and a $\KT$ different
from those of Eqs.(\ref{CME42})(\ref{CME43}) and which would have had an explicit
dependence on $t$. Consequently also the associated supersymmetric
charges $(Q^t_{\scriptscriptstyle D},\overline{Q}^t_{\scriptscriptstyle D},
Q^t_{\scriptscriptstyle K},\overline{Q}^t_{\scriptscriptstyle K})$, having extra terms depending on
$t$, would be different from those of Eqs.(\ref{CME44})-(\ref{CME47}). 
Being these charges different,
also their superspace representations will be different from those 
given in Eqs.(\ref{CME73})-(\ref{CME76}). The difference at the level of superspace is crucial
because it involves $t$ which is part of superspace.

Let us then start this over-all process by first building the explicitly $t$-dependent
$\DT$ and $\KT$ from the following operators $D$ and $K$:

	\begin{align}
	\label{CME78}
	H &=  H_{0}\\
	\label{CME79}
	D & =  t H+D_{0}\\
	\label{CME80}
	K & =  t^{2}H+2tD_{0}+K_{0}
	\end{align}

\noindent
from which we get:
	\begin{align}
	\label{CME81}
	\DT & = t\HT+\DT_{0}\\
	\label{CME82}
	\KT & = t^{2}\HT+2t\DT_{0}+\KT_{0}.
	\end{align}

\noindent
It is easy to understand why these relations hold by remembering the
manner we got the Lie-derivatives out of the superpotentials.
The explicit form of $\DT$ in terms of \quattrova
can be obtained from (\ref{CME81}) once we insert the $\HT$ and $\DT_{0}$
whose explicit form we already had in Eqs.(\ref{CME39}) and (\ref{CME42}). The same for
$\KT$. Let us now turn to the form of the
associated fermionic charges which we will indicate as
$(\QDT,\QKT,\QBDT,\QBKT)$ where the index ``$({\s\ldots})^{t}$" is to indicate
their explicit dependence on $t$. As it is shown in formula (\ref{C-1}) of 
Appendix \ref{app:super1}, the $\QD$ and $\QK$ can be written using the charges $\ND$
and $\NK$ of (\ref{C-2}) and the $\Qb$. As it is only the $N_{(\ldots)}$ and not the
$\Qb$ which pull in quantities like $D,K$ which may depend 
explicitly on time, we should only concentrate on the $N_{(\ldots)}$. From 
their definition (see Eq.(\ref{C-2})):

	\begin{equation}
	\label{CME83}
	\ND=c^{a}\partial_{a}D;~~~~~~~
	\NK=c^{a}\partial_{a}K
	\end{equation}

\noindent we see that applying the operator $c^{a}\partial_{a}$ on both sides 
of Eqs.(\ref{CME79})(\ref{CME80}), we get:

	\begin{align}
	\label{CME84}
	\NDT & = t\NH+\ND\\
	\label{CME85}
	\NKT & = t^{2}\NH+2t\ND+\NK.
	\end{align}
	
\noindent The next step is to write the $\QDT$ and $\QKT$. As they are given in
formula (\ref{C-1}), using that equation and (\ref{CME84})(\ref{CME85}) above
we get:

	\begin{align}
	\label{CME86}
	\QDT & = \Qb-2\gamma N^t_{\scriptscriptstyle D}\\
	\label{CME87}
	\QKT & = \Qb-\alpha N^t_{\scriptscriptstyle K}.
	\end{align}

\noindent In a similar manner, via Eq.(\ref{C-9}) and applying the operator
${\bar c}_{a}\omega^{ab}\partial_{b}$ to Eqs.(\ref{CME79}) and (\ref{CME80}), we get $\QBDT$
and $\QBKT$:

	\begin{align}
	\label{CME88}
	\QBDT & = \QBb+2\gamma\overline{N}^t_{\scriptscriptstyle D}\\	
	\label{CME89}
	\QBKT & = \QBb+\alpha\overline{N}^t_{\scriptscriptstyle K}.
	\end{align}

\noindent We shall not write down explicitly the expressions of $(\QDT,\QKT,\QBDT,\QBKT)$
in terms of \quattrova because we have already 
in Eqs.(\ref{CME44})-(\ref{CME47}) and (\ref{C-2})(\ref{C-9}) the expressions\footnote{Only be careful
in using the $D_{0}$ and $K_{0}$ in Eqs.(\ref{C-1}) and (\ref{C-9}).} of the various charges
$(\QD,\QK,\QBD,\QBK,\ND,\NK, {\overline\NK},{\overline\ND})$ which make up,
according to Eqs.(\ref{CME86})-(\ref{CME89}), the new time dependent charges.
The next step is to obtain the superspace version of $(\QDT,\QKT,\QBDT,\QBKT)$.
Following a procedure identical to the one explained in detailed in the
Appendix for the time-independent charges it is easy to get them and, via
their anticommutators, to derive the superspace version of $\DT$ and $\KT$. All these
operators are listed in the table below:
\begin{center}
{\bf TABLE 9}
\end{center}
\begin{small}
\[
\begin{array}{|rcl|}
\hline && \\
\HCT&=&\displaystyle i\frac{\partial}{\partial t};\\
\DCT&=&\displaystyle i t\frac{\partial}{\partial t}-\frac{i}{2}\sigma_3;\\
\KCT&=&\displaystyle it^2\frac{\partial}{\partial t}-it\sigma_3-i\sigma_-;\\
\mathscr{Q}^t_{\scriptscriptstyle D}&=&\displaystyle-\frac{\partial}{\partial\theta}-
2\gamma~{\bar\theta}t\frac{\partial}{\partial
t}+\gamma~{\bar\theta}\sigma_3;\\
\ov{\mathscr{Q}}^t_{\scriptscriptstyle 
D}&=&\displaystyle\frac{\partial}{\partial\bar{\theta}}+2\gamma~\theta
t\frac{\partial}{\partial t}-\gamma{\theta}\sigma_3;\\
\mathscr{Q}^t_{\scriptscriptstyle 
K}&=&\displaystyle-\frac{\partial}{\partial\theta}-\alpha~{\bar\theta}t^2\frac{\partial}{\partial
t}+\alpha~t{\bar\theta}\sigma_3+\alpha~{\bar\theta}\sigma_{-};\\
\ov{\mathscr{Q}}^t_{\scriptscriptstyle
K}&=&\displaystyle\frac{\partial}{\partial\bar{\theta}}+\alpha\theta~
t^2\frac{\partial}{\partial t}-\alpha t~\theta\sigma_3-\alpha\theta\sigma_{-}; \\
\mathscr{H}&=&\displaystyle\bar{\theta}\theta\frac{\partial}{\partial t};\\
\mathscr{D}&=&\displaystyle\bar{\theta}\theta~(t\frac{\partial}{\partial
t}-\frac{1}{2}\sigma_3);\\
\mathscr{K}&=&\displaystyle\bar{\theta}\theta~(t^2\frac{\partial}{\partial
t}-t\sigma_3-\sigma_{-}).\\
&& \\
\hline
\end{array}
\]
\end{small}
\noindent In the previous table $\sigma_{3}$ and $\sigma_{-}$ are the Pauli
matrices:

	\begin{equation}
	\label{CME90}
	\sigma_{3}  = 
	\left(\begin{array}{lr}
	1 & 0\\
	0 & -1
	\end{array}
	\right);~~~~~~~
	\sigma_{-}  = 
	\left(\begin{array}{lr}
	0 & 0\\
	1 & 0
	\end{array}
	\right).
	\end{equation}

\noindent The reasons for the presence of these two-dimensional matrices
has already been explained in the paragraph below Eq.(\ref{CME77}).

The last three operators listed in {\bf TABLE 9} are the
superspace version of the old $(H,D,K)$. To get this representation
we used again and again the rules given by Eq.(\ref{CPI48}). As their representation
looks quite unusual, we have reported the details of their
derivations in Appendix \ref{app:super2}.

\section{Exact Solution of the Supersymmetric Model} 

The original conformal mechanical model was solved exactly
in Eq.(2.35) of Ref.\cite{DFF}. The solution is given by the relation:
	\begin{equation}
	\label{CME99}
	q^{2}(t)=2t^{2}H-4t D_{0}+2K_{0}.
	\end{equation}

\noindent As $(H,D_{0},K_{0})$ are constants of motion, once their values are assigned we
stick them in Eq.(\ref{CME99}), and we get a relation between ``$q$" (on the LHS of 
(\ref{CME99})) and ``$t$" on the RHS. This is the solution of the equation
of motion with ``initial conditions" given by the values we assign
to the constants of motion $(H,D_{0},K_{0})$. The reader may object
that we should give only 
two constant values (corresponding to the initial conditions $(q(0),\,{\dot q}(0))$) and not three.  
Actually the three values assigned to $(H,D_{0},K_{0})$ are not arbitrary because, as it was
proven in Eq.(2-36) of Ref.\scite{DFF}, these three quantities are linked
by the constraint:
	\begin{equation}
	\label{CME100}
	\left(HK_{0}-D_{0}^{2}\right)={g\over 4}
	\end{equation}

\noindent where ``$g$" is the coupling which entered the original Hamiltonian (see Eq.(\ref{CME8})
of the present paper). Having one constraint among the three constants of motion brings them
down to two.

\noindent The proof of the relation (\ref{CME99}) above is quite simple. On the RHS, as the
$(H,D_{0},K_{0})$ are constants of motion, we can replace them with 
their time dependent expression $(H,D,K)$ (see Eqs.(\ref{CME78})-(\ref{CME80})), which are
explicitly:

	\begin{align}
	\label{CME101}
	H & = {1\over 2}\left({\dot q}^{2}(t)+{g\over q^{2}(t)}\right)\\
	\label{CME102}
	D & = tH-{1\over 2}q(t){\dot q}(t)\\
	\label{CME103}
	K & = t^{2}H-t~q(t){\dot q}(t)+{1\over 2}q^{2}
	\end{align}

\noindent Inserting these expressions in the RHS of Eq.(\ref{CME99}) we get immediately
the LHS. From Eqs.(\ref{CME101})-(\ref{CME103}) it is also easy to see the relations
between the initial conditions $(q(0),\,{\dot q}(0))$ and the constants
$(H,D_{0},K_{0}$); in fact we have:

	\begin{align}
	\label{CME104}
	H & = H(t=0)={1\over 2}\left({\dot q}^{2}(0)+{g\over q^{2}(0)}\right)\\
	\label{CME105}
	D & = D_{0}=-{1\over 2}\left(q(0){\dot q}(0)\right)\\
	\label{CME106}
	K & = K_{0}={1\over 2}q^{2}(0).
	\end{align}

\noindent From the relations above we see that, inverting them,
we can express ($q(0),\,{\dot q}(0)$) in term of $(H,D_{0},K_{0})$.
The constraint (\ref{CME100}) is already involved in the expression of 
$(H,D_{0},K_{0})$ in terms of $(q(0),\,{\dot q}(0))$.

What we want to do in this section is to see whether a relation
analogous to (\ref{CME101}) exists also for our supersymmetric extension
or in general whether the supersymmetric system can be solved exactly.
The answer is {\it yes} and it is based on a very simple trick.

Let us first remember Eq.(\ref{CPI59}) which told us how $\HT$ and $H$ are related:

	\begin{equation}
	\label{CME107}
	i\int H(\Phi)~d\theta d{\bar\theta}=\HT.
	\end{equation}

\noindent The same relation holds for $\DT_{0}$ and $\KT_{0}$ with respect to $D_{0}$
and $K_{0}$ as it is clear from the explanation given in the paragraph above 
Eqs.(\ref{CME42})(\ref{CME43}) that:

	\begin{align}
	\label{CME108}
	& i\int D_{0}(\Phi)~d\theta d{\bar\theta}= \DT_{0}\\
	\label{CME109}
	& i\int K_{0}(\Phi)~d\theta d{\bar\theta} = \KT_{0}.
	\end{align}

\noindent Of course the same kind of relations holds for the explicitly time-dependent
quantities of Eqs.(\ref{CME79})-(\ref{CME82}):

	\begin{align}
	\label{CME110}
	& i\int D(\Phi)~d\theta d{\bar\theta} = \DT\\
	\label{CME111}
	& i\int K(\Phi)~d\theta d{\bar\theta}=  \KT.
	\end{align}

\noindent Let us now build the following quantity:

	\begin{equation}
	\label{CME112}
	2t^2H(\Phi)-4tD(\Phi)+2K(\Phi).
	\end{equation}

\noindent This is functionally the RHS of Eq.(\ref{CME99}) with
the superfield $\Phi^{a}$ replacing the normal phase-space variable
$\v^{a}$. It is then clear that the following relation holds:

	\begin{equation}
	\label{CME113}
	\left(\Phi^{q}\right)^{2}=2t^2H(\Phi)-4tD(\Phi)+2K(\Phi).
	\end{equation}

\noindent The reason it holds is because, in the proof of the analogous one in $q$-space
(Eq.(\ref{CME99})), the only thing we used was the functional form of the $(H,D,K)$
that was given by Eqs.(\ref{CME101})-(\ref{CME103}). So that relation holds irrespective
of the arguments, $\v$ or $\Phi$, which enter our functions provided that
the functional form of them remains the same. 

\noindent Let us first remember the form of $\Phi^{q}$ which appears\footnote{The
index $({\s\ldots})^{q}$ is not a substitute for the index ``$a$" but it
indicates, as we said many times before, the first half of the indices ``$a$". Let us remember
in fact that the first half of the $\v^{a}$ are just the configurational variables
$q^{i}$ which in our case of a 1-dim. system is just one variable.}
on the LHS of Eq.(\ref{CME113}):

	\begin{equation}
	\label{CME114}
	\Phi^{q}(t,\theta,\bar\theta)=q(t)+\theta~c^{q}(t)+{\bar\theta}\omega^{qp}{\bar c}_{p}(t)+i
	{\bar\theta}\theta~\omega^{qp}\lambda_{p}(t).
	\end{equation}

\noindent Let us now expand in $\theta$ and ${\bar\theta}$ the LHS and RHS of
Eq.(\ref{CME113}) and compare the terms with the same power of $\theta$ and ${\bar\theta}$.

The RHS is:
	\begin{equation}
	\label{CME115}
	\left(\Phi^{q}\right)^{2}=q^{2}(t)+\theta[2q(t)c^{\scriptscriptstyle q}(t)]+{\bar\theta}
	[2q(t){\bar c}_{\scriptscriptstyle p}(t)]+
	{\bar\theta}\theta[2iq(t)\lambda_{\scriptscriptstyle p}(t)+2
	c^{\scriptscriptstyle q}(t){\bar c}_{\scriptscriptstyle p}(t)].
	\end{equation}
	
\noindent The LHS is instead:
	\begin{multline}
	\label{CME116}
	2t^2H(\Phi)-4tD(\Phi)+2K(\Phi) = 2t^2H(\v)-4tD(\v)+2K(\v)
	+ \theta\big[ 2t^2 N^t_{\scriptscriptstyle H}+ \\
	-4t\ND^{t}+2\NK^{t}\big]
	- {\bar\theta}\big[2t^2{\overline{N}^t_{\scriptscriptstyle H}}-4t{\overline\ND}^{t}+
	2{\overline\NK}^{t}\big]
	+ i\theta{\bar\theta}\big[2t^2\HT-4t\DT+2\KT\big],
	\end{multline}

\noindent where the $\NH,{\overline\NH}$,$\ND^{t},\NK^{t}$ are defined in
Eqs.(\ref{CME84})(\ref{CME85}), while the $\overline{N}^{t}_D$,
$\overline{N}^t_K$ are the time-dependent versions\footnote{By ``time-dependent
versions" we mean that they are related to the time independent ones
in the same manner as the $N^{t}$-functions were via Eqs.(\ref{CME84})(\ref{CME87}).} of the
operators defined in Eqs.(\ref{C-10}) and (\ref{C-11}). It is a simple exercise to
show that all these functions\break $(H,D,K,N_{(\ldots)},{\overline
N}_{(\ldots)},\HT,\KT,\DT)$ are constants of motion
in the enlarged space\break \quattrova.

If we now compare the RHS of Eq.(\ref{CME115}) and Eq.(\ref{CME116}) and equate terms with the
same power of $\theta$ and ${\bar\theta}$, we get (by writing the $N$ and
${\overline N}$ explicitly):

	\begin{align}
	\label{CME117}
	q^{2}(t) & = 2t^2H(\v)-4tD(\v)+2K(\v);\\
	2q(t)c^{q}(t) & = \left[2t^2{\partial H\over\partial\v^{a}}-4t{\partial
	D\over\partial\v^{a}}+2{\partial K\over\partial\v^{a}}\right]c^{a};\label{CME118}\\
	2q(t){\bar c}_{p}(t) & = \left[2t^2{\partial H\over\partial\v^{a}}-4t{\partial
	D\over\partial\v^{a}}+2{\partial K\over\partial\v^{a}}\right]\omega^{ab}
	{\bar c}_{b};\label{CME119}\\
	i2q(t){\lambda_{\scriptstyle p}}(t)+2 c^{q}(t){\bar
	c}_{p}(t)& = -i\left[2t^2\HT-4t\DT+2\KT\right].\label{CME120}
	\end{align}

\noindent We notice immediately that Eq.(\ref{CME117}) is the same as the one of the original
paper\scite{DFF} and solves the motion for ``$q$". Given this solution
we plug it in Eq.(\ref{CME118}) and, since on the RHS we have the $N$-functions which
are constants, once these constants are assigned we get the motion of
$c^{\scriptscriptstyle q}$. Next
we assign three constant values to the ${\overline N}$-functions
which appear on the RHS of Eq.(\ref{CME119}), then we plug in the solution for $q$ given by 
Eq.(\ref{CME117}) and so we get the trajectory for ${\bar c}_{p}$. Finally we do the
same in Eq.(\ref{CME119}) and get the trajectory of $\lambda_{\scriptscriptstyle p}$.

The solution for the momentum-quantities 
$(p,c^{\scriptscriptstyle p},{\bar c}_{\scriptscriptstyle
q},\lambda_{\scriptscriptstyle q})$ can be obtained via their
definition in terms of the previous variables.

The reader may be puzzled by the fact that in the space \quattrova we have 8
variables but we have to give 12 constants of
motion: $(H,D_{0},K_{0},N_{(\ldots)},$ ${\overline N}_{(\ldots)},\HT,\KT,\DT)$
to get the solutions from Eqs.(\ref{CME117})-(\ref{CME120}). The point is that, like in the case
of the standard conformal mechanics\scite{DFF}, we have constraints
among the constants of motion. We have already one constraint and it is
given by Eq.(\ref{CME100}). The others can be obtained in the following manner:
let us apply the operator $c^{\scriptscriptstyle a}
\partial_{\scriptscriptstyle a}$ on both sides of Eq.(\ref{CME100}) and what we get is the following relation:
	\begin{equation}
	\label{CME121}
	\NH K_{0}+\NK H-2\ND D_{0}=0
	\end{equation} 

\noindent which is a constraint for the $N$-functions. Let us now do the same applying
on both sides of Eq.(\ref{CME100}) the operator ${\bar c}_{a}\omega^{ab}
\partial_{\scriptscriptstyle b}$. What we get is:
	\begin{equation}
	\label{CME122}
	{\overline \NH}K_{0}+{\overline \NK}H-2{\overline \ND}D_{0}=0
	\end{equation}

\noindent which is a constraint among the ${\overline N}$-functions. Finally let us
apply  $\Qb$ to Eq.(\ref{CME122}) and we get:
	\begin{equation}
	\label{CME123}
	i\HT K_{0}+i\KT H-21{\DT}D_{0}-{\overline \NH}
	\NK-{\overline \NK}\NH+2{\overline \ND}\ND=0
	\end{equation}

\noindent which is a constraint among the $\HT,\DT,\KT$.

So we have 4 constraints (\ref{CME123})(\ref{CME122})(\ref{CME121})(\ref{CME100}) which bring down
the constants of motion to be specified in \quattrova  from 12 to 8
as we expected.

\chapter*{Conclusions}
\addcontentsline{toc}{chapter}{\numberline{}Conclusions}
\setcounter{chapter}{6}
\setcounter{section}{0}
\markboth{}{Conclusions}

\noindent
In this thesis we have continued the analysis of the formalism of the Classical Path Integral
introduced in the literature few years ago. The CPI-Hamiltonian exhibits many symmetries, some of
which have already been understood geometrically. In this thesis we have continued this project and we have
focused on some other symmetries which seem to have some interesting implications in several issues like
--- for instance --- the study of the geometry of the classical trajectories, the analysis of the ergodic
character of a classical system, the quantization mechanism etc. 

The first symmetry which we have analyzed is the classical $N=2$ supersymmetry. This symmetry is
{\it universal}, in the sense that it does not depend on the particular form of the Lagrangian of
the system at hand. To understand its geometrical meaning we have built an extension of the
CPI-Lagrangian, which exhibits the supersymmetry as a {\it local} symmetry. This model is therefore a
{\it gauge} theory and consequently only a subset of the overall Hilbert space turns out to be {\it
physical}. In particular, we have discovered that this physical Hilbert space is in one-to-one
correspondence with the {\it equivariant forms} derived from the action of the 1-parameter Hamiltonian
evolution group on the phase space. Our goal was to exploit this correspondence to study the geometry
of the space of the classical trajectories. In fact there is an important formula which relates the
equivariant cohomology $H_{\s G}({\cal M})$ built from the action of a group $G$ on a manifold $\cal M$ to the de-Rham
cohomology of the quotient space $H({\cal M}/G)$. The latter is the main instrument to analyze the
geometry of $H({\cal M}/G)$ which in our case ($G$ is the Hamiltonian evolution group) is the space of the
classical trajectories. 
However there is still some work to be done in this direction because the correspondence above is valid
only if the action of the group $G$ on $\cal M$ is free, which means that there are no fixed points. This
is not always true in our case but we hope to find other
useful tools for studying the space of classical orbits.

Besides the relation to the space of classical trajectories, there is also another point where the
classical susy can play an important role. In fact, in the literature, there are some attempts to relate
the classical supersymmetry to the ergodicity character of a classical system.
What was proved in Ref.\cite{Ergo} was that if susy is unbroken the system is ergodic ($\text{\it 
susy unbroken}\Longrightarrow\text{\it ergodicity}$), but
the authors could not prove the reverse, basically because their analysis should have been restricted to the
constant-energy surfaces, where the ergodic characterization makes sense. Here we have tried to make a step ahead
along this direction and we have modified the CPI-Lagrangian inserting by hand the fixed-energy
constraint. What we have noticed is that the $N=2$ supersymmetry reduces to an $N=1$ and hopefully it is
this supersymmetry which should be used as an instrument to test the ergodic phase of the system at hand.
However this is still an open problem.   

There is another symmetry which we have analyzed in detail and is strictly tied to the classical supersymmetry. In fact
in the superspace $(t,\t,\tb)$ its generators $D_{\s H}$ and $\ov{D}_{\s H}$ are represented by the
covariant derivatives associated to the classical susy charges $\QH$ and $\QBH$. Here we have shown that
the transformation generated by $D_{\s H}$ and $\ov{D}_{\s H}$ is very similar to the well known
$\kappa$-symmetry introduced in the literature 20 years ago in the context of the relativistic
superparticles. Following the same lines as for the susy, we have built a local extension of the
CPI-Lagrangian which has $D_{\s H}$ and $\ov{D}_{\s H}$ as Noether's charges associated to the gauge
symmetry. We have noticed that in our nonrelativistic case there is no problem (differently from the
relativistic case) in separating 1st-class from 2nd-class constraints.  Moreover the physical Hilbert
space associated to this gauge theory is either the space of the Gibbs states (describing the canonical 
ensemble) or the space of the distributions built of constants of motion only, according to the form of
the coupling we choose in the gauge Lagrangian.

There is also a further universal symmetry of the CPI formalism which we have discovered in this work and which
--- we think --- can be crucial in the understanding of the quantization process. Differently from 
the previous symmetries we have mentioned so far, this one is not canonical and the analytic
form of the generator can be found only in the superspace $(t,\t,\tb)$. Nevertheless this symmetry (we have
called it ${\mathscr Q}_{\s S}$) is very interesting because its effect is a rescaling of the overall
CPI-action, and this seems to play a crucial role in the quantization process. In fact the rescaling of
the classical action is a transformation which is a universal symmetry at the classical level, but turns
out to be lost when one goes to the quantum domain due to the presence of $\hbar$ setting a scale for the
action. Along this direction we have also analyzed the possibility that the $\mathscr{Q}_{\s
S}$-symmetry can be interpreted as an extension to the phase space $\MT$ of a transformation which
rescales the action already at the level of the ordinary phase space $\cal M$. Actually this
transformation does exist (and we called it ``MSA"), but it is not a ``true" symmetry. 
The reason why we called the $\mathscr{Q}_{\s S}$-transformation a symmetry was that it leaves
invariant the equations of motion of the system at hand. This is obviously true provided that the Lagrange
equations are mapped by the transformation ($\mathscr{Q}_{\s S}$ in this case) to new Lagrange equations.
This is no longer true in the case of the MSA transformation, which is anholonomic (it transforms
coordinates and differentials in an independent manner) and therefore does not preserve the Lagrange
equations. However we hope to shed some further light in this direction in the future, because we think
that the rescaling of the action is something which is playing a crucial role in
passing from the classical to the quantum regime of a physical system.

In the last chapter of the thesis we have studied a simpler kind of rescaling, which is known as
``superconformal transformation". In the literature this name is used to denote a composition of a
supersymmetry plus a conformal transformation. In this thesis we have studied a new kind of
superconformal algebra obtained by applying the CPI-formalism to a model (known in the literature as
``Conformal Mechanics") which exhibits a nonrelativistic conformal invariance. The universal susy of the
CPI combines with the generators of the conformal algebra leading to a new kind of superconformal algebra.
The main difference between the latter and the other models present in the literature is that in our
superconformal algebra the even part is represented faithfully (and not spinorially) on the odd part.   
We hope that this particular model may be a useful playground to tackle the more general problem of the
rescaling induced by $\mathscr{Q}_{\s S}$, but this still remains an open question.

\appendix

\chapter{Local Susy: Mathematical Details}
\markboth{A. Local Susy: Mathematical Details}{}

\section{Susy and time-derivative}
\label{app:intSusy}

\noindent
In deriving Eqs. (\ref{LS4}) and (\ref{LS5}), or even
in checking the global symmetry under $\QH$,
we had to work out things involving the variation of the kinetic
piece of $\LT$, i.e.:
\begin{equation}
\label{eq:B-uno}
	[\e\QH,\l_{a}\dot{\v}^{a}+i\bc_{a}{\dot
	c}^{a}]=(\delta\l_{a})\dot{\v}^{a}+\l_{a}
	\frac{d}{dt}(\delta\v^{a})+i(\delta\bc_{a})\dot{c}^{a}+i\bc_{a}\frac{d}{dt}(\delta
	c^{a})
\end{equation}

\noindent In this step we have interchanged the variation ``$\delta$" with
the time derivative ${d\over dt}$. If we actually do the time derivative
of a variation (for example of $\varphi^{a}$), we get

\begin{equation}
\label{eq:B-due}
\begin{array}{ll}
	\begin{array}{rl}
	\displaystyle\frac{d}{dt}(\delta\v^{a}) & = \displaystyle\frac{d}{dt}[\e\QH,\v^{a}]
	\vspace{.3cm}\\
	& = [\frac{d}{dt}(\e\QH),\v^{a}] + [\e\QH,\frac{d\v^{a}}{dt}]\vspace{.15cm} \\
	& = [\e\QH,\frac{d\v^{a}}{dt}]\vspace{.2cm} \\
	& = \delta\left(\displaystyle\frac{d\v^{a}}{dt}\right),
	\end{array}
\end{array}
\end{equation}

\noindent and if $\e$ is a global parameter the third equality in the equation
above holds (and as a result  we can interchange the variation
with the time derivative)  only if we use the fact that  ${d\QH\over dt}=0$. 
We have supposed the same thing in the case of the local variations
(\ref{LS4})(\ref{LS5}), and the only
extra term appearing with respect to Eq.(\ref{eq:B-due}) is the one containing
the ${\dot\epsilon}$.  
Using the conservation of $\QH$
means that we have assumed that the equations of motion hold.
Actually it is better not to assume that. In fact, if we
make this assumption, then the $\LT$ itself, of which we are checking the
invariance
via the variations above, would be zero. This is due to the fact that $\LT$ is
proportional
to the equations of motion and checking the invariance of
something that is zero is a nonsense.  

To avoid that problem the trick to use is to define the following
integrated charge:

\begin{equation}
\label{eq:B-tre}
	\widetilde{\QH}=\int_{0}^{T}\QH(t)dt,
\end{equation}

\noindent where $0$ and $T$ are the endpoints of the interval over which we
consider our motion.

It is then easy to check that all the steps done in Eq.(\ref{eq:B-due}),
once we replace $\QH$ with ${\widetilde\QH}$, can go 
 through without assuming the conservation of $\QH$. 
In fact  the ${d\widetilde\QH\over dt}$ in the third step of equation 
(\ref{eq:B-due})
is zero not because of the conservation of $\QH$ but because ${\widetilde\QH}$
is independent of $t$. Moreover the variation $\delta$ generated
by the ${\widetilde\QH}$ is the same as the one generated by $\QH$. This is so
because in checking the variation induced by ${\widetilde\QH}$
we have to use the non-equal-time commutators given by the path-integral
(\ref{CPI7}) which are:

\begin{equation}
\label{eq:B-quattro}
[\v^{a}(t),\lambda_{b}(t^{\prime})]=i\delta^{a}_{b}
\delta(t-t^{\prime})
\end{equation}

\noindent and similarly for the $c^{a}$ and ${\ov c}_{a}$.

This charge was introduced before in the literature\scite{Blau} in order to
handle things in an abstract ``loop space". In our case we need 
that charge  for the much simpler reasons explained above.

\section{Combination of two Susy transformations}
\label{app:combSusy}

\noindent In this Appendix we will show what happens when we combine
two Susy transformations.

Let us define the following two transformations $G_{\e_{\s 1}}$,
$G_{\e_{\s 2}}$ on any of the variables \quattrova (which we will collectively 
indicate with $O$).

\begin{equation}
\label{eq:C-uno}
	\begin{array}{l}
	\delta_{1}O\equiv[G_{\e_{\s 1}}, O]\equiv[\ov\e_{1}\QBH + \e_{1}\QH, O]
	\\
	\delta_{2}O\equiv[G_{\e_{\s 2}}, O]\equiv[\ov\e_{2}\QBH + \e_{2}\QH, O] 
	\end{array}
\end{equation}

\noindent where the infinitesimal parameters $\e_{\s 1}$,$\e_{\s 2}$ are time
dependent.

Combining two of these transformations we get

\begin{equation}
\label{eq:C-due}
	[\delta_{1},\delta_{2}]O=[G_{\e_{\s 1}},[G_{\e_{\s 2}},O]]-
	[G_{\e_{\s 2}},[G_{\e_{\s 1}},O]].
\end{equation}

\noindent Applying the Jacobi identity to the  RHS of (\ref{eq:C-due}), we get

\begin{equation}
\label{eq:C-tre}
	[\delta_{1},\delta_{2}]O=[[G_{\e_{\s 1}},G_{\e_{\s 2}}],O].
\end{equation}

\noindent 
It is  easy to find out that $[G_{\e_{\s 1}}, G_{\e_{\s 2}}]$ is given by:
\begin{equation}
\label{eq:C-quattro}
	[G_{\e_{\s 1}},G_{\e_{\s
	2}}]=2i\beta(\ov{\e}_{1}\e_{2}+\e_{1}\ov{\e}_{2})\HT;
\end{equation}

\noindent inserting (\ref{eq:C-quattro}) in (\ref{eq:C-tre}), we get

\begin{equation}
\label{eq:C-cinque}
	\begin{array}{rl}
	[\delta_{1},\delta_{2}]O & 
	= 2i\beta(\ov{\e}_{1}\e_{2}+\e_{1}\ov{\e}_{2})[\HT,O] \vspace{.2cm} \\
	&= 2\beta(\ov{\e}_{1}\e_{2}+\e_{1}\ov{\e}_{2})\displaystyle\frac{dO}{dt}.
	\end{array}
\end{equation}

\noindent So we see from here that the composition of two local Susy transformations
produces a local time-translation with parameter ${\ov\e}_{\s 1}\e_{\s 2}+
\e_{\s 1}{\ov\e}_{\s 2}$.

It is also instructive to do the composition of two {\it finite} 
Susy transformations, the first ($G_{\s 1}$) with parameter $\e_{\s 1}$ 
and the other ($G_{\s 2}$) with parameter $\e_{\s 2}$.
The transformed variable $O^{\prime}$ has the expression:

\begin{equation}
\label{eq:C-sei}
	O'= \mbox{e}^{iG_{\s 1}}\mbox{e}^{iG_{\s 2}}\; O
	\;\mbox{e}^{-iG_{\s 2}}\mbox{e}^{-iG_{\s 1}}.
\end{equation}

\noindent Using the Baker-Hausdorff identity on the RHS above, we obtain:

\begin{equation}
\label{eq:C-sette}
	\begin{array}{rl}
	O' & = \mbox{e}^{[iG_{1} + iG_{2} -\frac{1}{2}[G_{1},G_{2}]]}\;
	O \; \mbox{e}^{[-iG_{1} - iG_{2} +\frac{1}{2}[G_{1},G_{2}]]} \\
	&= \mbox{e}^{i[\ov{\e_{1}}\QBH + \e_{1}\QH + \ov{\e_{2}}\QBH + \e_{2}\QH
	-\beta(\ov{\e_{1}}{\e_{2}} + \e_{1}\ov{\e_{2}})\HT]} \; O \;
	\mbox{e}^{-i[\ov{\e_{1}}\QBH +
	\e_{1}\QH + \ov{\e_{2}}\QBH + \e_{2}\QH -\beta(\ov{\e_{1}}{\e_{2}} +
	\e_{1}\ov{\e_{2}} )\HT]} \\
	&= \mbox{e}^{i\ov{\gamma}\QBH + i\gamma\QH + i\Delta t \HT}\; O
	\;\mbox{e}^{-i\ov{\gamma}\QBH-i\gamma\QH - i\Delta t\HT},
	\end{array}
\end{equation}

\noindent where $\ov\gamma$, $\gamma$ and $\Delta t$ are respectively

\begin{equation}
\label{eq:C-8}
	\left\{
	\begin{array}{l}
	\gamma=\e_{1} + \e_{2} \\
	\ov{\gamma} = \ov{\e_{1}} + \ov{\e_{2}} \\
	\Delta t = -\beta(\ov{\e_{1}}\e_{2}+\e_{1}\ov{\e_{2}}).
	\end{array}
	\right.
\end{equation}

\noindent So we see from Eq.(\ref{eq:C-sette}) that the composition of two
finite local Susy transformations
is a local Susy plus a local time-translation.

We will write down here how the variables \quattrova transform under a local
time
translation:

\begin{equation}
\label{eq:C-nove}
\left\{
\begin{array}{l}
\delta^{\s
loc}_{\HT}\v^{a}=[\eta(t)\HT,\v^{a}]=-i\eta~\omega^{an}\partial_{n}\HT\\
\delta^{\s
loc}_{\HT}\lambda_{a}=[\eta(t)\HT,\lambda_{a}]=i\eta~\partial_{a}\HT\\
\delta^{\s
loc}_{\HT}c^{a}=[\eta(t)\HT,c^{a}]=-i\eta~\omega^{an}\partial_{n}\partial_{l}Hc^{l}\\
\delta^{\s loc}_{\HT}{\bar c}_{a}=[\eta(t)\HT,{\bar c}_{a}]=i\eta~{\bar
c}_{m}\omega^{mn}\partial_{n}\partial_{a}H\;.
\end{array}
\right.
\end{equation}

\section{First and Second class constraints}
\label{app:constr}

\noindent
In this Appendix we  analyze the question of how the transformations
(\ref{LS14}) are generated by our first class constraints
(\ref{LS15})(\ref{LS20}). This is a delicate issue
which is explained in detail on page 75 {\it ff} of Ref.\scite{Teitel}. In fact, naively
the transformation on $g$ contained in (\ref{LS14}) apparently
cannot be obtained by doing the commutator of $g$ with the
proper gauge generators. The authors of Ref.\scite{Teitel} are 
aware of similar problems and they 
suggested the following approach. First let us build an extended action
defined in the following way:
\begin{equation}
\label{eq:D-uno}
S_{ext.}=\int dt [\LTloc +\pipsi{\dot \psi}+\pipsibar{\dot{\ov\psi}}+
\pigi{\dot g}-U^{(i)}G_{i}]
\end{equation}

\noindent
where the $G_{i}$ are all the six first class constraints (\ref{LS24})
(and not just the primary ones)
and the $U^{(i)}$ the relative Lagrange multipliers. A general gauge
transformation
of an observable $O$ will be:

\begin{equation}
\label{eq:D-due}
\delta O=[{\ov\e} {\QBH}+\e \QH
+\eta\HT+{\ov\alpha}\pipsi+{\alpha}\pipsibar+\beta\pigi,O]
\end{equation}
\noindent
where $({\ov\e},\e,\eta, {\ov\alpha},{\alpha},\beta)$ are six infinitesimal gauge
parameters
associated to the six generators $G_{i}$.
If we consider the Lagrange multipliers $U^{i}$ as functions of the basic
variables,
then they will also change under the gauge transformation above. As we do not
know
the exact expression of the $U^{i}$ in terms of the basic variables, we will
formally indicate their gauge variation as $\delta U^{i}$. Using this notation
it is then a simple but long calculation to show that the gauge variation
of the action $S_{ext.}$ is:

\begin{equation}
\label{eq:D-tre}
\begin{array}{rl}
\delta S_{ext}=\displaystyle\int dt & \left[
i{\dot\e}\QH-i{\dot{\ov\e}}\QBH-i{\dot\eta}\HT-2i\HT
(\e\psi+{\ov\e}{\ov\psi})+\right. \\
&+{\ov\alpha}\QBH+\alpha\QH-i\beta\HT+\pipsi{\dot{\ov\alpha}}+\pipsibar
{\dot\alpha}-i\pigi{\dot\beta}+ \\
&-\delta U^{\s (2)}\pipsi-\delta U^{\s (1)}\pipsibar-\delta U^{\s(3)}\pigi-
\delta U^{\s (4)}\QH+ \\
&\left. -\delta U^{\s (5)}\QBH-\delta U^{\s(6)}\HT-U^{\s
(4)}{\ov\e}2i\HT-U^{\s(5)}\e 2i\HT\right],
\end{array}
\end{equation}

\noindent
where we have indicated  with $U^{\s(1)}$, for example, the 
Lagrange multiplier associated to the first of the constraints 
in (\ref{LS24}), with $U^{\s(2)}$ the one associated 
to the second and so on.

It is now easy to choose the variation of the Lagrange multipliers
in such a way to make $\delta S_{ext}=0$:

\begin{equation}
\label{eq:D-quattro}
\begin{array}{ll}
\left\{
\begin{array}{l}
\delta U^{\s(1)}=-{\dot\alpha}\\
\delta U^{\s(2)}=-{\dot{\ov\alpha}}\\
\delta U^{\s(3)}=-i{\dot{\beta}};
\end{array}
\right.
\;\;&\;\;
\left\{
\begin{array}{l}
\delta U^{\s(4)}=\alpha-i{\dot\e}\\
\delta U^{\s(5)}={\ov\alpha}-i{\dot{\ov\e}}\\
\delta U^{\s(6)}=-i{\dot\eta}-i\beta-2i({\ov\e}{\ov\psi}+\e\psi)+2i({\ov\e}
U^{\s(4)}+\e U^{\s(5)}).
\end{array}
\right.
\end{array}
\end{equation}

\noindent
We can now  proceed as in Ref.\scite{Teitel} by restricting the Lagrange
multipliers to be only those of the primary constraints (\ref{LS15})

\begin{equation}
\label{eq:D-cinque}
U^{\s (4)}=U^{\s (5)}=U^{\s (6)}=0
\end{equation}

\noindent
which imply that also the gauge variations of these variables must be zero. From these two
conditions we get from Eq.({\ref{eq:D-quattro}) the following
relations among the six gauge parameters

\begin{equation}
\label{eq:D-sei}
\left\{
\begin{array}{l}
\alpha = i {\dot\e} \\
{\ov\alpha} = i{\dot{\ov\e}} \\
\beta = -{\dot\eta}-2({\ov\e}{\ov\psi}+\e\psi).
\end{array}
\right.
\end{equation}
\noindent
As a consequence  the general gauge variation of an observable $O$
given in Eq.(\ref{eq:D-due}) becomes

\begin{equation}
\label {eq:D-sette}
\delta O =[{\ov\e}\QBH+\e\QH+\eta\HT+i{\dot{\ov\e}}\pipsi+
i{\dot\e}\pipsibar+(-{\dot\eta}-2({\ov\e}{\ov\psi}+\e\psi))\pigi,O]
\end{equation}
\noindent
and applying it on the three variables $\gaug$
we get:

\begin{equation}
\label{eq:D-otto}
\left\{
\begin{array}{l}
\delta\psi=[i{\dot{\ov\e}}\pipsi,\psi]=i{\dot{\ov\e}}\\
\delta{\ov\psi}=[i{\dot\e}\pipsibar,{\ov\psi}]=i{\dot\e}\\
\delta g =[-({\dot\eta}+2({\ov\e}{\ov\psi}+\e\psi)\pigi,g]=i{\dot\eta}+2i({\ov\e}
{\ov\psi}+\e\psi).
\end{array}
\right.
\end{equation}

\noindent
This is exactly the transformation (\ref{LS14}) obtained here
from the generators $G_{i}$ of Eq.(\ref{LS24}).  This concludes
the explanation of how the variations (\ref{LS14}), derived
from a pure Lagrangian variation, could be obtained via the {\it canonical}
gauge generators $G_{i}$.
\section{How to gauge away $\gaug$}
\label{app:gauge}

\noindent
In this Appendix, for purely pedagogical reasons, we will show
how to gauge away the $\gaug$.

The infinitesimal transformations are given in Eq.(\ref{LS14})
and the first thing to do is to build  finite transformations out of the
infinitesimal ones. If we
start from a configuration $(\psi_{\scriptscriptstyle 0}(t),
{\ov\psi}_{\scriptscriptstyle 0}(t),g_{\scriptscriptstyle 0}(t))$, after
one step we arrive at $(\psi_{\scriptscriptstyle 1}(t),
{\ov\psi}_{\scriptscriptstyle 1}(t),g_{\scriptscriptstyle 1}(t))$
which are given by:
 
\begin{equation}
\label{eq:E-uno}
	\left\{
	\begin{array}{l}
	\psi_{1}(t)=\psi_{0}(t)+i\dot{\ov{\e}}(t) \\
	\bpsi_{1}(t)=\bpsi_{0}(t)+i\dot{\e}(t)  \\
	g_{1}(t)=g_{0}(t)+i\dot{\eta}(t)+2i(\e(t)\psi_{0}(t)+{\ov\e}(t)\bpsi_{0}(t)).
	\end{array}
	\right.
\end{equation}

\noindent
It is not difficult to work out what we get
after $N$ steps:
\begin{equation}
\label{eq:E-due}
	\left\{
	\begin{array}{l}
	\psi_{\s N}(t)=\psi_{0}(t)+iN\dot{\ov{\e}}(t) \\
	\bpsi_{\s N}(t)=\bpsi_{0}(t)+iN\dot{\e}(t)  \\
	g_{\s N}(t)=g_{0}(t)+iN\dot{\eta}(t)+2i(N\e(t)\psi_{0}(t)+N{\ov\e}(t)\bpsi_{0}(t))-
	N(N+1)(\e\dot{\ov\e}+{\ov\e}\dot{\e}).
	\end{array}
	\right.
\end{equation}

\noindent
Taking now the limit $N\rightarrow\infty$, but with the conditions:

\begin{equation}
\label{eq:E-tre}
	\left\{
	\begin{array}{l}
	N\e(t)\longrightarrow\Delta(t) \\
	N{\ov\e}(t)\longrightarrow\ov{\Delta}(t) \\
	N\eta(t)\longrightarrow\Delta_{g}(t) ,
	\end{array}
	\right.
\end{equation}

\noindent
where the various $\Delta(t)$ are finite quantities, we get
that a finite transformation has the form:
\begin{equation}
\label{eq:E-quattro}
	\left\{
	\begin{array}{l}
	\psi(t)=\psi_{0}(t)+i\dot{\ov{\Delta}}(t) \\
	\bpsi(t)=\bpsi_{0}(t)+i\dot{\Delta}(t)  \\
	g(t)=g_{0}(t)+i\dot{\Delta}_{g}(t)+2i(\Delta(t)\psi_{0}(t)+\ov{\Delta}(t)\bpsi_{0}(t))-
	(\Delta\dot{\ov\Delta}+{\ov\Delta}\dot{\Delta}).
	\end{array}
	\right.
\end{equation}

\noindent From the equation above it is easy to see that with the following
choice of $\Delta$'s

\begin{equation}
\label{eq:E-cinque}
	\left\{
	\begin{array}{l}
	\ov{\Delta}(t)=i\displaystyle\int_{0}^{t}\psi_{0}(\tau)d\tau \\
	\Delta(t)=i\displaystyle\int_{0}^{t}\ov{\psi}_{0}(\tau)d\tau   \\
	\Delta_{g}(t)=\displaystyle\int_{0}^{t}d\tau[ig_{0}(\tau)-2(\Delta\psi_{0}+{\ov\Delta}
	\bpsi_{0})-
	i(\Delta\dot{\ov\Delta}+{\ov\Delta}\dot{\Delta})]
	\end{array}
	\right.
\end{equation}

\noindent
we can bring the $\gaug$ to zero.
We should anyhow be careful and see whether there are no obstructions
to this construction. Actually, after Eq.(\ref{LS5})
we said that, in order that the surface terms disappear,
we needed to require that $\e(t)$ and ${\ov\e}(t)$ be zero at the
end-points $(0,T)$ of integration. From
Eq.(\ref{eq:E-tre}) one sees that this implies:

\begin{equation}
\label{eq:E-sei}
	\begin{array}{l}
	\Delta(0)=\ov{\Delta}(0)=0 \\
	\Delta(T)=\ov{\Delta}(T)=0.
	\end{array}
\end{equation}

\noindent 
While the first condition is easily satisfied, as can be seen from
Eq.(\ref{eq:E-cinque}), the second one would imply:

\begin{equation}
\label{eq:E-sette}
	\begin{array}{l}
	\Delta(T)=i\displaystyle\int_{0}^{T}\ov{\psi}_{0}(\tau)d\tau = 0  \\
	\ov{\Delta}(T)=i\displaystyle\int_{0}^{T}\psi_{0}(\tau)d\tau = 0.
	\end{array}
\end{equation}

\noindent
This is a condition which is not satisfied by any initial configuration
$\psi_{\scriptscriptstyle 0}$, ${\ov\psi}_{\scriptscriptstyle 0}$ but only
by special ones. So we can say that, if we want transformations
which do not leave surface terms, then it may be impossible to gauge
away $(\psi,{\ov\psi})$. Not to have surface terms may turn out to
be an important issue in some contexts. Anyhow this problem  does not
arise in the time-reparametrization transformation because, as we see
from Eq.(\ref{LS11}), that transformation does not generate surface
terms.
\section{$\Qb$, $\QBb$ and Physical States}
\label{app:Qb}

\noindent
In this Appendix we will show how the constraints ({\ref{LS46}})
affect the Hilbert space of the system. We know that the {\it physical} states
should be annihilated by the constraints:

\begin{equation}
\label{eq:F-uno}
\left\{
\begin{array}{l}
\,[\Pi_{\s\alpha}+{\Qb}]\mid\mbox{phys}\rangle=0\\
\,[\Pi_{\s{\ov\alpha}}+{\QBb}]\mid\mbox{phys}\rangle=0
\end{array}
\right.
\end{equation}
\noindent
and so this seems to restrict the original Hilbert space of the system.
On the other hand we have proved that the system obeying these constraints
and with Lagrangian (\ref{LS44}) has the same number of degrees of
freedom as the original system with Lagrangian (\ref{CPI6}) and moreover
they seem equivalent. If that is so then the Hilbert
space of the {\it physical} states should be equivalent or isomorphic to the
original Hilbert space. This is what we are going to prove in what follows.

Let us first solve the constraint (\ref{eq:F-uno}). The wave-functions
$\Psi({\s\ldots})$ of the system will depend not only on the $(\varphi^{a},c^{a})$
but also on the gauge-parameters $\alpha(t)$ and ${\ov\alpha}(t)$. So equation
(\ref{eq:F-uno}) takes the form:

\begin{equation}
\label{eq:F-due}
\left\{
\begin{array}{l}
\displaystyle\frac{\partial\Psi(\varphi,c;\alpha,{\ov\alpha})}{\partial\alpha}=-\Qb
\Psi(\varphi,c;\alpha,{\ov\alpha}) \vspace{.2cm}\\
\displaystyle\frac{\partial\Psi(\varphi,c;\alpha,{\ov\alpha})}{\partial{\ov\alpha}}=-\QBb
\Psi(\varphi,c;\alpha,{\ov\alpha})
\end{array}
\right.
\end{equation}
\noindent
whose solution is

\begin{equation}
\label{eq:F-tre}
\Psi(\varphi,c;\alpha,{\ov\alpha})=\exp[-\alpha\Qb-{\ov\alpha}\QBb]~\psi(\varphi,c)
\end{equation}
\noindent
where $\psi(\varphi,c)$ is a state of the Hilbert space of the
old system  with Lagrangian (\ref{CPI6}) and the $\Qb$ and $\QBb$ 
should be interpreted as the differential operator associated
to the relative charges via the substitutions (\ref{CPI40}) and (\ref{CPI44}); the same
for all the Hamiltonians which we will use from now on. To prove that the
two systems are equivalent we should prove that there is an isomorphism
in Hilbert space between the solutions of the two Koopman-von Neumann\footnote{
By Koopman von Neumann equation we mean the analog of the Liouville
equations built via  the full $\HT$ and not via  just its bosonic part.}  
equations, the first one associated to the old Hamiltonian (\ref{CPI8}) 
and the second to the
Hamiltonian of the Lagrangian (\ref{LS44}). This last one is
the {\it primary}\scite{Vinc} Hamiltonian:

\begin{equation}
\label{eq:F-quattro}
\HT_{\s P}\equiv \HT_{can}+\mu(\Pi_{\s\alpha}+\Qb)+{\ov\mu}
(\Pi_{\s{\ov\alpha}}+\QBb)
\end{equation}
where $\mu$ and ${\ov\mu}$ are Lagrange multipliers and $\HT_{can}$
is the {\it canonical}\scite{Vinc} Hamiltonian associated to the Lagrangian
(\ref{LS44}).

The Koopman-von Neumann equation for this system is:

\begin{equation}
\label{eq:F-cinque}
\HT {\s P}\Psi(\varphi,c,t;\alpha,{\ov\alpha})=i
\frac{\partial\Psi(\varphi,c,t;\alpha,{\ov\alpha})}{\partial t} 
\end{equation}

\noindent which can be rewritten as:
\begin{equation}
[\HT +\mu(\Pi_{\s\alpha}+\Qb)+{\ov\mu}
(\Pi_{\s{\ov\alpha}}+\QBb)] \Psi(\varphi,c,t;\alpha,{\ov\alpha})= i
\frac{\partial\Psi(\varphi,c,t;\alpha,{\ov\alpha})}{\partial t}.  
\end{equation}
\noindent
Since $\Psi(\varphi,c,t;\alpha,{\ov\alpha})$ is annihilated by the constraints
(\ref{eq:F-uno}) we get:

\begin{equation}
\label{eq:A}
\HT\Psi(\varphi,c,t;\alpha,{\ov\alpha})=i
\frac{\partial\Psi(\varphi,c,t;\alpha,{\ov\alpha})}{\partial t}; 
\end{equation}

\noindent now we use (\ref{eq:F-tre}) in (\ref{eq:A}) and this yields:

\begin{equation}
\label{chicco}
\HT \exp [-\alpha\Qb -{\ov\alpha}\QBb]\psi(\v,c,t)= i
\frac{\partial}{\partial
t}\left[\exp(-\alpha\Qb-{\ov\alpha}\QBb)\psi(\varphi,c,t)\right].
\end{equation}

\noindent The last step is to work out the derivatives in Eq.(\ref{chicco}); we
obtain:

\begin{equation}
\exp[-\alpha\Qb-{\ov\alpha}\QBb]~\HT\psi(\varphi,c,t) =  
\exp[-\alpha\Qb-{\ov\alpha}\QBb]~i\frac{\partial\psi(\v,c,t)}{\partial t}
\end{equation}

\noindent which holds if and only if:

\begin{equation}
\HT\psi=i\frac{\partial\psi}{\partial t}
\end{equation}

\noindent and this concludes the proof that the two systems have not only the
same number of degrees of freedom but also the same Hilbert space.
\section{Details about Eq.(\ref{beta})}
\label{app:eq}

\noindent In this Appendix we provide details regarding the derivation of Eq.(\ref{beta}).
Consider first the dependence on $g$. From 

\begin{equation}
\mid\mbox{phys}\rangle = Q_{\scriptscriptstyle (1)} |\chi\rangle
\;\;\;\;\mbox{and}\;\;\;\;
\Pi_{\s g}\mid\mbox{phys}\rangle = 0
\end{equation}

\noindent we infer that 

\begin{equation}
\Pi_{\s g}Q_{\s (1)} |\chi\rangle =  Q_{\s (1)}\Pi_{\s g}|\chi\rangle = 0,
\end{equation}

\noindent which means that
\begin{equation}
\label{ker}
\Pi_{\s g}|\chi\rangle \in \ker Q_{\scriptscriptstyle (1)}. 
\end{equation}

\noindent If we represent $\Pi_{\s g}$ as $\Pi_{\s g}=-i\frac{\partial}{\partial
g}$,  Eq.(\ref{ker}) implies:

\begin{equation}
\displaystyle\frac{\partial}{\partial g}|\chi\rangle =
\sum_{m}f_{m}(g,\alpha)|\zeta_{m}\rangle
\end{equation}

\noindent where $|\zeta_{m}\rangle$ form a basis of $\ker Q_{\s (1)}$. Solving
this last differential equation we get 

\begin{equation}
\label{chi}
|\chi\rangle = |\chi_{0};\alpha\rangle + \sum_{m}\left[\int
dg~f_{m}(g,\alpha)\right]|\zeta_{m}\rangle
\end{equation}

\noindent where $= |\chi_{0};\alpha\rangle$ does not depend on $g$ anymore. By
the same line of reasoning we 
can prove that $\Pi_{\s \alpha}|\chi\rangle \in \ker Q_{\scriptscriptstyle
(1)}$, which in turn implies that 
$\Pi_{\s \alpha}|\chi_0;\alpha\rangle \in \ker Q_{\scriptscriptstyle (1)}$. We
can repeat the 
previous steps and we arrive at the relation:

\begin{equation}
|\chi_{0};\alpha\rangle = |\chi_{0}\rangle + \sum_{m}\left[\int d\alpha
~l_{m}(g,\alpha)\right]|\zeta_{m}
\rangle
\end{equation}
    
\noindent which, substituted in Eq.~(\ref{chi}), yields:

\begin{equation}
|\chi\rangle = |\chi_{0}\rangle + \sum_{m}\left[\int d\alpha
~l_{m}(g,\alpha)\right]|\zeta_{m}\rangle + 
\sum_{m}\left[\int dg ~f_{m}(g,\alpha)\right]|\zeta_{m}\rangle \equiv
|\chi_{0}\rangle + 
|\zeta;\alpha,g\rangle,  
\end{equation}

\noindent as we claimed in Eq.~(\ref{beta}).
\chapter{CPI and $\kappa$-symmetry: Mathematical Details}
\markboth{B. CPI and $\kappa$-symmetry: Mathematical Details}{}

\label{app:kappa}

\noindent
In this Appendix we show the standard procedure to build the Dirac Brackets for the 
$2^{\text{nd}}$-class constraints listed in Eq.(\ref{1-7}). They are explicitly:

	\begin{equation}
	\begin{cases}
	\Pi_{\s p}^{\mu} = 0 & (a)\\
	(\Pi_{\s x})_{\mu}-p_{\mu} = 0 & (b) \vspace{.1cm}\\ 
	D^{\s a}\equiv(\Pi_{\s\zb})^{\s a}+\displaystyle\frac{i}{2}(\pslsh\,\z)^{\s a} = 0
	&(c) \vspace{.15cm}\\
	\ov{D}_{\s a}\equiv(\Pi_{\s\z})_{\s a} +\displaystyle\frac{i}{2}(\zb\pslsh)_{\s a} = 0
	& (d), 
	\end{cases}\label{K-1}
	\end{equation} 
\noindent
and we can arrange them in the following row:
	\begin{equation}
	\phi_i=\big(\Pi_{\s p}^{\mu},(\Pi_{\s x})_{\mu}-p_{\mu},D^{\s a},\ov{D}_{\s a}\big). 
	\end{equation} 
\noindent
Then, in order to construct the Dirac Brackets associated to these constraints,
we need the matrix $\Delta_{ij}$ defined as follows:
 	\begin{equation}
	\Delta_{ij}=\big[\phi_i,\phi_j\big]_{\s PB}.
	\end{equation}
\noindent 
The Poisson Brackets among all the constraints are found after a long but easy calculation:
	\begin{align}
	\label{K-4}
	& [(\Pi_{\s p})^{\mu},(\Pi_{\s x})_{\nu}-p_{\nu}]_{\s PB}=\delta^{\mu}_{\nu} \\	
	& [(\Pi_{\s p})^{\mu},D^{\s a}]_{\s PB}=-\displaystyle\frac{i}{2}(\gamma^{\mu})^{\s
	a}_{\s b}\z^{\s b} \\
	& [(\Pi_{\s p})^{\mu},\ov{D}_{\s b}]_{\s PB}=-\displaystyle\frac{i}{2}\zb_{\s
	d}(\gamma^{\mu})^{\s d}_{\s b} \\	
	& [(\Pi_{\s x})_{\nu}-p_{\nu},(\Pi_{\s p})^{\mu}]_{\s PB}=-\delta^{\mu}_{\nu} \\
	& [(\Pi_{\s x})_{\nu}-p_{\nu},D^{\s a}]_{\s PB}=0 \\
	& [(\Pi_{\s x})_{\nu}-p_{\nu},\ov{D}_{\s b}]_{\s PB}=0 \\
	& [D^{\s a},(\Pi_{\s p})^{\mu}]_{\s PB}=\displaystyle\frac{i}{2}(\gamma^{\mu})^{\s
	a}_{\s b}\z^{\s b} \\
	& [D^{\s a},(\Pi_{\s x})_{\nu}-p_{\nu}]_{\s PB}=0 \\
	& [D^{\s a},\ov{D}_{\s b}]_{\s PB}=i(\pslsh)^{\s a}_{\;\s b} \\
	& [\ov{D}_{\s b},(\Pi_{\s p})^{\mu}]_{\s PB}=\displaystyle\frac{i}{2}\zb_{\s
	d}(\gamma^{\mu})^{\s d}_{\s b} \\
	& [\ov{D}_{\s b},(\Pi_{\s x})_{\nu}-p_{\nu}]_{\s PB}=0 \\
	& [\ov{D}_{\s b},D^{\s a}]_{\s PB}=i(\pslsh)^{\s a}_{\;\s b}=i(\pslsh^{\; \s T})_{\s
	b}^{~\s a} 
	\end{align}

\noindent 
and therefore the matrix $\Delta_{ij}$ takes the form: 
 	\begin{equation}
	\Delta_{ij}=\big[\phi_i,\phi_j\big]_{\s PB}=
	\begin{pmatrix} 
	\label{K-5}
	0 & 1 & -\frac{i}{2}(\gamma^{\mu}\z)^{\s T} & -\frac{i}{2}\zb\gamma^{\mu}
	\vspace{.1cm}\\
	-1 & 0 & 0 & 0 \vspace{.1cm}\\
	\frac{i}{2}\gamma^{\mu}\z & 0 & 0 & i\pslsh \vspace{.1cm}\\
  	\frac{i}{2}(\zb\gamma^{\mu})^{\s T} & 0 & i\pslsh^{\;\s T} & 0
	\end{pmatrix} .
 	\end{equation}

\noindent
This is an even supermatrix and there are many ways to find the inverse. After a 
little algebra one finds that the inverse has the following form:
 
 	\begin{equation}
	(\Delta^{-1})^{ij}=
	\begin{pmatrix} 
	\label{K-6}
	0 & -1 & 0 & 0 \vspace{.1cm}\\
	1 & 0 & \frac{1}{2}\zb\gamma^{\mu}\pslsh^{\:-1} &
	\frac{1}{2}(\pslsh^{\:-1}\gamma^{\mu}\z)^{\s T} \vspace{.1cm}\\
	0 & \frac{1}{2}(\zb\gamma^{\mu}\pslsh^{\:-1})^{\s T} & 0 & -i(\pslsh^{\;\s T})^{\:-1}
	\vspace{.1cm}\\
  	0 & \frac{1}{2}\pslsh^{\:-1}\gamma^{\mu}\z  & -i\pslsh^{\:-1} & 0
	\end{pmatrix}. 
 	\end{equation}

\noindent
As a consequence, the Dirac Brackets derived from the previous constraints are:
	\begin{equation}
	\label{K-7}
	\begin{split}
	[A,B]_{\s DB}&=[A,B]_{\s PB}+ \\ 
	& -[A,(\Pi_{\s p})^{\mu}]_{\s PB}(-\delta^{\mu}_{\nu})\; [(\Pi_{\s
	x})_{\nu}-p_{\nu},B]_{\s PB}+ \\	
	& -[A,(\Pi_{\s x})_{\nu}-p_{\nu}]_{\s PB}\;\delta^{\nu}_{\mu}\;[(\Pi_{\s
	p})^{\mu},B]_{\s PB}+  \\
	& -[A,(\Pi_{\s x})_{\nu}-p_{\nu}]_{\s
	PB}\;\displaystyle\frac{1}{2}(\zb\gamma^{\nu}\pslsh^{\:-1})_{\s a}\;[D^{\s a},B]_{\s
	PB}+ \\	
	& -[A,(\Pi_{\s x})_{\nu}-p_{\nu}]_{\s
	PB}\;\displaystyle\frac{1}{2}(\pslsh^{\:-1}\gamma^{\mu}\z)^{\s T}_{\s b}\;[\ov{D}_{\s
	b},B]_{\s PB}+ \\	
	& -[A,D^{\s a}]_{\s
	PB}\;\displaystyle\frac{1}{2}\big[(\zb\gamma^{\mu}\pslsh^{\:-1})^{\s T}\big]^{\s a}\;
	[(\Pi_{\s x})_{\nu}-p_{\nu},B]_{\s PB}+\\
	& -[A,D^{\s a}]_{\s PB}(-i)\big[(\pslsh^{~\s T})^{-1}\big]^{\s a}_{\s b}\;[\ov{D}_{\s
	b},B]+\\
	& -[A,\ov{D}_{\s b}]_{\s PB}\;\displaystyle\frac{1}{2}(\pslsh^{\:-1}\gamma^{\mu}\z)^{\s
	b}\;[(\Pi_{\s x})_{\nu}-p_{\nu},B]_{\s PB}+\\
	& -[A,\ov{D}_{\s b}]_{\s PB}(-i)(\pslsh^{\:-1})^{\s b}_{\;\s a}\;[D^{\s a},B];
	\end{split}
	\end{equation}
\noindent
and it is not a difficult exercise (even if rather long) to check that the previous expression
leads to
the same brackets as those in Eqs.(\ref{1-13})-(\ref{1-17}).
\chapter{The Conformal Group}
\markboth{C. The Conformal Group}{}

\noindent 
In this Appendix we give a brief review of the conformal group and its representations 
in various dimensions. We think this can be useful because of the big interest that
this group has recently gained in many fields of theoretical physics, from astrophysics to
statistical physics.
\section{Conformal Transformations}
\label{app:Conf}
\noindent
Consider a Riemannian (or Lorentzian\footnote{A metric $g$ is called ``Riemannian" if its signature is $n$
($n$ being the dimension of the manifold, while it is called ``Lorentzian" if the signature is $n-2$.})
manifold $\cal M$ with metric $g$; it is well known that the change of the metric under a transformation of
coordinates is given by:
	\begin{equation}
	\label{A-1}
	\begin{cases}
	x \longrightarrow y(x) \\
	g_{\mu\nu}(x) \longrightarrow g^{\prime}_{\mu\nu}(y) = \displaystyle\frac{\p x^{\alpha}}
	{\p y^{\mu}}\frac{\p x^{\b}}{\p y^{\nu}}g_{\alpha\beta}(x).
	\end{cases}
	\end{equation}
\vspace{.3cm}
\noindent
A transformation is called {\it conformal} if Eq.(\ref{A-1}) takes the form:
	\begin{equation}
	\label{A-2}
	g^{\prime}_{\mu\nu}(y) = \Omega(x) g_{\mu\nu}(x).
	\end{equation}

\noindent Now we want to analyze the form which the factor $\Omega(x)$ can have in order
to find out the generators of the group. First of all we consider the infinitesimal version of (\ref{A-1}):
 	\begin{equation}
	\label{A-3}
	\begin{cases}
	y^{\mu}= x^{\mu}+\e^{\mu}(x) \\
	g^{\prime}_{\mu\nu}(y) = g_{\mu\nu}(x)- g_{\mu\beta}(x)\p_{\nu}\e^{\b}-
	g_{\a\nu}(x)\p_{\mu}\e^{\a};
	\end{cases}
	\end{equation}
\vspace{.3cm}
\noindent
and if we write $\Omega(x)\equiv 1-\omega(x)$ (according to the infinitesimal character of the
transformation), from (\ref{A-2}) and (\ref{A-3}) we obtain:
	\begin{equation}
	\label{A-4}
	\omega(x)g_{\mu\nu}(x) = g_{\mu\beta}(x)\p_{\nu}\e^{\b}+
	g_{\a\nu}(x)\p_{\mu}\e^{\a}.
	\end{equation}
\noindent
Consider for simplicity the case $g_{\mu\nu}=\eta_{\mu\nu}$: the previous equation becomes:
	\begin{equation}
	\label{A-5}
	\omega(x)\eta_{\mu\nu} = \p_{\nu}\e_{\mu}(x)+ \p_{\mu}\e_{\nu}(x),
	\end{equation}
\noindent
from which it is immediately found that:
	\begin{equation}
	\label{A-6}
	\omega(x) = \frac{2}{d}\,\p_{\mu}\e^{\mu}
	\end{equation}
\noindent
where $d$ is the dimension of the space-time. If we apply $\p_{\rho}$ to Eq.(\ref{A-5}) we get:
	\begin{equation}
	\label{A-7}
	\p_{\rho}\omega(x)\eta_{\mu\nu} = \p^2_{\rho\nu}\e_{\mu}(x)+ \p^2_{\rho\mu}\e_{\nu}(x),
	\end{equation}
\noindent
from which, after some rearrangements and permutations of the indices, we arrive at:
	\begin{equation}
	\label{A-8}
	2\p_{\mu}\p_{\nu}\e_{\rho}(x)= \eta_{\mu\rho}\p_{\nu}\omega(x)+
	\eta_{\nu\rho}\p_{\mu}\omega(x)-\eta_{\mu\nu}\p_{\rho}\omega(x).
	\end{equation}
\noindent
Now we can multiply the previous equation by $\eta^{\mu\nu}$ and then take a differentiation
$\p_{\sigma}$; what we get is:
	\begin{equation}
	\label{A-9}
	2\Box\p_{\sigma}\e_{\rho}(x)= (2-d)\p_{\rho}\p_{\sigma}\omega(x),
	\end{equation}	
\noindent
which, since the RHS is symmetric under exchange $\rho\longleftrightarrow\sigma$, can be rewritten
as:
	\begin{equation}
	\label{A-10}
	\Box(\p_{\sigma}\e_{\rho}+\p_{\rho}\e_{\sigma})(x)= (2-d)\p_{\rho}\p_{\sigma}\omega(x).
	\end{equation} 
\noindent
Now we want to compare Eq.(\ref{A-9}) with the Laplacian $\Box$ of Eq.(\ref{A-5}) which is:
	\begin{equation}
	\label{A-11}
	\Box(\p_{\sigma}\e_{\rho}+\p_{\rho}\e_{\sigma})(x)= \eta_{\sigma\rho}\Box\omega(x);
	\end{equation}
\noindent
from the comparison between (\ref{A-11}) with (\ref{A-10}) we finally get the equations which
characterize the parameter $\omega$ of the an infinitesimal conformal transformation:
	
	\begin{equation}
	\label{A-12}
	\boxed{
	\begin{array}{lr}
	(2-d)\p_{\rho}\p_{\sigma}\omega(x) = \eta_{\sigma\rho}\Box\omega(x)  &~~~(a) \vspace{.2cm} \\
	(1-d)\,\Box\omega(x) = 0  &~~~(b) \vspace{.2cm} \\
	\text{where }~ \omega(x)=\displaystyle\frac{2}{d}\p_{\mu}\e^{\mu}(x). &	
	\end{array}}
	\end{equation}

\vspace{.25cm}
\noindent
We can now distinguish three cases: $d=1$, $d>2$ and $d=2$. 

\subsection{Case $d=1$}

\noindent 
This is obviously the simplest case. In fact from Eq.(\ref{A-12}) it is clear that no constraint is
imposed on $\omega(x)$ and therefore on $\e(x)$. As we could guess, we deduce that in $d=1$ every
coordinate transformation is conformal.
\subsection{Case $d>2$}

\noindent 
In this case we have that from (\ref{A-12}-a) and (\ref{A-12}-b) we obtain:
	\begin{equation}
	\label{A-13}
	\p_{\mu}\p_{\nu}\omega(x)=0,
	\end{equation}
\noindent
which implies that $\omega(x)$ can be at most a linear function of $x$:
	\begin{equation}
	\label{A-14}
	\omega(x)=A+B_{\mu}x^{\mu},
	\end{equation}
\noindent
with $A$ and $B_{\mu}$ are independent of $x$. Next, as a consequence of Eq.(\ref{A-6}) we have that
$\e_{\mu}(x)$ can be at most a quadratic function of $x$:
	\begin{equation}
	\label{A-15}
	\e_{\mu}(x)=a_{\mu}+b_{\mu\nu}x^{\nu}+c_{\mu\nu\rho}x^{\nu}x^{\mu}.
	\end{equation}
\noindent
After some long and boring steps \cite{DiFrancesco} making use of some of the previous equations, we can
arrive at the following parametrization of an infinitesimal conformal transformation in $d>2$ dimensions:

	\begin{equation}
	\label{A-16}
	\boxed{
	\begin{array}{lr}
	\e_{\mu}(x)=a_{\mu}+(m_{\mu\nu}+f\eta_{\mu\nu})x^{\nu}+2x_{\mu}(h\cdot x)-h_{\mu}x^2 &(a)\\
	\omega(x)=f+4h\cdot x. &(b)	\\
	 \text{remember that } ~\Omega(x)=1-\omega(x) &
	\end{array}}
	\end{equation}	

\vspace{.3cm}	
\noindent
where all the parameters $a_{\mu}$, $m_{\mu\nu}=m_{\nu\mu}$, $f$ and $h_{\mu}$ are independent of
$x$ and are functions of the parameters entering Eqs.(\ref{A-14})(\ref{A-15}). If we exponentiate
the infinitesimal transformation to get the finite one we obtain that a generic {\it finite}
conformal transformation is a composition of the following transformations:

	\begin{align}
	& &&\text{\bf Conformal Group (d$>$2)} &&\nonumber \\
	&\text{Translations} && (x^{\prime})^{\mu}=x^{\mu}+a^{\mu} && (\Omega=1) \label{A-17} \\
	&\text{Dilations} && (x^{\prime})^{\mu}=\l x^{\mu} && (\Omega=\l^{-2}) \label{A-18} \\
	&\text{Lorentz} && (x^{\prime})^{\mu}=M^{\mu}_{\nu} x^{\nu} && (\Omega=1) \label{A-19} \\
	&\text{Special Conf.} && (x^{\prime})^{\mu}=\displaystyle\frac{x^{\mu}-
	h^{\mu}x^2}{1-2k\cdot x+k^2x^2} && (\Omega=1-2k\cdot x+k^2x^2). \label{A-20} 
	\end{align}
	
\noindent
We can proceed and find out the generators of the previous transformations:

	\begin{align}
	&\text{Translations} && P_{\mu}=-i\p_{\mu} \label{A-21} \\
	&\text{Dilations} && D=-ix^{\mu}\p_{\mu} \label{A-22} \\
	&\text{Lorentz} && L_{\mu\nu}=i(x_{\mu}\p_{\nu}-x_{\nu}\p_{\mu}) \label{A-23} \\
	&\text{Special Conf.} && K_{\mu}=-i(2x_{\mu}x^{\nu}\p_{\nu}-x^2\p_{\mu}) \label{A-24} 
	\end{align}

\noindent
which satisfy the following algebra:
	
\begin{center}{\bf\hspace*{-1.5cm} Conformal Algebra}\vspace{-.4cm}\end{center}	
	\begin{equation}\label{A-25}
	\boxed{\begin{array}{ll}
	\,[D,P_{\mu}]=iP_{\mu}  &(a) \\
	\,[D,K_{\mu}]=-iK_{\mu}   &(b)\\
	\,[K_{\mu},P_{\mu}]=2i(\eta_{\mu\nu}D-L_{\mu\nu})   &(c)\\
	\,[K_{\rho},L_{\mu\nu}]=i(\eta_{\rho\mu}K_{\nu}-\eta_{\rho\nu}K_{\mu})   &(d)\\	
	\,[P_{\rho},L_{\mu\nu}]=i(\eta_{\rho\mu}P_{\nu}-\eta_{\rho\nu}P_{\mu})   &(e)\\
	\,[L_{\mu\nu},L_{\r\si}]=i(\eta_{\nu\rho}L_{\mu\si}+\eta_{\mu\si}L_{\nu\rho}-
	\eta_{\mu\r}L_{\nu\si}-\eta_{\nu\si}L_{\mu\rho})   &(f)
	\end{array}}
	\end{equation}

\vspace{.25cm}
\noindent
It is easy to construct an isomorphism between the Conformal Group in $d$ dimensions and the groups
$SO(d+1,1)$ or $SO(d,2)$, depending on the signature of the metric $\eta_{\mu\nu}$: if $\eta_{\mu\nu}$ is
Euclidean we have $\text{\it Conf}(d)\cong SO(d+1,1)$, while $\text{\it Conf}(d)\cong SO(d,2)$ if
$\eta_{\mu\nu}$ is Lorentzian. 
In fact, if we define the following correspondences:
	\begin{align}
	\label{A-26}
	& J_{\mu\,\nu}=L_{\mu\nu} \\
	& J_{-1\, 0}=D \\
	& J_{-1\,\nu}=\displaystyle\frac{1}{2}(P_{\mu}-K_{\mu}) \\
	& J_{0\,\nu}=\displaystyle\frac{1}{2}(P_{\mu}+K_{\mu})
	\end{align}
\noindent
where $J_{ab}=-J_{ba}$ and $a,b \in (-1,0,1,2,\ldots,d)$, we easily see that the $J_{ab}$ operators satisfy
the $SO(d,2)$ algebra:
	\begin{equation}
	\label{A-27}
	[J_{ab},J_{cd}]=i(\eta_{ad}J_{bc}+\eta_{bc}J_{ad}-\eta_{ac}J_{bd}-\eta_{bd}J_{ac}).
	\end{equation}	
\subsection{Case $d=2$}

\noindent
We still have to discuss the conformal transformations in two dimensions. First of all, if we consider again
Eqs.(\ref{A-5})(\ref{A-6}) and we set $d=2$ we obtain the following equations:
	\begin{equation}
	\label{A-28}
	\p_{\a}\e^{\a}\eta_{\mu\nu} = \p_{\nu}\e_{\mu}(x)+ \p_{\mu}\e_{\nu}(x),
	\end{equation}	
\noindent
from which:
	\begin{equation}
	\label{A-29}
	\boxed{
	\begin{array}{rcl}
	\mu=\nu  &\longrightarrow & ~~~~ \p_1\e_1=\p_2\e_2 \\
	\mu\neq\nu &\longrightarrow & ~~~~ \p_1\e_2=-\p_2\e_1.
	\end{array}}
	\end{equation}	
\noindent
Eqs.(\ref{A-29}) are nothing but the Cauchy-Riemann equations for $\e_1$ and $\e_2$. This suggests to use a
complex formalism for describing the transformations at hand. Thus we define:
 	\begin{equation}
	\label{A-30}
	\begin{split}
	&z\equiv x_1+ix_2; \\
	&\ov{z}\equiv x_1-ix_2; \\
	&\e\equiv\e_1+i\e_2;
	\end{split}
	\end{equation}	 	
\noindent
consequently a generic conformal transformation takes the form:
	\begin{equation}
	\label{A-31}
	z^{\prime}=z+\e(z,\ov{z}).
	\end{equation}	 
\noindent
But the Cauchy-Riemann conditions (\ref{A-29}) impose the constraint\footnote{Note that Eq.(\ref{A-32}) is
not equivalent to the analiticity of $\e(z)$, because also the continuity of $\e(z)$ is required.}
 	\begin{equation}
	\label{A-32}
	\p_{\ov{z}}\,\e(z,\ov{z})=0,
	\end{equation}	 
\noindent
and then Eq.(\ref{A-31}) becomes:	
	\begin{equation}
	\label{A-33}
	z^{\prime}=z+\e(z).
	\end{equation}	 	
\noindent
Now suppose that $\e(z)$ is smooth enough to be expanded in Laurent series around $z=0$; in this case we can
write:
 	\begin{equation}
	\label{A-34}
	\e(z)=\sum_{n=-\infty}^{+\infty} c_n\, z^{n+1},
	\end{equation}	  
\noindent
and the effect of a conformal transformation on a (for example) scalar field is given by:
 	\begin{equation}
	\label{A-35}
	\begin{cases}
	\Phi^{\prime}(z^{\prime},\ov{z}^{\prime})=\Phi(z,\ov{z}) \\
	\delta\Phi(z,\ov{z}) = \displaystyle\sum_{n=-\infty}^{+\infty}(c_n l_n + \ov{c}_n \ov{l}_n)\Phi(z,\ov{z}), 
	\end{cases}
	\end{equation}	
\noindent
where we have introduced the following generators:	
 	\begin{equation}
	\label{A-36}
	\begin{cases}
	l_n\equiv -z^{n+1}\p_z \\
	\ov{l}_n\equiv -\ov{z}^{n+1}\p_{\ov{z}}. 
	\end{cases}
	\end{equation}	
\noindent
The algebra realized by the previous operators is easily found:

	\begin{equation}
	\label{A-37}
	\boxed{
	\begin{split}
	&[l_n,l_m]=(n-m) l_{n+m} \\
	&[\ov{l}_n,\ov{l}_m]=(n-m) \ov{l}_{n+m}\\
	&[l_n,\ov{l}_m]=0.
	\end{split}}
	\end{equation}

\vspace{.25cm}
\noindent
This algebra is nothing but a Virasoro algebra without central charge. Therefore the important message is
that while in $d=1$ and $d>2$ dimensions the Conformal Group has a finite number of generators, in $d=2$
there is an infinite number of generators. Furthermore there is another point to be discussed: so far we have
focused on {\it local} conformal transformations -- {\it local} in the sense that we have never worried
about the possibility of inverting the conformal map $z\longrightarrow z^{\prime}(z)$
(and therefore we have never cared about its domain of definition). If we restrict to the
conformal transformations which map the whole Riemann sphere (i.e. the complex plane plus $\infty$) onto
itself, we obtain the {\it Global} Conformal Group, which is easy to show to be formed by the following
transformations:
 	\begin{equation}
	\label{A-38}
	z^{\prime}=f(z)=\frac{az+b}{cz+d} ~~~~~\text{with: } ~ad-bc=1.
	\end{equation}	    
It is easy to check that these transformations form a group since every $f(z)$ can be associated to a
$2\times 2$ matrix $\big(\begin{smallmatrix} a&b \\c&d \end{smallmatrix}\big)$ in such a way that the
composition of two conformal transformations becomes the ordinary product between matrices:
$f\circ g =\big(\begin{smallmatrix} a&b \\c&d \end{smallmatrix}\big)\big(\begin{smallmatrix} e&f \\g&h
\end{smallmatrix}\big)$. This implies that the Global Conformal Group is isomorphic to $SL(2,C)$ which in
turn is isomorphic to $SO(3,1)$. In fact the Virasoro subalgebra which generates the Global Conformal Group
is given by:
  	\begin{equation}
	\label{A-39}
	G:\begin{cases}
	l_{-1};~~~\ov{l}_{-1}; \\
	l_{0};~~~~\ov{l}_{0}; \\
	l_{1};~~~~\ov{l}_{1};
	\end{cases}\cong SO(3,1). 
	\end{equation}
\noindent From 
this subalgebra we can extract another subalgebra taking --- for example --- the ``real" part of $G$:
  	\begin{equation}
	\label{A-40}
	G_{\text{real}}:\begin{cases}
	L_{-1}\equiv l_{-1}+\ov{l}_{-1}; \\
	L_{0}\equiv l_{0}+\ov{l}_{0}; \\
	L_{1}\equiv l_{1}+\ov{l}_{1},
	\end{cases} 
	\end{equation}
\noindent
and these operators generate the so-called Conformal Group in 0+1 dimensions, which we used in Section
\ref{sec:CME}. We can see that by using the following correspondence:
 	\begin{equation}
	\label{A-41}
	\begin{split}
	& L_{-1}\longleftrightarrow iH \\
	& L_{0}\longleftrightarrow iD \\
	& L_{1}\longleftrightarrow iK;
	\end{split}
	\end{equation} 
\noindent
if we compare the two algebras (that of $(L_{-1},L_{0},L_{1})$ (inherited by Eq.(\ref{A-37})) and that of
$iH,iD,iK$ (derived from Eqs.(\ref{CME11})-(\ref{CME13})), we see that it is precisely the same.

\chapter{General Relativity and CME}
\markboth{D. General Relativity and CME}{}

\section{The Reissner-Nordstr{\o}m metric}
\label{app:Reiss}

\noindent For our purposes we shall consider charged and static (i.e. non-rotating) black
holes and, as a starting point, we write down the metric of a spherically
symmetric space:
	\begin{equation}
	\label{B1}
	ds^2=-\text{e}^{2\alpha(r,t)}dt^2 + \text{e}^{2\beta(r,t)}dr^2 +
	r^2d\Omega^2,
	\end{equation}
\noindent
where:
	\begin{equation}
	\label{B2}
	d\Omega^2=\sin ^2\t\, d\v^2 +d\t^2.
	\end{equation}
 \noindent
As is well known, $\alpha(r,t)$ and $\beta(r,t)$ have to be determined by solving Einstein's equations.
To do that, we impose that the energy-momentum tensor $T_{\mu\nu}$ be determined only by the Maxwell 
tensor $F_{\mu\nu}$, because we are supposing to have a point particle with charge $Q$ and mass $M$ at 
the origin of the frame. The Einstein equations then read:
	\begin{equation}
	\label{B3}
	R_{\mu\nu}-g_{\mu\nu}R=8\pi G T_{\mu\nu}= 8\pi G \left[\frac{1}{4\pi} \left(F_{\mu\rho} 	
	F_{\nu}^{\rho}-\frac{1}{4} g_{\mu\nu} F_{\rho\sigma} F^{\rho\sigma}\right)\right],
	\end{equation}
\noindent
where $F_{\mu\nu}$ must obey the following Maxwell equations:
	\begin{align}
	& g^{\mu\nu}\nabla_{\mu}F_{\nu\rho}=0 \label{B4}\\
	& \nabla_{\mu}F_{\nu\rho}+\nabla_{\nu}F_{\rho\mu}+
	\nabla_{\rho}F_{\mu\nu}=0 \label{B5} 	\\
	& (\text{obviously: }~~\nabla_{\mu}V^{\alpha}\equiv\p_{\mu}V^{\alpha}+
 	\Gamma^{\alpha}_{\mu\sigma}V^{\sigma}). \label{B5b}
	\end{align}
\noindent
The spherical symmetry of the problem implies that the only non vanishing components are:
	\begin{equation}
	\label{B6}
	F_{tr}=-F_{rt}=-\frac{Q}{r^2}.	
	\end{equation}
\noindent
Then, the solution of the Einstein equations is:
	\begin{align}
	& ds^2=-\Delta dt^2 + \Delta^{-1}dr^2 + r^2d\Omega^2 \label{B7}\\
	& \text{with }~~\Delta=1-\frac{2GM}{r}+\frac{GQ^2}{r^2}, \label{B8}	
	\end{align}
\noindent
which is just the Reissner-Nordstr{\o}m metric. The singularities of this metric are immediately found:
	\begin{align}
	& r=0 & \Rightarrow & ~~~\text{essential singularity} \\
	& \Delta=0 & \Rightarrow & ~~~r_{\pm}=GM \pm \sqrt{G^2M^2-GQ^2}~~~\text{(event horizons).}
	\end{align}
\noindent
The presence of one, two or no event-horizon depends on the sign of $( G^2M^2-GQ^2)$ 
(from now on we put $G=1$ and use the so-called {\it geometric units}). Then we can distinguish three cases:
	\begin{enumerate}
	\item $M^2<Q^2$: {\it NAKED SINGULARITY} \\
	This possibility is ruled out by the principle of cosmic censorship which assumes that a 
	gravitational collapse cannot lead to a naked singularity.  
	\item $M^2>Q^2$: {\it TWO REAL SEPARATE SOLUTIONS} \\
	They are situated respectively at:
		\begin{align}
		& r_- = M-\sqrt{M^2-Q^2} \\
		& r_+ = M+\sqrt{M^2-Q^2}. 
		\end{align}
	\noindent
	There are two event horizons, the first inside the second one. The only peculiar feature 	
	of the resulting geometry is that in the region between $r_-$ and $r_+$ you are not 	
	allowed to reverse your motion: if your direction is $r_+\longrightarrow r_-$, you must 	
	keep it until you reaches $r_-$ and vice versa.
	\item $M^2=Q^2$: {\it TWO COINCIDENT SOLUTIONS} \\
	This is called the {\it extreme} Reissner-Nordstr{\o}m black hole. Its event horizon is situated at:
		\begin{equation}
		r_-=r_+=M
		\end{equation}     
	\noindent
	and the metric (\ref{B7}) in this case reduces to:
	\begin{equation}
	ds^2=-\left(1-\frac{M}{r}\right)^2 dt^2 + \left(1-\frac{M}{r}\right)^{-2}dr^2 +
	r^2d\Omega^2.
	\end{equation}
	\end{enumerate}
\section{Killing Vectors}
\label{app:Killing}
\noindent
In this section we make a brief introduction about the Killing Vectors and their meaning in General Relativity. 
Consider a Riemannian (or Lorentzian) manifold $\cal M$ and a vector field $X=X^{\mu}\p_{\mu}$ on it. If the following 
transformation of coordinates:
	\begin{equation}
	\label{B9}
	x^{\mu}\longrightarrow y^{\mu}=x^{\mu}+\e X^{\mu}
	\end{equation}
\noindent
is an {\it isometry} for the metric g (i.e. g is invariant under this change), then $X$ is called a 
{\it Killing Vector} of the metric at hand. Let us work out an explicit condition for Killing Vectors. 
Under a general change of coordinates the metric coefficients $g_{\mu\nu}$ change according to:
	\begin{equation}
	\label{B10} 
	g^{\prime}_{\mu\nu}(y)= \frac{\p x^{\alpha}}{\p y^{\mu}}\frac{\p x^{\beta}}{\p 	
	y^{\nu}}\,g_{\alpha\beta}(x),
	\end{equation} 
\noindent
in such a way that the isometry condition becomes:
	\begin{equation}
	\label{B11} 
	g^{\prime}_{\mu\nu}(y)= g_{\mu\nu}(y)= \frac{\p x^{\alpha}}{\p y^{\mu}}\frac{\p 	
	x^{\beta}}{\p y^{\nu}}\,g_{\alpha\beta}(x).
	\end{equation} 
\noindent
If we use Eq.(\ref{B9}) in Eq.(\ref{B11}), after few steps we arrive at:
	\begin{equation}
	\label{B12}
      \p_{\mu} g_{\alpha\beta}(x) X^{\mu}+ g_{\mu\beta}(x)\p_{\alpha}X^{\mu}+
      g_{\alpha\nu}(x)\p_{\beta}X^{\nu}= 0.
	\end{equation}
\noindent
We can rewrite the previous formula in two equivalent forms:
	\begin{align}
	\label{B13}
	& {\mathscr L}_{\s X}g=0 \\
	& \nabla_{\mu}X_{\nu}-\nabla_{\nu}X_{\mu}=0, \label{B14}
	\end{align} 
\noindent
where ${\mathscr L}_{\s X}$ is the {\it Lie derivative} along $X$ and $\nabla_{\mu}$ is the usual
covariant derivative introduced in Eq.(\ref{B5b}). 

There are some properties satisfied by Killing Vectors:
	\begin{enumerate}
	\item If $\xi_1$ and $\xi_2$ are Killing vectors, also $\alpha\xi_1+\beta\xi_2$ and
 	$\big[\xi_1,\xi_2\big]$ (where the commutator is the {\it Lie bracket}) are Killing
 	vectors.
	\item One can show that an $N$-dimensional Riemannian manifold can have at most
 	$N(N+1)\over 2$ independent Killing vectors (when this happens the manifold is called {\it
 	maximally symmetric}). For example the 4-dim. Minkowski space-time admits 10 Killing
  	vectors, which correspond to the generators of the Poincar\'e group. 
	\item If $\xi=\xi^{\mu}\p_{\mu}$ is a Killing vector, then the product:
		\begin{equation}
		C(\l):=\xi^{\mu}\frac{dx_{\mu}(\l)}{d\l}\equiv\xi^{\mu}\dot{x}_{\mu}(\l)
		\end{equation}
	\noindent
	is constant along a geodesic $x_{\mu}(\l)$. In fact we have: 
		\begin{multline}
		\frac{DC(\l)}{D\l}:=\dot{x}^{\nu}(\l)\nabla_{\nu}C(\l)=\dot{x}^{\nu}(\dot{x}^{\mu}\nabla_{\nu}\xi_{\mu}
		+ \xi_{\mu}\nabla_{\nu}\dot{x}^{\mu}(\l))= \\
		=\dot{x}^{\nu}\dot{x}^{\mu}\nabla_{\nu}\xi_{\mu}+
		\xi_{\mu}\dot{x}^{\nu}\nabla_{\nu}\dot{x}^{\mu}=0,
		\end{multline}
	\noindent
	where in the last step the first term vanishes because of (\ref{B14}) and the second is zero because 
	$x(\l)$ is a geodesic and therefore satisfies:
		\begin{equation}
		\frac{D\dot{x}^{\mu}}{D\l}=\dot{x}^{\nu}\nabla_{\nu}\dot{x}^{\mu}=0.  
		\end{equation}
	\item Consider the action of a point particle in a gravitational field:
		\begin{equation}
		S=\sqrt{g_{\mu\nu}\dot{x}^{\mu}\dot{x}^{\nu}};
		\end{equation}
	\noindent
	since $p_{\alpha}=\displaystyle\frac{\dot{x}^{\alpha}}{\sqrt{g_{\mu\nu}\dot{x}^{\mu}
	\dot{x}^{\nu}}}$ and $g_{\mu\nu}\dot{x}^{\mu}\dot{x}^{\nu}$ is constant along a geodesic, property
	3 implies that $p_{\alpha}\xi^{\alpha}$ is conserved along a geodesic. This law of conservation is a 
	generalization of what is well known in Classical Mechanics. In fact, if $g_{\mu\nu}(x)$ is independent 
	of a coodinate $x^{\tau}$, then ${\mathscr L}_{\zeta}g=0$ where $\zeta=\frac{\p}{\p x^{\tau}}$, which 
	means that $\zeta$ is a Killing vector. According to property 3, we have that $p_{\mu}\zeta^{\mu}=
	p_{\tau}$ is constant along a geodesic, and so we conclude that the momentum conjugate to the cyclic 
	coordinate $x^{\tau}$ is conserved.   
	\end{enumerate}
\section{De Sitter and Anti-De Sitter Space-Time}
\label{app:Desitt}
\noindent
In the previous Section we have defined what a Killing vector is and when a manifold is called {\it maximally
symmetric}. In general one can show that a maximally symmetric manifold is characterized by the following
equation:
  	\begin{equation}
  	\label{AdS1}
	R_{\s \alpha\beta\gamma\delta}=\frac{1}{N(N-1)}R
	(g_{\s \alpha\gamma}g_{\s\beta\delta}-g_{\s\alpha\delta}g_{\s\beta\gamma})
	\end{equation}
\noindent 
where $R_{\s\alpha\beta\gamma\delta}$ is the curvature tensor and $R$ is the Ricci scalar. Moreover, we can prove
that from Eq.(\ref{AdS1}) together with the Bianchi identities 
  	\begin{equation}
  	\label{AdS2}
	\nabla_{\s\alpha}\left(R^{\s\alpha}_{\s\gamma} - \frac{1}{2}\delta^{\s\alpha}_{\s\gamma}\right) = 0
	\end{equation}
\noindent 
it follows that $R=const$. In fact from Eq.(\ref{AdS1}) we easily get:
   	\begin{equation}
  	\label{AdS3}
	R_{\s\alpha\gamma}=\frac{1}{N}R g_{\s\alpha\gamma};
	\end{equation}
\noindent 
and if we insert (\ref{AdS3}) in (\ref{AdS2}) we get:
  	\begin{equation}
  	\label{AdS4}
	\left(\frac{1}{N}-\frac{1}{2}\right)\p_{\s\gamma}R = 0.
	\end{equation}
\noindent
The previous equation tells us that if $N>2$ (even if it can be shown --- in a more elaborate way --- that the
same result holds also for $N=2$) we have that $R$ is constant. Thus the maximally symmetric spaces are uniquely
specified by a curvature constant $R$ and by the number of positive (or negative) eigenvalues of the metric
$g_{\s\alpha\beta}$. Now let us focus on the case $N=4$ and consider again Eq.(\ref{AdS3}) which becomes:
   	\begin{equation}
  	\label{AdS5}
	R_{\s\alpha\gamma}=\frac{1}{4}R g_{\s\alpha\gamma},
	\end{equation}
\noindent
from which we obtain:
   	\begin{equation}
  	\label{AdS6}
	G_{\s\alpha\beta}=R_{\s\alpha\beta}-\frac{1}{2}g_{\s\alpha\beta}R=-\frac{1}{4}g_{\s\alpha\beta}R.
	\end{equation}
\noindent
The space with $R=0$ is the Minkowski space-time, the space with $R>0$ is called {\it De Sitter} space-time and
that with $R<0$ is called {\it Anti-De Sitter} space-time.
\subsection{De Sitter ($dS$) space-time}
\noindent
We can visualize this space as the following hyperboloid in $\mathbb{R}^5$:
	\begin{equation}
	\label{DS}
	-v^2+w^2+x^2+y^2+z^2=\alpha^2
	\end{equation}
\noindent
with metric:
	\begin{equation}
	\label{DS2}
	ds^2=-dv^2+dw^2+dx^2+dy^2+dz^2.
	\end{equation}
\noindent 
We can deduce from (\ref{DS2}) that the topology of this space is $\mathbb{R}\times\mathbb{S}^3$
($\mathbb{R}$ is parameterized by $v$ and $\mathbb{S}^3$ by $w,x,y,z$) and we can introduce a new set of
coordinates $(t,\chi,\t,\phi)$ defined by the following relations:
	\begin{align}
	\label{DS3}
	& v:=\alpha\sinh(\alpha^{-1}t); & & w:=\alpha\cosh(\alpha^{-1}t)\cos\chi; \\
	& x:=\alpha\cosh(\alpha^{-1}t)\sin\chi\cos\t; & &
	y:=\alpha\cosh(\alpha^{-1}t)\sin\chi\sin\t\cos\phi; \\
	& z:=\alpha\cosh(\alpha^{-1}t)\sin\chi\sin\t\sin\phi. && \label{DS3b}
	\end{align}
\noindent
With this choice of variables we obtain that the metric (\ref{DS2}) takes the form:
	\begin{equation}
	\label{DS4}
	ds^2=-dt^2+\alpha^2\cosh^2(\alpha^{-1}t)\big[d\chi^2+\sin^2\chi(d\t^2+\sin^2\t\;d\phi^2)\big].
	\end{equation}
\subsection{Anti-De Sitter ($AdS$) space-time}

\noindent
The Anti-De Sitter space $AdS_4$ can be represented as the following hyperboloid
	\begin{equation}
	\label{ADS}
	-u^2-v^2+x^2+y^2+z^2=1
	\end{equation}
\noindent
embedded in $\mathbb{R}^5$ with the following metric:
	\begin{equation}
	\label{ADS2}
	ds^2=-du^2-dv^2+dx^2+dy^2+dz^2.
	\end{equation}
\noindent
As we did in the previous case, we can introduce the variables $(t,\chi,\t,\phi)$ defined as in
(\ref{DS3})-(\ref{DS3b}) and we see that the metric (\ref{ADS2}) becomes
	\begin{equation}
	\label{ADS3}
	ds^2=-dt^2+\cos^2 t\big[d\chi^2+\sinh^2\chi(d\t^2+\sin^2\t\;d\phi^2)\big].
	\end{equation}
The topology of the metric (\ref{ADS2}) is given by $\mathbb{S}^1\times\mathbb{R}^3$ ($\mathbb{S}^1$
is parameterized by ($u,v$) while $\mathbb{R}^3$ by ($x,y,z$)) and the symmetry group is $SO(3,2)$,
which leaves invariant the LHS of (\ref{ADS}). In Appendix \ref{app:Conf} we have seen that
$SO(3,2)\cong\text{\it Conf}(3)$
and by the same strategy it is possible to show that $SO(2,1)\cong\text{\it Conf}(1)$, where $SO(2,1)$ is the
symmetry group of the $AdS_2$ space.
\chapter{Superspace Formulation: Mathematical Details}
\markboth{E. Superspace Formulation: Mathematical Details}{}

\section{Details of the derivation of Eqs.(\ref{CME73})-(\ref{CME76})}
\label{app:super1}

\noindent
In this Appendix we are going to show the detailed calculations leading to\break
Eqs.(\ref{CME73})-(\ref{CME76}).
The reader may have noticed the  similarity between the charge
$\QH$ (\ref{CPI28}) and the $\QD$,$\QK$ of Eqs.(\ref{CME44})-({\ref{CME47}).
We say ``similarity" because all of them are made of
two pieces, the first is the $\Qb$ for all of them. It is easy to show
that also the second pieces can be put in a similar form. Like for $\QH$ the second
piece had the form $\NH=c^{a}\partial_{a}H$, it is
easy to show that also $\QD$, and $\QK$ can be put in the form:

	\begin{equation}
	\label{C-1}
	\QD=\Qb-2\gamma~\ND;~~~~~~~\QK=\Qb-\alpha~\NK
	\end{equation} 

\noindent
where $\ND$ and $\NK$ are respectively:

	\begin{equation}\label{C-2}
	\ND=c^{a}\partial_{a}D_0;~~~~~~~
	\NK=c^{a}\partial_{a}K_0
	\end{equation}

\noindent
with the $D_0$ and $K_0$ given\footnote{Actually we take the classical
version of (\ref{CME17}) as we are doing Classical Mechanics.} by 
Eqs.(\ref{CME17})(\ref{CME18}).
\noindent
So all the three operators $(\NH,\ND,\NK)$ could be put in the general form:

	\begin{equation}\label{C-3}
	N_{\scriptscriptstyle X}=c^{a}\partial_{a}X
	\end{equation}

\noindent
where $X$ is either $H,D_0$ or $K_0$. In the case of $D_0$ and $K_0$,  $X$ is quadratic
in the variables~$\v^{a}$:

	\begin{equation}
	\label{C-4}
	X={1\over 2}X_{ab}\v^{a}\v^{b}
	\end{equation}

\noindent
where $X_{ab}$ is a constant $2\times 2$ matrix.

In order to find ${\widehat N}_{\scriptscriptstyle X}$ (that is the superspace
version of ${ N}_{\scriptscriptstyle X}$) we should use  Eq.(\ref{CPI48}) where 
$Q$ is now our operator $N_{\scriptscriptstyle X}$. From the expression of 
~$N_{\scriptscriptstyle X}$ we get for ~$\delta\Phi^{a}(t,\theta,{\bar\theta})$ of Eq.(\ref{CPI48}):

	\begin{equation}\label{C-5}
	\delta\Phi^{a}(t,\theta,{\bar\theta})={\bar\theta}\omega^{ab}({\bar\varepsilon}
	\partial_{b}X)+i{\bar\theta}\theta\omega^{ab}(i\bar\varepsilon c^{d}\partial_{d}\partial_{b}X)
	\end{equation}

\noindent
where ${\bar\varepsilon}$ is the anticommuting parameter associated to the transformation.
\noindent
Given the form of $X$ (see Eq.(\ref{C-4}) above), we get for (\ref{C-5}):

	\begin{equation}\label{C-6}
	\delta\Phi^{a}(t,\theta,{\bar\theta})={\bar\theta}{\bar\varepsilon}\omega^{ab}X_{bd}[\phi^{d}+
	\theta c^{d}].
	\end{equation}

\noindent
Note that, using superfields,  the above expression  can be written as:

	\begin{equation}\label{C-7}
	\delta\Phi^{a}(t,\theta,{\bar\theta})=-{\bar\varepsilon}\omega^{ab}X_{bd}\bar{\theta}\Phi^{d}(t,\theta,
	{\bar\theta}).
	\end{equation}

\noindent
So we obtain from Eq.(\ref{CME68}) that the superspace expression of $N_{\scriptscriptstyle X}$ is

	\begin{equation}\label{C-8}
	({\mathscr N}_{\scriptscriptstyle X})^a_d=\omega^{ab}X_{bd}{\bar\theta}.
	\end{equation}

\noindent The same kind of analysis we have done here for the $\QD$ and $\QK$ can be done also
for the $\QBD$ and $\QBK$. They can be written as:

	\begin{equation}\label{C-9}
	\QBD=\QBb+2\gamma~{\overline N}_{\scriptscriptstyle D};~~~~~ \QBK=\QBb+\alpha~{\overline
	N}_{\scriptscriptstyle K};
	\end{equation}
\noindent
with 

	\begin{equation}\label{C-10}
	{\overline N}_{D}={\bar c}_{a}\omega^{ab}\partial_{b}D;~~~~~~~~{\overline N}_{K}={\bar
	c}_{a}\omega^{ab}\partial_{b}K;
	\end{equation}

\noindent
and the superspace representation of the ${\overline N}_{\scriptscriptstyle X}$
turns out to be:

	\begin{equation}\label{C-11}
	({\ov{\mathscr N}}_{\scriptscriptstyle X})^a_b=\omega^{ac}X_{cb}~\theta.
	\end{equation}

\noindent Remembering the form of the $D_0$ and $K_0$ functions in their classical version 
(see Eqs.(\ref{CME17})(\ref{CME18})) and comparing it with the general form of $X$ of
Eq.(\ref{C-4}) above,
we get from Eqs.(\ref{C-2})(\ref{C-3})  that the matrices $X_{ab}$  associated to $D$ and $K$ 
are\footnote{We will call this
form of $X_{ab}$ as $D_{ab}$, and the one associated to $K_0$ as $K_{ab}$, to stick to the
conventions of Eqs.(\ref{CME73})-(\ref{CME76}).} exactly those of Eq.(\ref{CME77}). 
This is precisely what we wanted to prove.

\section{Representation of H,D,K in superspace}
\label{app:super2}

\noindent
In this Appendix we will reproduce the calculations which provide
the superspace representations of the operators $(H,D,K)$ contained in {\bf TABLE 9}.
We will start first with the operators at time $t=0$ which are listed in 
Eqs.(\ref{CME16})-(\ref{CME18}). Using Eq.(\ref{CPI48}) let us first do the variations
$\delta_{\scriptscriptstyle(H,D,K)}\Phi^{a}$. As $(H,D_{0},K_{0})$ 
contain only $(\v^{a})$ their action will affect
only the $\lambda_{a}$ field contained in the superfield $\Phi$:

	\begin{align}
	\delta_{\scriptscriptstyle H}\lambda_{q}& = \varepsilon
	[H,\lambda_{q}]=-i\varepsilon{g\over q^{3}}\\
	\delta_{\scriptscriptstyle H}\lambda_{p}& = \varepsilon
	[H,\lambda_{p}]=i\varepsilon p\\
	\delta_{\scriptscriptstyle D_{0}}\lambda_{q}& = \varepsilon
	[D_{0},\lambda_{q}]=-{i\over 2}\varepsilon p\\
	\delta_{\scriptscriptstyle D_{0}}\lambda_{p}& = \varepsilon
	[D_{0},\lambda_{p}]=-{i\over 2}\varepsilon q\\
	\delta_{\scriptscriptstyle K_{0}}\lambda_{q}& = \varepsilon
	[K_{0},\lambda_{q}]=i\varepsilon q\\
	\delta_{\scriptscriptstyle K_{0}}\lambda_{p}& = \varepsilon
	[K_{0},\lambda_{p}]=0.
	\end{align}

\noindent Considering that the two superfields are:

	\begin{align}
	\Phi^{q}& = q+\theta~c^{q}+{\bar\theta}{\bar c}_{p}+i{\bar\theta}\theta\lambda_{p};\\
	\Phi^{p}& = p+\theta~c^{p}-{\bar\theta}{\bar c}_{q}-
	i{\bar\theta}\theta\lambda_{q};
	\end{align}

\noindent
it is very easy to see that the $\mathscr{O}$-operators in the
RHS of Eqs.(\ref{CPI48}) can only be the following:

	\begin{align}
	{\mathscr H} & =  {\bar\theta}\theta{\partial\over\partial t};\label{h}\\
	{\mathscr D}_{0} & =  -\frac{1}{2}{\bar\theta}{\theta}\sigma_{3};\\
	{\mathscr K}_{0} & =  -{\bar\theta}\theta \sigma_{-}. \label{k}
	\end{align}

\noindent
Next we should pass to the representation of the time-dependent
operators which are related to the time-independent ones by
Eqs.(\ref{CME78})-(\ref{CME80}). Also for the  superspace representation 
there will be the same relations between the two set of operators,
that means:

	\begin{align}
	{\mathscr H} &=  {\mathscr H}_{0};\\
	{\mathscr D} & = t {\mathscr H}+{\mathscr D}_{0};\\
	{\mathscr K} & = t^{2}{\mathscr H}+2t{\mathscr D}_{0}+{\mathscr K}_{0}.
	\end{align}
\noindent
Using the above relations and the expressions obtained
in Eqs.(\ref{h})-(\ref{k}), it is easy to reproduce the last three
operators contained in {\bf TABLE 9}.
\newpage
\markboth{Bibliography}{Bibliography}
\addcontentsline{toc}{chapter}{\numberline{}Bibliography}


\begin{thebibliography}{99}

\bibitem{Marsd}
R.Abraham and J.Marsden, "{\it Foundations of Mechanics}" Benjamin, New
York 1978;
\bibitem{Planck}
A.A.Abrikosov (jr.), E.Gozzi, Nucl.Phys. B (Proc.Supp.) {\bf vol.88} 369 (2000); 
\bibitem{Alv}
E.Alvarez, Phys. Rev.D {\bf 2} 320 (1984);
\bibitem{Alvarez}
L. Alvarez-Gaum\'e, Jour. Phys. {\bf A16} (1983) 4177; \\
L. Alvarez-Gaum\'e and E. Witten, Nucl. Phys. {\bf B234} (1984) 269; \\
L. Alvarez-Gaum\'e, Bonn Summer School Lectures (1984), unpublished.
\bibitem{Avez}
V.I.Arnold and A.Avez ``{\it Ergodic problems of classical mechanics}",
W.A.Benjamin Inc. (1968);
\bibitem{Baul}
L.Baulieu, B.Grossman, R.Stora, Phys.Lett. {\bf 180B} 95 (1986); \\
L.Beaulieu et al., Phys. Lett. {\bf 275B} 315 (1992);ibid.{\bf 275B} 323 (1992);\\
L.Alvarez-Gaum\'e and L.Baulieu, Nucl. Phys. {\bf B212}, 255 (1982); \\
F.R.Ore, Jr. P.Van Nieuwenhuizen, Nucl. Phys. {\bf B204}, 317 (1982);  
\bibitem{Baye}
D. Baye, Phys. Rev. Lett. {\bf 58} (1987) 2738.
\bibitem{Berline}
N.Berline, E.Getzler, M.Vergne, ``{\it Heat Kernels and Dirac Operators}"
Springer-Verlag, Berlin 1996.
\bibitem{Blau}
M.Blau, E.Keski-Vakkuri, A.Niemi, Phys.Lett. {\bf 264 B} 92 (1990);
\bibitem{Brin}
L.Brink et al.Phys.Lett.B {\bf 64} 435 (1976);
\bibitem{Cartan}
H. Cartan, {\it Colloque de Topologie (Espaces Fibr\'es)}, Brussels (1950), CBRM 15-56;
\bibitem{KAL}
P.Claus et al. Phys. Rev.Lett. 81 (1998) 4553, hep-th/9804177.
\bibitem{ColMand}
S. Coleman and J. Mandula, Phys. Rev. {\bf 159} (1967) 1251.
\bibitem{Cooper}
F. Cooper and B. Freedman, Ann. Phys. {\bf 146} (1983) 262.
\bibitem{DFF}
V.de Alfaro, S.Fubini and G.Furlan, Nuovo Cimento, 34A (1976) 569.
\bibitem{DeAz}
J.~A.~de Azcarraga and J.~Lukierski, Phys.\ Lett.\ B {\bf 113} (1982) 170;
\bibitem{chiccode}
E. Deotto, hep-th/0109103;
\bibitem{DG}
E.~Deotto and E.~Gozzi,
Int.\ J.\ Mod.\ Phys.\ A {\bf 16} (2001) 2709;
\bibitem{DFG}
E.Deotto, G.Furlan, E.Gozzi, Phys. Lett.{\bf B 481} 315 (2000); \\
E.Deotto, G.Furlan, E.Gozzi, Jour. Math. Phys. {\bf 41} 8083 (2000);
\bibitem{Hilb}
E.Deotto, E.Gozzi, D.Mauro, {\it  work in progress};
\bibitem{DiFrancesco}
P. Di Francesco, P. Mathieu and D. Senechal, ``Conformal Field Theory", 
New York, NY, Springer-Verlag, 1996;
\bibitem{Vinc}
P.A.M.Dirac, Canad.Jour. Math. {\bf 2} 129 (1950); \\
A.Hanson, T.Regge and C.Teitelboim , ``{\it Constrained Hamiltonian
Systems}", Accademia Nazionale dei Lincei 1976; \\
K.Sundermeyer, ``{\it Constrained Dynamics}", Springer-Verlag Berlin,
Heidelberg, New York 1982;
\bibitem{Duit}
J.J.Duistermaat and G.J.Heckman, Inv. Math. {\bf 69} 259 (1982);{\bf 72} 153
(1983); \\
M.F. Atiyah and R. Bott, Topology {\bf 23} 1 (1984);
\bibitem{Pavao}
L.D. Faddeev, Theor. Math.Phys.{\bf 1} 1 (1969); \\
P. Senjanovic, Ann.Phys.{\bf 100} 227 (1976);
\bibitem{Kleinert}
P.~Fiziev and H.~Kleinert,
``Anholonomic Transformations of Mechanical Action Principle'', gr-qc/9605046;\\
H.~Kleinert,
Lectures presented at the 1996 Cargese Summer School on FUNCTIONAL INTEGRATION: BASICS AND APPLICATIONS,
quant-ph/9612040;
\bibitem{FUB}
S. Fubini and E. Rabinovici, Nucl. Phys. B245 (1984) 17;
V.P.Akulov and A.I.Pashnev, Theor.Mat.Fiz. 56 (1983) 344.
\bibitem{Fuji}
K.~Fujikawa, Phys.Rev.D21:2848,1980, Erratum-ibid.D22:1499,1980; 
\bibitem{Gelfand}
Y.A. Gel'fand and E.P. Likhtman, JETP Lett. {\bf 13} (1971) 452 (English p. 323);
\bibitem{Gerv}
J.L.~Gervais and B.~Sakita, Nucl. Phys. B34, 427 (1971);
\bibitem{Goldstein}
H.~Goldstein, ``Classical Mechanics" 2nd ed., Addison Wesley (1980);
\bibitem{GozEn}
E.~Gozzi, Phys. Rev. {\bf D}28 (1983) 1922; \\
E.~Gozzi, Phys. Rev. {\bf D}30 (1984) 1218;
\bibitem{GozSusy}
E. Gozzi, Phys.\ Lett.\ B {\bf 130} (1983) 183; \\
E. Gozzi, Phys.\ Lett.\ B {\bf 129} (1983) 432;
\bibitem{GozMSA}
E. Gozzi, Phys.\ Lett.\  {\bf 158B} (1985) 489
[Erratum-ibid.\ B {\bf 386} (1985) 495];
\bibitem{Ennio}
E. Gozzi, Phys. Lett. {\bf B201} (1988) 525; \\
E. Gozzi, M. Reuter and W.D. Thacker, Phys. Rev. D {\bf 40} 3363 (1989);\\
E. Gozzi, M. Reuter and W.D. Thacker, Phys. Rev. D {\bf 46} 757 (1992); 
\bibitem{Ergo}
E. Gozzi, M. Reuter, Phys.Lett.{\bf 233B} 383 (1989); \\
E. Gozzi, Prog. Theor. Phys. (Suppl.) {\bf 111}, 115, (1993); \\
E. Gozzi et al.,  Chaos, Solitons and Fractals 
{\bf 2} 441 (1992); ibid. {\bf 4} 653 (1994); ibid. {\bf 4} 1117 (1994); 
\bibitem{Geom}
E. Gozzi, M. Reuter, Phys. Lett. B {\bf 240} (1,2) 137 (1990); \\
E. Gozzi, M. Regini, Phys.Rev.D {\bf 62} 067702 (2000) (hep-th/9903136); \\
E. Gozzi, D. Mauro, Jour.Math.Phys {\bf 41} no.4 (2000) 1916 (hep-th/9907065);
\bibitem{Teitel}
M. Hanneaux, C. Teitelboim, ``Quantization of Gauge Systems", Princeton
Univ. Press, Princeton, N.J. 1992;
\bibitem{Iachello}
F. Iachello, Phys. Rev. Lett. {\bf 44} (1980) 772;\\
F. Iachello, Physica {\bf D15} (1985) 85; \\
R. Casten, Physica {\bf D15} (1985) 99.
\bibitem{Hull}
L. Infeld and T.E. Hull, Rev. Mod. Phys. {\bf 23} (1951) 21.
\bibitem{KAC}
V.G.Kac, "{\it Lie Superalgebras}" Advances in Mathematics 26 (1977) 8;
\bibitem{Kallosh} 
R. Kallosh, Phys. Rev. D{\bf 56} 3515 (1997); \\
M. Hatsuda and K. Kamimura, Nucl Phys. {\bf B490} 145 (1998);  
\bibitem{Koop}
B.O.Koopman, Proc.Nat.Acad.Sci. USA {\bf 17}, 315 (1931); \\
J.von Neumann, Ann.Math. {\bf 33},587 (1932); 
\bibitem{Kostel}
A. Kosteleck\'y and M.M. Nieto, Phys. Rev. Lett. {\bf 53} (1984) 2285; \\
A. Kosteleck\'y, Phys. Rev. {\bf A32} (1985) 1293, 3243; \\
A. Kosteleck\'y, Phys. Rev. Lett. {\bf 56} (1986) 96. 
\bibitem{Landau}
L.D. Landau and E.M. Lifshits, Course of theoretical physics. v.1: Mechanics. 3rd ed.,
London, Pergamon, 1976;
\bibitem{MAL}
J.M.Maldacena, Adv.Theor.Math.Phys. 2 (1998) 231.
\bibitem{Moshe}
M. Moshe and N. Sakai, Phys.\ Rev.\ D{\bf 62}:086004 (2000);
\bibitem{Niem}
A.Niemi and O.Tirkkonen, Jour. Math. Phys.{\bf 35} 6418 (1994);
\bibitem{Parisi}
G. Parisi and N. Sourlas, Nucl. Phys. {\bf B206} (1982) 321.
\bibitem{Salom}
P. Salomonson and J.W. van Holten, Nucl. Phys. {\bf B196} (1982) 509.
\bibitem{Schulman}
L.Schulman, ``{\it Techniques and Applications of Path Integration}", Wiley, New York NY 1981;
\bibitem{Schw}
A.Schwarz, in ``{\it Topics in statistical and theoretical physics}" ed.
R.L.Dobrushin, R.A.Minlos, M.A.Shubin and M.Vershik (AMS, Providence, RI, 1996);
\bibitem{Siegel}
W.~Siegel, Phys.\ Lett.\ B {\bf 128} (1983) 397;
\bibitem{Sorokin}
D.~P.~Sorokin, V.~I.~Tkach, D.~V.~Volkov and A.~A.~Zheltukhin,
Phys.\ Lett.\ B {\bf 216} (1989) 302; \\
D.~P.~Sorokin, V.~I.~Tkach, D.~V.~Volkov, Mod. Phys. Lett. A {\bf 4} (1989) 901;
\bibitem{Stora}
R.~Stora, ``Exercises in equivariant cohomology'', 
published in *Cargese 1996, Quantum fields and quantum space time* 265-279; hep-th/9611114;\\
R.~Stora, ``Exercises in equivariant cohomology and topological theories'', 
published in *Saclay 1996, The mathematical beauty of physics* 51-66; hep-th/9611116;
\bibitem{Bill}
W.D.Thacker, Jour. Math. Phys.{\bf 38} 2389 (1997);
\bibitem{Volkov}
D.V. Volkov and V.P. Akulov, JETP Lett.16 (1972) {\bf 621} (English p. 438);
\bibitem{Wess}
J. Wess and B. Zumino, Nucl. Phys. {\bf B70} (1974) 39;
\bibitem{WittenDG}
E. Witten, J. Diff. Geom. 17 (1982) 661;
\bibitem{Witten}
E. Witten, Nucl. Phys. {\bf B188} (1981) 513;
\bibitem{Witten2}
E. Witten, Nucl. Phys. {\bf B202} (1982) 253;
\bibitem{Topol}
E.Witten, Comm.Math.Phys. {\bf 117} 353 (1988); \\
S. Ouvry, R.Stora, P.Van Baal, Phys. Lett. {\bf B220} 159 (1989).
\end{thebibliography}
\end{document}